\documentclass[a4paper,11pt]{article}
\pdfoutput=1
\usepackage{jheppub}
\usepackage{subfig}
\usepackage{amsmath,amssymb,bm,graphicx,epsf,colordvi}
\usepackage{slashed}
\def\bea#1\eea{\begin{align}#1\end{align}} 
\usepackage{comment}
\usepackage{diagbox}
\usepackage{mathtools}
\usepackage{lipsum}
\usepackage{soul}
\usepackage{enumitem}
\usepackage{braket}
\usepackage{bm}
\allowdisplaybreaks 
\usepackage{color}
\usepackage[dvipsnames]{xcolor}

\usepackage{mathrsfs}
\usepackage{appendix}
\usepackage{bm}
\usepackage{lipsum}

\newcommand{\bef}{\begin{figure}[hbt]\centering}
\newcommand{\eef}{\end{figure}}

\usepackage{mathtools}
\usepackage{booktabs}
\usepackage{graphicx}
\graphicspath{ {./figure/} }
\usepackage{caption}
\usepackage{lipsum}

\newcommand{\beq}{\begin{equation}}
\newcommand{\eeq}{\end{equation}}
\def\bea#1\eea{\begin{align}#1\end{align}}

\def \be  {\begin{equation}}
\def \ee  {\end{equation}}
\def \ba  {\begin{eqnarray}}
\def \ea  {\end{eqnarray}}

\newcommand{\rmd}{\mathrm{d}}

\allowdisplaybreaks

\usepackage{graphicx}
\usepackage{subcaption} 

\def\cE{\mathcal{E}}

\usepackage{multicol}
\usepackage[dvipsnames]{xcolor}

\def\Fig#1{Fig.~{\ref{#1}}}
\DeclareRobustCommand{\Sec}[1]{Sec.~\ref{#1}}
\DeclareRobustCommand{\App}[1]{App.~\ref{#1}}
\DeclareRobustCommand{\Eq}[1]{Eq.~(\ref{#1})}

\usepackage{stackengine}

\title{Imaging the Wakes of Jets with Energy-Energy-Energy Correlators}

\author[a]{Hannah Bossi,}
\affiliation[a]{Massachusetts Institute of Technology, Cambridge, MA, 02139}

\author[b, c]{Arjun Srinivasan Kudinoor,}
\affiliation[b]{Center for Theoretical Physics, Massachusetts Institute of Technology, Cambridge, MA 02139}
\affiliation[c]{DAMTP, University of Cambridge, Cambridge, CB3 0WA, UK}

\author[d]{Ian Moult,}
\affiliation[d]{Department of Physics, Yale University, New Haven, CT 06511}

\author[e]{Daniel Pablos,}
\affiliation[e]{
IGFAE, Universidade de Santiago de Compostela, E-15782 Galicia-Spain}

\author[d]{Ananya Rai,}

\author[b]{Krishna Rajagopal}

\preprint{MIT-CTP-5739}

\abstract{As the partons in a high energy jet propagate through the droplet of quark-gluon plasma (QGP) produced in a heavy-ion collision they lose energy to, kick, and are kicked by the medium. The resulting modifications to the parton shower encode information about the microscopic nature of QGP.  
A direct consequence, however, is that the momentum and energy lost by the parton shower are gained by the medium and, since QGP is a strongly coupled liquid, this means that the jet excites a wake in the droplet of QGP.  After freezeout, this wake becomes soft hadrons with net momentum in the jet direction meaning that what an experimentalist later reconstructs as a jet includes hadrons originating from both the modified parton shower and its wake. This has made it challenging to find experimental observables that provide an unambiguous view of the dynamical response of a droplet of QGP to a jet shooting through it.
Recent years have seen significant substantial advances in the theoretical and experimental understanding of the substructure of jets, in particular, using correlation functions, $\langle \cE(\vec n_1)\cdots \cE(\vec n_k)\rangle$,  of the energy flux operator in proton-proton collisions and, recently, in heavy-ion collisions. 
So far, such studies have focused primarily on the two-point correlator, which allows for the identification of the angular scale of the underlying dynamics. 
Higher-point correlators hold the promise of mapping out the dynamics themselves.
In this paper we perform the first study of the shape-dependent three-point energy-energy-energy correlator in heavy-ion collisions.
Using the Hybrid Model to simulate the interactions of high energy jets with the QGP medium, we show that the three-point correlator presents us with a striking new opportunity.
We find that hadrons originating from wakes are the dominant contribution to the three-point correlator in the kinematic regime in which the three points are well-separated in angle, forming a roughly equilateral triangle. This equilateral region of the correlator
is far from the region populated by collinear vacuum emissions,
making it a canvas on which jet wakes are laid out, where experimentalists can map their shapes. 
Our work provides a key step towards the systematic use of energy correlators to image and unravel the dynamical response of a droplet of QGP that has been probed by a passing jet, and motivates numerous experimental and theoretical studies.
}

\begin{document}

\maketitle

\section{Introduction}\label{sec:intro}

The analysis of experimental measurements of the debris from the explosions of droplets of big bang matter produced in ultra-relativistic collisions of heavy ions at RHIC and the LHC
led to the discovery that this primordial fluid, conventionally called quark-gluon plasma (QGP), is a strongly coupled fluid~\cite{PHENIX:2004vcz,BRAHMS:2004adc,PHOBOS:2004zne,STAR:2005gfr,Gyulassy:2004zy}.
This provides a unique opportunity to study the properties and dynamics of matter governed by quantum chromodynamics (QCD) and indeed there have been a
plethora of QGP-related QCD phenomena that have been analyzed with increasing incisiveness and precision over the past two decades via theoretical and experimental investigations of heavy-ion collisions. 
(For reviews, see Refs.~\cite{Muller:2006ee,Casalderrey-Solana:2007knd,dEnterria:2009xfs,Wiedemann:2009sh,Jacak:2012dx,Muller:2012zq,Heinz:2013th,Shuryak:2014zxa,Akiba:2015jwa,Romatschke:2017ejr,Connors:2017ptx,Busza:2018rrf,Nagle:2018nvi,Cao:2020wlm,Schenke:2021mxx,Cunqueiro:2021wls,Apolinario:2022vzg,Harris:2023tti}.)
Studies of the bulk properties of QGP have led to numerous remarkable insights, building upon the discovery that QGP behaves as a nearly perfect fluid, and have prompted
numerous theoretical developments in the study of strongly coupled QFTs and hydrodynamics. 

As in many other domains of physics, after the discovery and bulk characterization of a new form of strongly coupled matter, many of the most important routes to further understanding involve interrogating it with some high-energy probe (think microscopy or scattering experiments going back to Rutherford),
studying how it deflects or reshapes the probe and how
it responds to being probed. For the droplets of QGP produced in a heavy-ion collision the only possible probes are those produced in the same collision, including high energy jets and heavy quarks. We shall focus here on jets, although the extension of the methods that we are investigating and shall present here to (jets containing) heavy quarks holds promise as well.
In recent years, there have been rapid advances in the modeling and experimental exploration of jets in heavy-ion collisions that plow through a droplet of QGP and
how they differ from jets in proton-proton collisions that form in vacuum. At the same time, there have been rapid advances
in the formulation and calculation of new, more differential, more targeted, jet observables.
The analysis of the modification to the ``shape'' (in both energy and angle) of jets that have propagated through QGP provides unique opportunities to probe the microscopic structure of QGP and to reveal its dynamical response to the passage of a jet.
The internal shape and structure of jets is referred to as jet substructure; the experimental analysis of jet substructure and its {\it ab initio} calculation from the collinear dynamics of QCD
has been extensively developed for jets in vacuum. See Refs.~\cite{Dasgupta:2013ihk,Larkoski:2013eya,Larkoski:2017jix,Asquith:2018igt,Marzani:2019hun} for reviews.  

To extract information about the underlying microscopic interactions of quarks and gluons in a jet with those in a droplet of QGP 
from macroscopic measurements of the energy flux within high energy jets, in a way that allows us to infer conclusions about 
the microscopic structure of QGP, and to understand how a strongly coupled liquid emerges from asymptotically free quarks and gluons
is a highly non-trivial problem.
This is due to the complexity of the perturbative parton shower and its interaction with the strongly coupled medium, as well as the non-perturbative hadronization process. 
The goal of fully exploiting the information about QGP and its structure and dynamics encoded in jets in heavy-ion collisions
will require both theoretical and experimental developments and is already invigorating  interactions between the theoretical and experimental communities. Our goal in this paper is 
only one element of this larger ambition, albeit one that presents substantial opportunity and challenge and thus motivation.

The most direct consequence of the passage of a jet through a droplet of QGP is that the partons in the jet shower lose momentum and energy.  Where does this momentum and energy end up? Very close (in time and space) to the jet parton itself, this question is hard to answer. But, noting that the droplet of QGP through which the jet passes is a strongly coupled liquid, and reflecting upon the fact that this hydrodynamic liquid itself forms very rapidly (within 1 fm$/c$) after the initial collision~\cite{Heinz:2001xi,Heinz:2002un,Kolb:2003dz,Heinz:2004pj,Chesler:2008hg,Chesler:2009cy,Chesler:2010bi,Shen:2010uy,Heller:2011ju,Shen:2012vn,Heller:2012je,Heller:2012km,vanderSchee:2012qj,Heller:2013oxa,Kurkela:2015qoa,Chesler:2015lsa,Chesler:2015fpa,Heller:2016gbp,Kurkela:2018vqr,Kurkela:2018wud,Brewer:2019oha,Brewer:2022vkq,Rajagopal:2024lou} (for reviews, see Refs.~\cite{Casalderrey-Solana:2011dxg,Busza:2018rrf,Schlichting:2019abc,Berges:2020fwq}), the most natural answer to this question is that the momentum and energy lost by the jet is carried by hydrodynamic excitations of the droplet of QGP, namely by a wake that the jet excites in the droplet.
The earliest versions of this idea go back decades~\cite{Casalderrey-Solana:2004fdk} and from a theoretical perspective this is well established both in strongly
coupled gauge theory plasmas where rigorous calculations can be done via holography~\cite{Chesler:2007an,Gubser:2007ga,Gubser:2007ni,Gubser:2007xz,Chesler:2007sv,Chesler:2014jva,Chesler:2015nqz,Rajagopal:2016uip,Brewer:2017fqy,Brewer:2018mpk} and in 
many approaches to modeling
heavy-ion collisions~\cite{Ruppert:2005uz,Renk:2005si,Casalderrey-Solana:2006lmc,Betz:2008ka,Neufeld:2008fi,Li:2010ts,He:2015pra,Casalderrey-Solana:2016jvj,Cao:2016gvr,Chen:2017zte,Tachibana:2017syd,He:2018xjv,Park:2018acg,Chang:2019sae,Casalderrey-Solana:2020rsj,Chen:2020tbl,Cao:2020wlm,Yang:2021qtl,Yang:2022nei,Luo:2023nsi}.
However, by momentum conservation the soft hadrons originating from a jet wake (as the droplet of QGP with the wake in it freezes out) carry momentum in the same direction as the parton shower that excited the wake. Thus, in a detector they are part and parcel of what an experimentalist calls a jet. For this reason, the experimental investigation of this idea has proved to be a challenge.
The question we ask here -- and answer affirmatively! --  is whether the new tools that have been developed (with different goals in mind) for the analysis of jet substructure in vacuum can be deployed on jets in heavy-ion collisions in such a way as to yield a clean arena in which to visualize jet wakes unambiguously and explore their dynamics.

The theoretical problem that we are faced with in heavy-ion collisions is not so different in spirit from that in cosmology. In cosmology, we wish extract from subtle correlations in the cosmic microwave background (CMB), the microscopic details of the underlying inflationary model that created these correlations. In heavy-ion collisions, our goal is to extract, from the high multiplicities of hadrons produced in the collision and in the jets therein,
features of the underlying microscopic interactions of  QGP and of its dynamics as it responds to a jet. In the context of cosmology, following the seminal calculation of the three-point function~\cite{Maldacena:2002vr} as a function of angles on the sky, there has been tremendous effort to understand the structure of correlation functions in the CMB, and to map out the ``shapes" of inflationary models. See, e.g., Refs.~\cite{Babich:2004gb,Arkani-Hamed:2015bza,Arkani-Hamed:2018kmz}.
In the context of collider physics experiments one can measure correlations in the energy flux $\langle \cE(\vec n_1)\cdots \cE(\vec n_k)\rangle$ as a function of the angles of the $\vec n$ vectors. These observables were proposed early on in the study of $e^+e^-$ collisions~\cite{Basham:1978bw,Basham:1977iq,Ellis:1978ty,Basham:1979gh,Basham:1978zq} and were given operator definitions, and shown to be interesting observables in generic theories, in Ref.~\cite{Hofman:2008ar}. More recently, they were 
introduced~\cite{Dixon:2019uzg,Chen:2020vvp,Komiske:2022enw} as practical jet substructure observables, which have since been proposed as tools by which to 
study many aspects of high 
energy QCD~\cite{Chen:2019bpb,Chen:2020adz,Chen:2021gdk,Lee:2022ige,Chen:2022swd,Craft:2022kdo,Lee:2023npz,Lee:2023tkr,Lin:2024lsj}, the Standard Model \cite{Holguin:2022epo,Holguin:2023bjf,Xiao:2024rol}, and nuclear physics \cite{Liu:2022wop,Liu:2023aqb,Cao:2023oef,Li:2023gkh,Liu:2024kqt,Guo:2024jch,Devereaux:2023vjz,Chen:2024nfl,Chen:2024bpj,Andres:2022ovj,Andres:2023xwr,Andres:2023ymw,Barata:2023vnl,Barata:2023zqg,Yang:2023dwc,Barata:2023bhh,Barata:2024nqo,Andres:2024ksi}.  
The experimental measurement of these observables at hadron colliders is now underway~\cite{Fan:2023,Tamis:2023guc,CMS:2024mlf}, and has already yielded the most precise determination of the strong coupling constant from measurements of jet substructure~\cite{CMS:2024mlf}.

Energy correlator observables are particularly appealing in the context of heavy-ion collisions, where, much like in the cosmological context, we do not understand the underlying microscopic model of the interactions of high energy quarks and gluons with  QGP. A natural approach is to therefore map out the ``shapes" in the space of correlators that result from different models for elements of the dynamics of jets in  QGP. A primary advantage of the energy correlator observables, in a general context, is that due to their operator definition, they can be computed at both weak and strong coupling \cite{Hofman:2008ar}. This allows one to understand their shape in different toy QFTs, to develop a qualitative understanding and intuition, and to facilitate the interpretation of experimental measurements. In the specific case of the dynamics of jets in QGP, 
there are a variety of different physical effects which are expected to modify parton showers in medium, including energy loss, medium-induced radiation, elastic and inelastic scattering at larger angles, and the possibility that nearby partons in the shower may or may not be resolved by the medium.
Each of these modifications of the parton shower should 
modify suitably crafted energy correlator observables in appropriate kinematic regions. 
In this context, the fact that at the same time that the parton shower is modified via its passage through  QGP, the momentum and energy that it loses excite a wake in the droplet of QGP that later hadronizes into soft hadrons could be viewed as an annoyance. The soft hadrons from its wake that are reconstructed as a part of any jet in a heavy-ion collision will obscure some of aspects of energy correlator observables that have motivated their study in jets in vacuum, and risk being a confounding effect in efforts to tease out signatures originating from the different physical effects that lead to modifications of parton showers in medium.  
Our goal in this work is to find a region in the space of energy correlators for jets in heavy-ion collisions where hadrons coming from the (modified) parton shower make much less of a contribution than hadrons coming from the wake that the parton shower excited in the droplet of QGP.  Imaging the wake of jets in heavy-ion collisions is a long-held goal that has proved elusive; we shall show that the energy-energy-energy three-point correlator provides us with a kinematic region in which this can be done -- in correlator space -- precisely because it is far from the region populated by the collinear emissions that dominate parton showers.

There have been a number of studies of the energy correlators using different theoretical models for the interactions of high energy partons with a droplet of QGP and consequent modifications to the parton showers for jets in heavy-ion collisions~\cite{Andres:2022ovj,Andres:2023xwr,Andres:2023ymw,Barata:2023vnl,Barata:2023zqg,Yang:2023dwc,Barata:2023bhh,Barata:2024nqo,Andres:2024ksi}. To date, these studies have focused on the two-point energy correlator, which is a function of a single angle.  All studies find an enhancement of the correlator at large angles, with a scale set by the droplet of QGP. However, since the qualitative behavior of the two-point correlator is only sensitive to the scale of the dynamics, all these models give a qualitatively similar behavior. And indeed we shall find that the contribution from jet wakes occupies a similar large angle region in the two-point correlator to 
that identified previously as the place where effects originating from medium-induced modifications of the parton shower are to be found.
This motivates pursuing a broad program of mapping out the ``shapes" of higher-point correlators which are specified by more angular variables
and identifying the regions in correlator space where different theoretical effects make large contributions.

In this paper we take an initial step in this broader direction by performing the first study of the three-point energy-energy-energy correlator inside high energy jets produced in heavy-ion collisions.  We do so in a model that incorporates the soft hadrons produced when the droplet of QGP that contains a wake excited by a jet that has passed through it freezes out, and in which we can turn this wake off and on, and in which jets can be reconstructed and their energy correlators ``measured'' as is done in the analysis of experimental data. Experimentalists can never turn the wake off -- as doing so violates momentum conservation.  But a model in which this can be done can be used to identify the regions in correlator space that are dominated by hadrons originating from jet wakes.
The Hybrid Model~\cite{Casalderrey-Solana:2014bpa,Casalderrey-Solana:2015vaa,Casalderrey-Solana:2016jvj,Hulcher:2017cpt,Casalderrey-Solana:2018wrw,Casalderrey-Solana:2019ubu,Hulcher:2022kmn} 
is well-suited for these purposes.
As has been emphasized by its authors since they first introduced it~\cite{Casalderrey-Solana:2016jvj}, the Hybrid Model implementation of the hadrons originating from jet wakes is crude and in particular yields too many hadrons with $p_{\rm T}<2$~GeV/$c$ and too few hadrons with $p_{\rm T}$ between 2 and 4 GeV/$c$ than seems to be required by comparison to data that was then available.  Our goal, however, is simply to identify a region in the space of energy-energy-energy correlators where jet wakes make a dominant contribution. This defines the arena in which future comparisons between experimental data and model predictions (for example predictions after an improved treatment of the wake that builds upon the work in Ref.~\cite{Casalderrey-Solana:2020rsj} is implemented in the Hybrid model) can reveal the shape of jet wakes.  Looking even further ahead, we can imagine that the comparison of such studies in the collisions of smaller nuclei to those in PbPb collisions may one day tell us how the wakes of jets evolve, as they will have done so for a longer time in larger collision systems.

We shall use our Hybrid Model study to show that jet wakes
are imprinted in a large modification 
to the energy-energy-energy correlator in a regime in which all three angles between the three directions in the energy correlator are comparable, forming a roughly equilateral triangle. We perform numerous tests of this interpretation that go beyond just turning the wake off and on, including changing the energy weights of the correlator, changing the overall angular scale of the equilateral triangle made by the three points in the correlator,
studying correlations between and among hadrons originating from the parton shower or the wake and comparing jets produced in association with a high energy photon ($\gamma$-jets, where there may be only one jet in the event) with inclusive jets in which each event contains at least two jets each with a wake.
We also show that the angular coordinates typically used to specify the two angles that define the energy-energy-energy correlator are well-suited to the collinear regime but that a different set of coordinates is better suited to obtaining a faithful visual representation of the correlator in the equilateral region.
We also highlight some of the non-trivial aspects of translating this investigation into a realistic experimental measurement, and show that making representative choices in this regard does not compromise the central messages of our study.
In a larger sense, our investigation serves as a powerfully illustrative example of one specific case of mapping an underlying physical model of nontrivial dynamics (the dynamics of how a droplet of QGP responds to a jet)
into the shape of an energy correlator (the energy-energy-energy correlator in the equilateral region) that we believe will be the first of many, as comparisons between future such studies and experimental data serve to unravel the dynamics of jets in QGP and ultimately the microscopic structure of QGP itself.

An outline of this paper is as follows. In \Sec{sec:eec_review} we define the different energy correlator observables that we will study in this paper, and discuss their basic theoretical properties and motivation. In \Sec{sec:coords}, we introduce angular coordinates for parametrizing the three-point energy correlator. The three points (three directions) in the correlator define a triangle, whose shape is specified by two angles.  The coordinates that we introduce yield a faithful representation of the shape of the three-point correlator and its modification in heavy-ion collisions in the regime where the triangle defined by the three points is roughly equilateral; the coordinates used in previous literature were designed to best represent the correlator in the regime where two of the three points are nearly collinear.
In \Sec{sec:wake_review}, we review the Hybrid Model for jets propagating in strongly coupled QGP, and discuss the details of our simulations.  In \Sec{sec:proj}, we study the two-point and the three-point {\it projected} correlator, as well as their ratios, and show how both the effects of parton energy loss and the distribution of soft hadrons originating from the wake are imprinted onto the ratio of the projected two- and three-point correlators. In \Sec{sec:shape}, we perform the first study of the full shape-dependent three-point correlator for jets propagating through QGP, showing that the hadrons originating from the wake that the jets excite in the droplet of QGP imprints itself in the three-point correlator in a shape-dependent manner, and is the dominant contribution to the correlator in the equilateral region. In \Sec{sec:exp} we discuss challenges associated with performing a realistic experimental measurement of the energy correlators in heavy ion experiments, and provide some evidence that the compromises that will need to be made as experimental analyses are carried out will not modify our central conclusions.
We conclude and look ahead in \Sec{sec:conc}. We defer a fully detailed discussion of several important points to the Appendices.

\section{Projected  Correlators and Full, Shape-Dependent, Energy Correlators}\label{sec:eec_review}

Energy correlators are a class of observables which measure correlations in the asymptotic flux of energy. They are defined as correlation functions, $\langle \Psi|\cE(\vec n_1) \cE(\vec n_2)\cdots \cE(\vec n_k)|\Psi \rangle$ of the energy flow operator
\cite{Sveshnikov:1995vi,Tkachov:1995kk,Korchemsky:1999kt,Bauer:2008dt,Hofman:2008ar,Belitsky:2013xxa,Belitsky:2013bja,Kravchuk:2018htv}
\begin{align}
    \cE(\vec n_1) = 
    \lim_{r\rightarrow \infty} \int \mathrm{d}t \,r^2 n_1^i \,
    T_{0i}(t,r\vec{n}_1)\,.
    \label{eq:def}
\end{align}
This detector, which is specified by a three vector, $\vec n_1$, detects any particles whose momentum is along the direction specified by the unit vector $\vec n_1$, and measures their energy. Correlation functions of these operators are functions of the angles between the vectors $\vec n_i$ on the celestial sphere. In the hadron collider environment, 
where the detectors are many orders of magnitude larger than the $\sim 10$ fm length scales over which the dynamics of a jet propagating through a droplet of QGP and the wake that results play out, the angles on the celestial sphere are simply the azimuthal and polar angles specifying the directions of the particles (of order 30,000 hadrons in the case of a central -- i.e.~almost head-on -- PbPb collision at the LHC) seen by the detector.
It is common to use the boost invariant angle $\Delta R=\sqrt{\Delta \eta^2 + \Delta \phi^2}$, with $\phi$ the azimuthal angle and $\eta$ the pseudorapidity, as well as to replace the energy with the transverse momentum, $p_{\rm T}$.

Energy correlator observables were first introduced to study the structure of the asymptotic energy flux in $e^+e^-$ colliders in  Refs.~\cite{Basham:1978bw,Basham:1977iq,Ellis:1978ty,Basham:1979gh,Basham:1978zq}. These authors also computed the two-point correlator (also referred to as the Energy-Energy Correlator or EEC) at leading order.  The energy correlators have numerous elegant theoretical properties, which have made them the focus of intense study from the theory community, leading to many recent advances in their calculation and in understanding how they characterize the response of the vacuum in a quantum field theory to a perturbation at one point in spacetime, as via an elementary collision. These studies have been performed in a wide variety of quantum field theories, at both weak and strong coupling, in conformal theories and in asymptotically free theories.
A remarkable property of the energy correlators in this context is that they can be computed using two different approaches. On the one hand, they are directly related to correlation functions of local operators involving the stress tensor of the theory. On the other they can be computed at weak coupling using perturbative form factors or scattering amplitudes.  This has lead to numerous calculations exploring their structure in different states. In perturbation theory, the two-point correlator has been computed at next-to-leading order (NLO) in QCD \cite{Dixon:2018qgp,Luo:2019nig}, and NNLO in $\mathcal{N}=4$ super Yang-Mills (SYM) \cite{Belitsky:2013ofa,Henn:2019gkr}. The three-point correlator has been computed in both the collinear limit \cite{Chen:2019bpb}, and at generic angles \cite{Yan:2022cye,Yang:2022tgm,Yang:2024gcn}, and the four-point correlator was recently computed in $\mathcal{N}=4$ SYM \cite{Chicherin:2024ifn}. The energy correlators have also been computed in $\mathcal{N}=4$ SYM at strong coupling \cite{Hofman:2008ar} using AdS/CFT \cite{Maldacena:1997re,Gubser:1998bc,Witten:1998qj}, as well as more recently in non-trivial backgrounds modelling confinement \cite{Csaki:2024joe}. They have also been computed in heavy states in $\mathcal{N}=4$ SYM \cite{Chicherin:2023gxt} using correlation functions of $1/2$ BPS operators, and in generic large charge states \cite{Firat:2023lbp} using an effective field theory approach. Furthermore, they can be computed non-perturbatively in conformal field theories using the light-ray operator product expansion  \cite{Kravchuk:2018htv,Kologlu:2019mfz,Chang:2020qpj}. These diverse calculations have enabled an understanding and intuition for the structure of higher point correlators at both weak and strong coupling. We believe that the ability to compute the correlators at both weak and strong coupling is crucial for fully interpreting these observables when measured in heavy-ion collisions, but it is also important to note that their application to heavy-ion collisions introduces new elements. A heavy-ion collision starts off with many elementary collisions but that is only the beginning. The many (tens of thousands) partons created at the earliest moment interact strongly with each other, and in particular the partons originating from one elementary collision at one spacetime point interact with those originating from many others,
so as to
form a droplet of QGP which then evolves hydrodynamically. Furthermore, the high energy partons in a jet shower interact with the medium long
after the elementary collision from which the shower originates.
Developing the calculational tools to relate the dynamical processes in heavy-ion collisions to (experimental measurements of) energy correlators in this setting is just 
beginning~\cite{Andres:2022ovj,Andres:2023xwr,Andres:2023ymw,Barata:2023vnl,Barata:2023zqg,Yang:2023dwc,Barata:2023bhh,Barata:2024nqo,Andres:2024ksi}.
Our addition is to focus on how to use energy correlator observables to ``see'' the wakes that the jets excite in the medium through which they propagate.

Without further orientation, 
energy correlator observables beyond the two-point correlator may appear to be complicated functions of multiple angles on the celestial sphere that are hard to interpret.
We shall refer to the largest angle among the $N$ vectors that specify an $N$-point correlator as $R_{\rm L}$. In a two-point correlator $R_{\rm L}$ is the only relevant angle, but for higher-point correlators $R_{\rm L}$ serves as one way to characterize the angular scale of the correlator but its full specification of the $N$ vectors involves additional angles, which by definition are smaller than $R_{\rm L}$. In the case
of the three-point correlator, the 3 vectors form a triangle whose longest side is $R_{\rm L}$.
We are interested in using energy correlators to characterize jets as reconstructed in heavy-ion collisions, which, as noted, include hadrons coming from the parton shower and hadrons coming from the wake. We shall look at jets reconstructed with the anti-$k_{\rm T}$ algorithm~\cite{Cacciari:2008gp,Cacciari:2011ma} with $R=0.8$, and shall always choose $R_{\rm L}<R$.
We want to first look at how the two and three-point correlators depend on $R_{\rm L}$ without regard for the shape of the triangle that fully specifies the three-point correlator. That is, we shall first focus on the ``projected three-point correlator''. We shall then turn to analyzing the full, shape-dependent, three-point correlator.

The projected correlators were introduced in Ref.~\cite{Chen:2020vvp} for the purpose of focusing on the scaling behavior of the $N$-point correlators as a function of the overall angular scale set by $R_{\rm L}$. 
The projected $N$-point correlator, which we denote as ENC$(R_{\rm L})$ is defined by integrating out all the shape information of the correlator, keeping only $R_{\rm L}$ fixed:
\begin{align}
  \label{eq:projection}
\text{ENC}(R_{\rm L}) \equiv \left(\prod_{k=1}^N \int \!  d\Omega_{\vec{n}_k} \right) \delta (R_{\rm L} - \Delta \hat R_{\rm L})
\cdot \frac{1}{(E_{\rm jet})^{(n*N)}} \, \langle  
{\cal E}^{n}(\vec{n}_1) {\cal E}^{n}(\vec{n}_2)  \ldots 
{\cal E}^{n}( \vec{n}_N)\rangle, 
\end{align}
where $n$ represents the energy weighting scheme used. Most of the results in this paper use $n=1.0$, and this should be assumed unless otherwise specified.
In the case of the two-point energy-energy correlator, the projected correlator (E2C) is the same as the full correlator (EEC).
The first nontrivial example is the three-point correlator, where the projected correlator (E3C) depends only on $R_{\rm L}$ whereas the full characterization of the energy-energy-energy correlator (EEEC) requires specifying two additional angles that define the shape of a triangle whose sides can be written in increasing length as $R_{\rm S}$, $R_{\rm M}$, and $R_{\rm L}$.

The scaling behavior of the ENC projected correlators as a function of $R_{\rm L}$ in elementary collisions in QCD is
governed by the dynamics of the parton shower (at larger angles within the jet radius $R$), by
hadronization (at smaller angles), with the transition between these two regimes clearly visible at an angular scale that is of order the ratio between $\Lambda_{\rm QCD}$ and the jet $p_{\rm T}$.
These features of the projected correlators makes them ideal tools to perform precision tests of QCD dynamics in elementary collisions with experimental data. 

The E2C has been experimentally measured in proton-proton collisions at both the LHC~\cite{CMS:2024mlf,Fan:2023} and RHIC~\cite{Tamis:2023guc}, showing good agreement with both Monte Carlo generators and theoretical calculations~\cite{Lee:2022ige}. 
Recent theoretical work has shown that the two-point correlator is also sensitive to the
modification of the parton shower that results from its propagation through QGP in a heavy-ion collision~\cite{Andres:2022ovj,Andres:2023xwr,Andres:2023ymw, Barata:2023vnl,Barata:2023zqg,Yang:2023dwc,Barata:2023bhh,Barata:2024nqo,Andres:2024ksi}, leading to an increased interest in measuring the E2C in heavy-ion collisions.

The E3C has been measured in proton-proton collisions both by CMS~\cite{CMS:2024mlf} and by ALICE~\cite{Fan:2023}. In addition, the ratio of the E3C to the E2C probes the strong coupling constant, $\alpha_{\rm s}$. The recent determination
of $\alpha_{\rm s}$ by the CMS collaboration~\cite{CMS:2024mlf} via their measurements of the E3C/E2C ratio in proton-proton collisions represents the most precise extraction of $\alpha_{\rm s}$ using jet substructure techniques to date. 


\begin{figure}[t]
    \centering
    \begin{minipage}{0.45\textwidth}
        \centering
        \includegraphics[width=\textwidth]{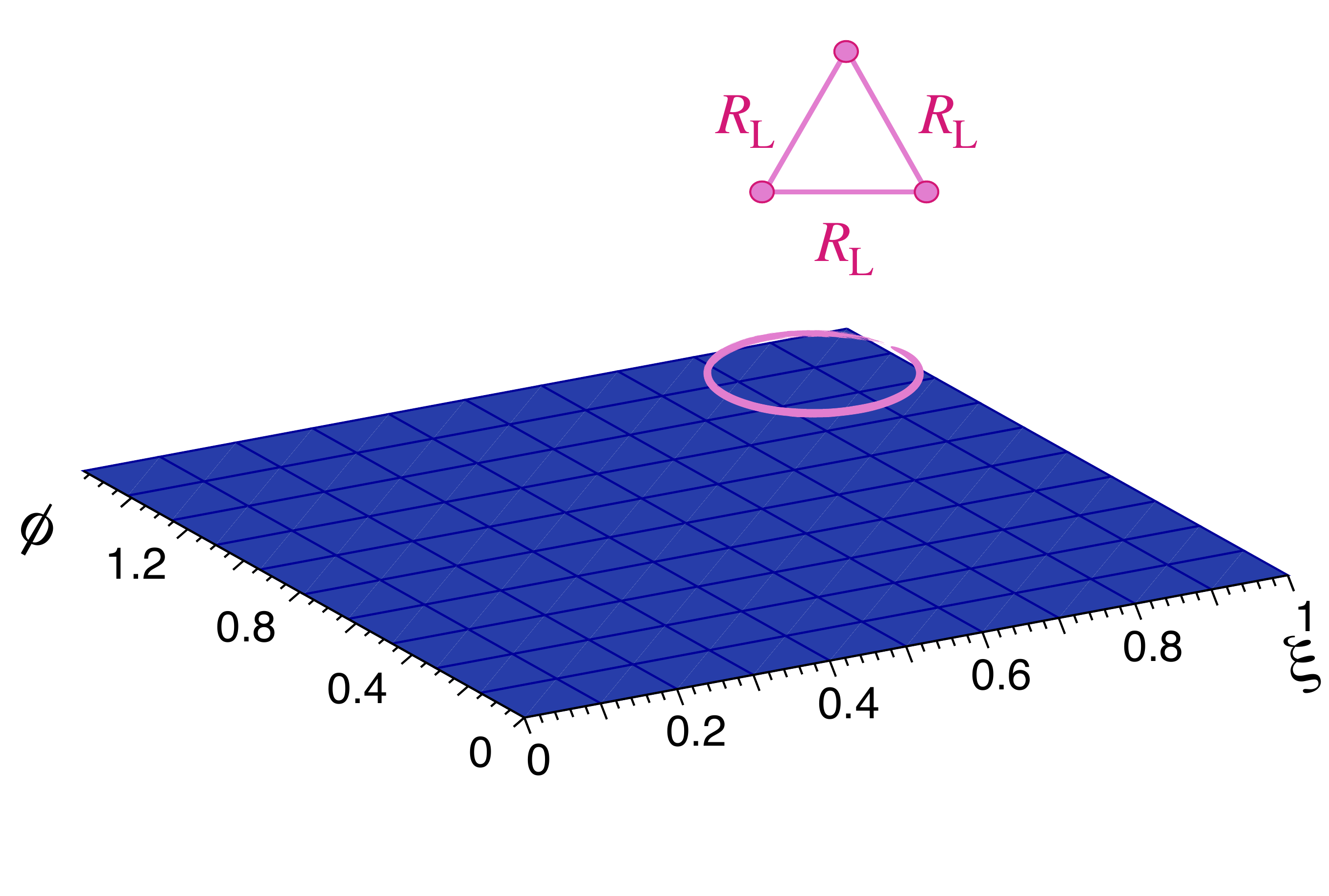}
    \end{minipage}
    \hfill
    \begin{minipage}{0.45\textwidth}
        \centering
        \includegraphics[width=\textwidth]{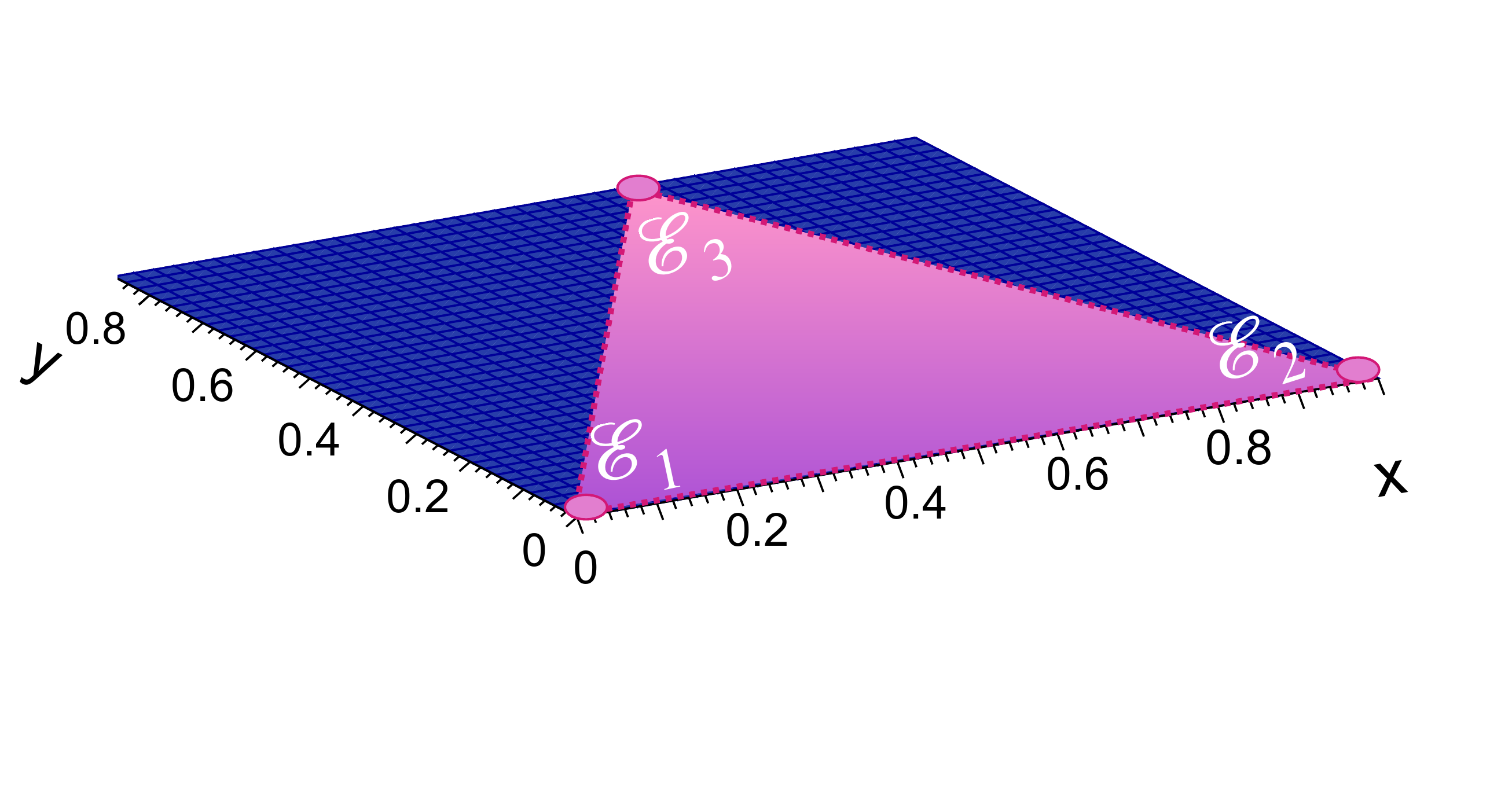}
    \end{minipage}
    \vspace{0.5em}
    \parbox[c]{0.9\textwidth}{
        \centering
       (a) Equilateral \\
    }
    \vspace{1em} 

    \begin{minipage}{0.45\textwidth}
        \centering
        \includegraphics[width=\textwidth]{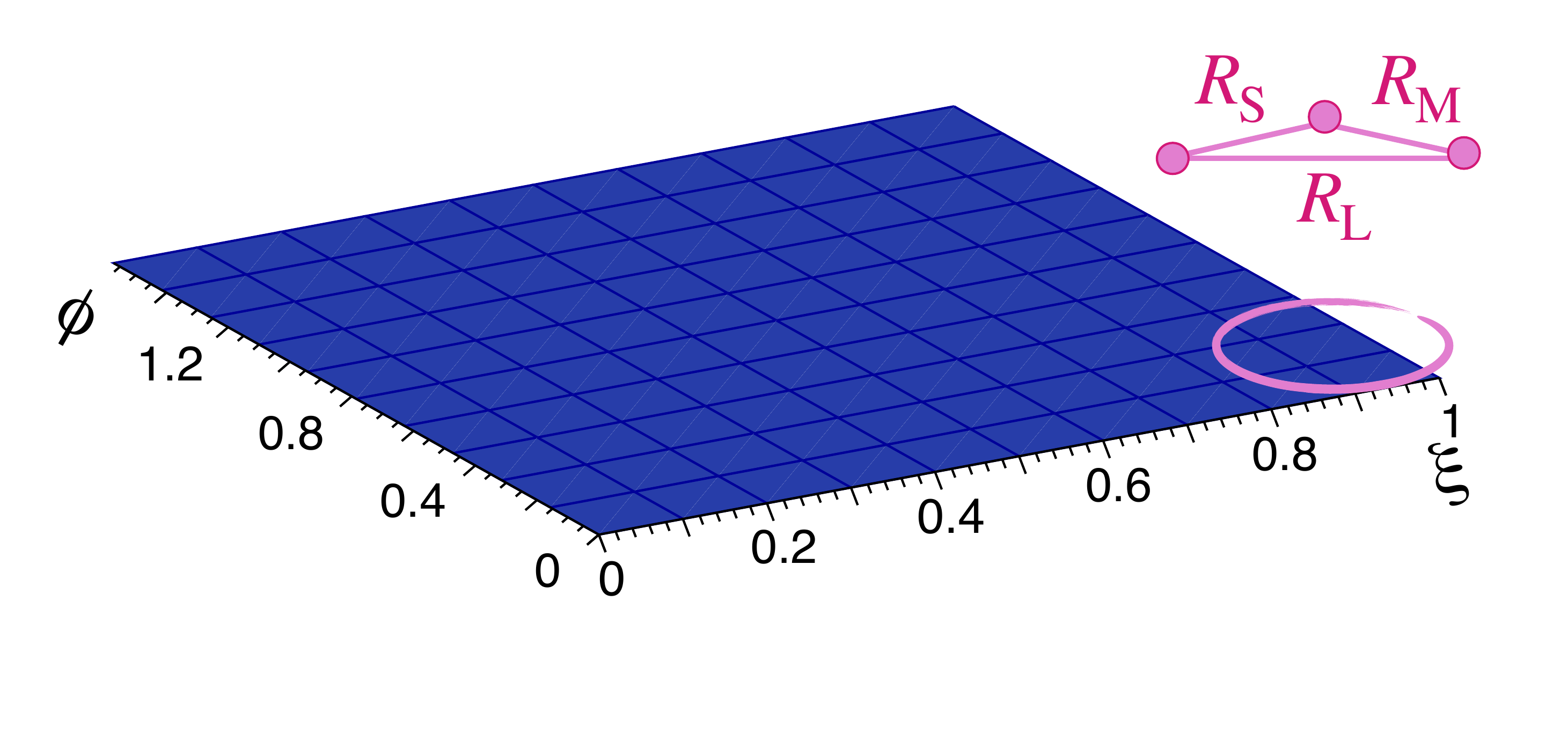}
    \end{minipage}
    \hfill
    \begin{minipage}{0.45\textwidth}
        \centering
        \includegraphics[width=\textwidth]{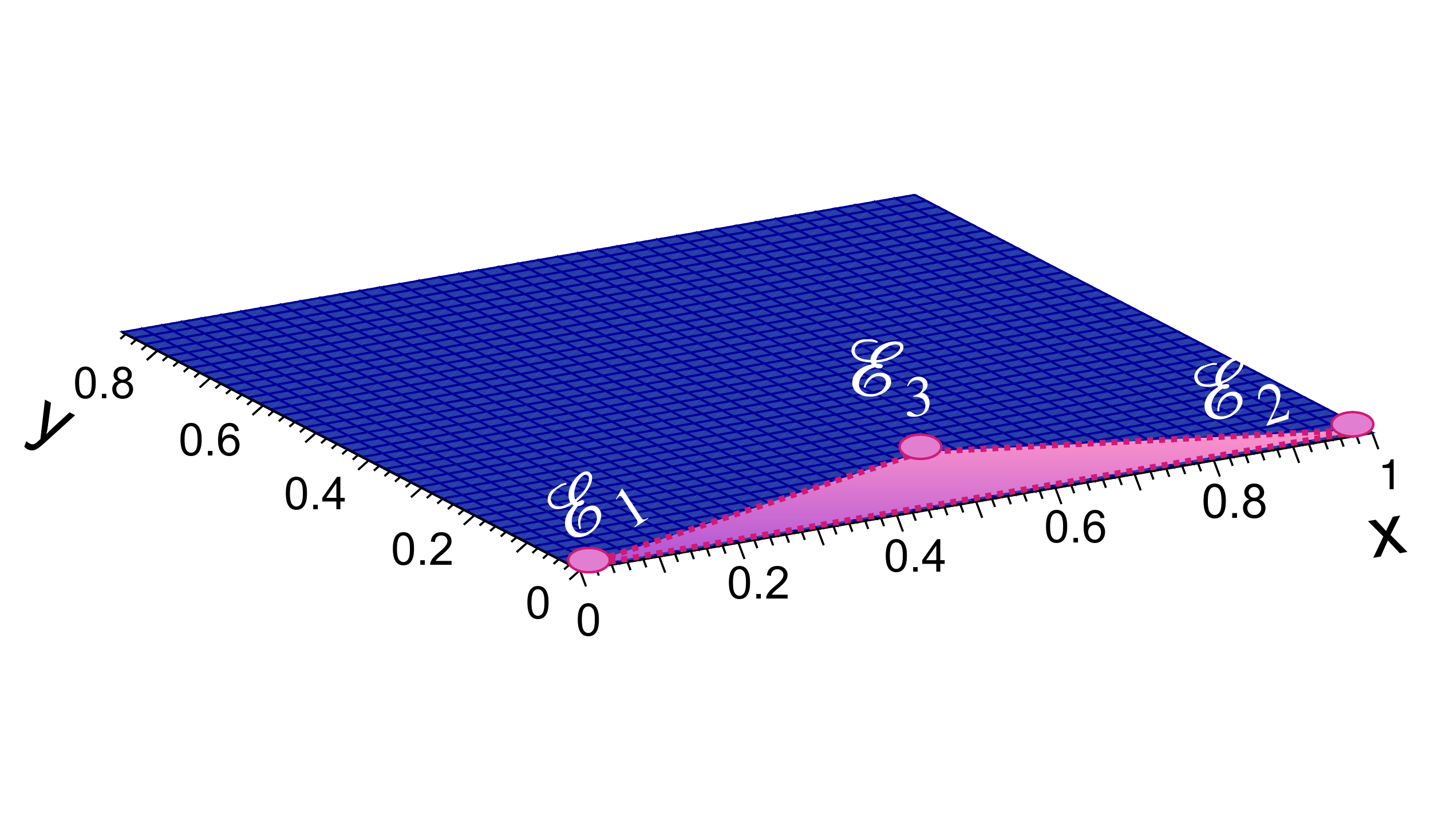}
    \end{minipage}
    \vspace{0.5em}
    \parbox[c]{0.9\textwidth}{
        \centering
       (b) Flattened \\
    }
    \vspace{1em} 

    \begin{minipage}{0.45\textwidth}
        \centering
        \includegraphics[width=\textwidth]{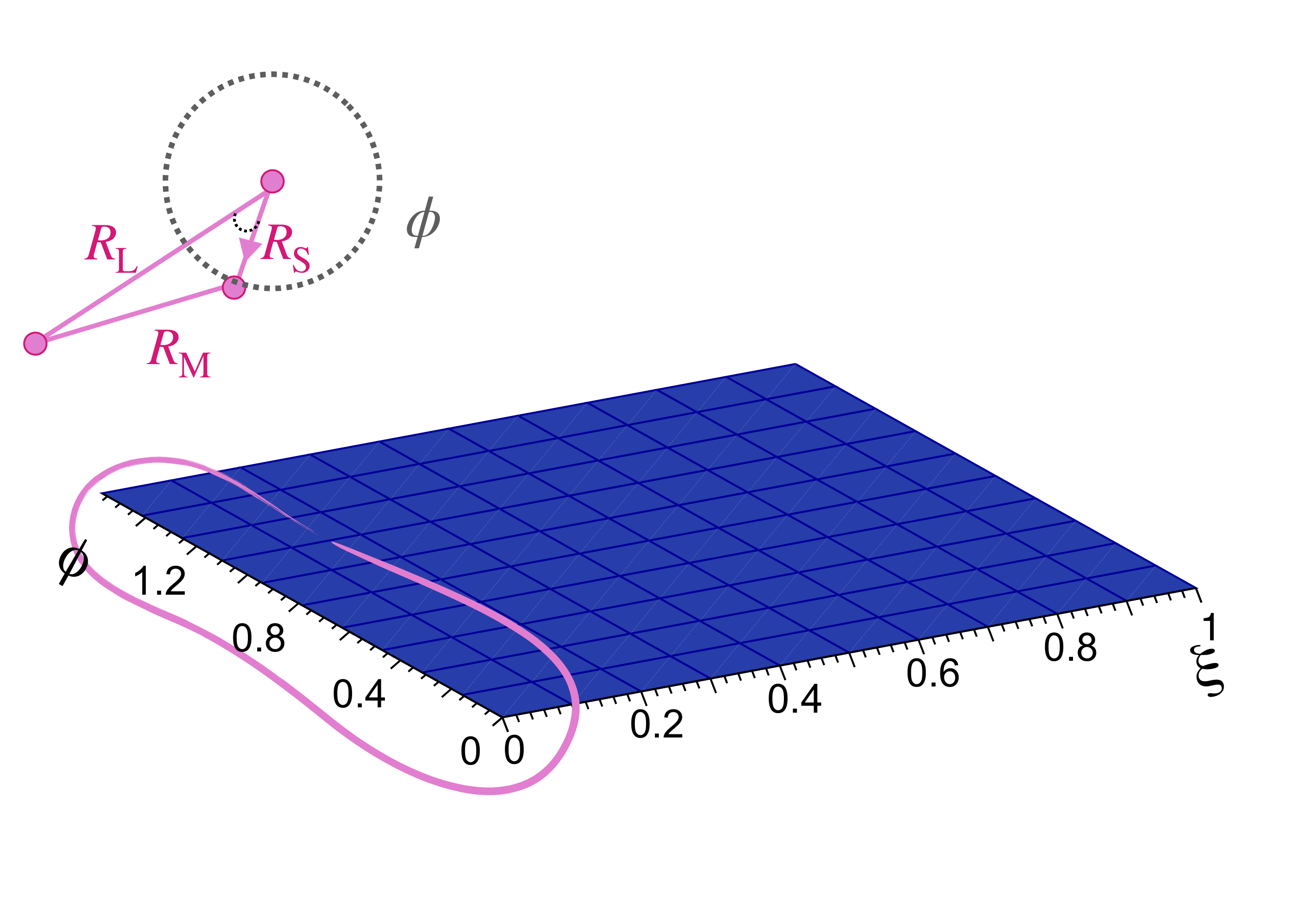}
    \end{minipage}
    \hfill
    \begin{minipage}{0.45\textwidth}
        \centering
        \includegraphics[width=\textwidth]{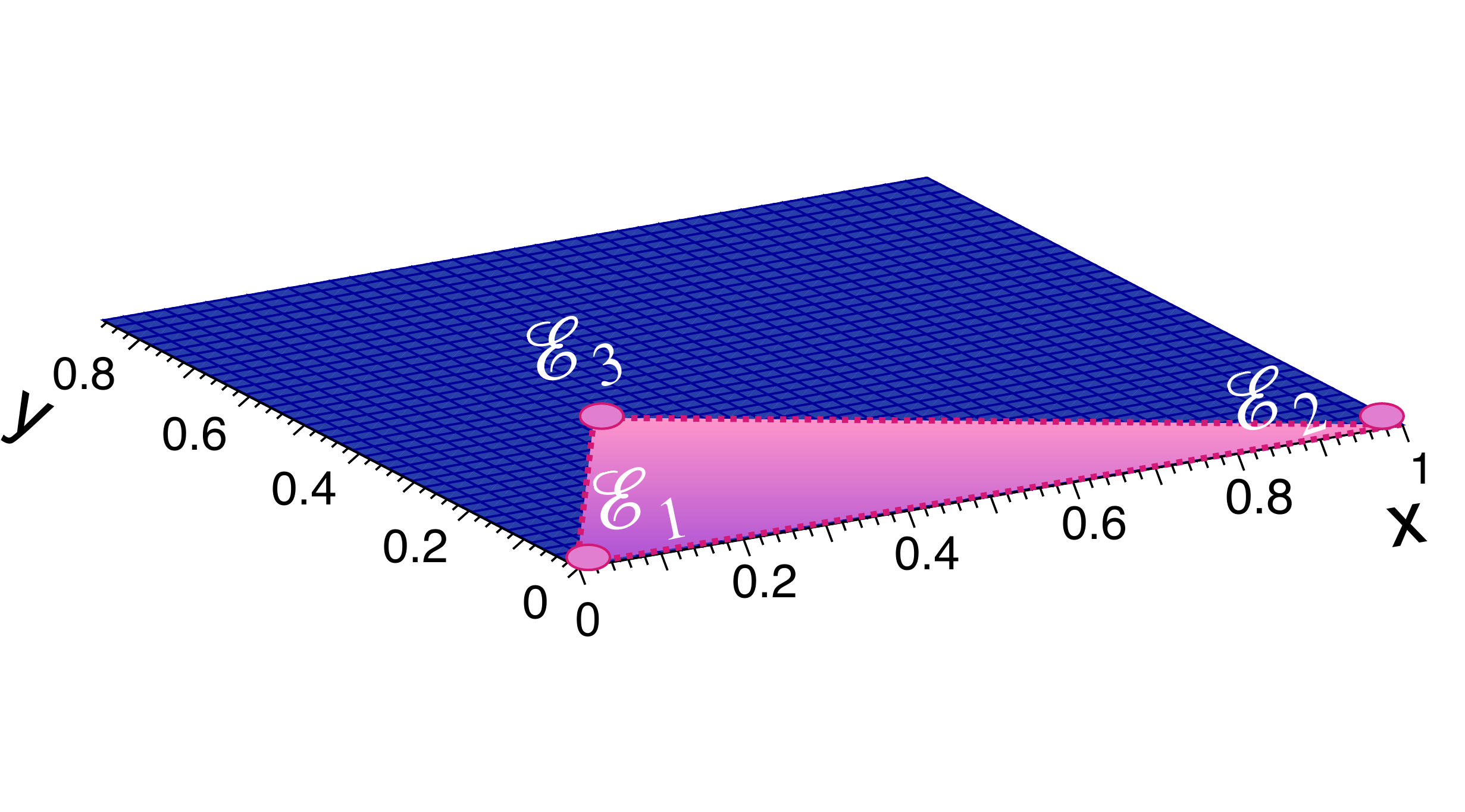}
    \end{minipage}
    \vspace{0.5em}
    \parbox[c]{0.9\textwidth}{
        \centering
       (c) Squeezed \\
    }
    \caption{Three possible shapes of the three-point energy-energy-energy correlator: equilateral, flattened and squeezed triangles. In the left and right columns we show where each shape is found in the $\xi$-$\phi$ and   $x$-$y$ coordinate systems, respectively. The collinear singularity of QCD guarantees that for jets in vacuum the EEEC is dominated by the squeezed triangle region.}
    \label{fig:EEEC_shapes}
\end{figure}

\begin{figure}[t]
\begin{center}
\subfloat[]{\includegraphics[width = 0.53\textwidth]{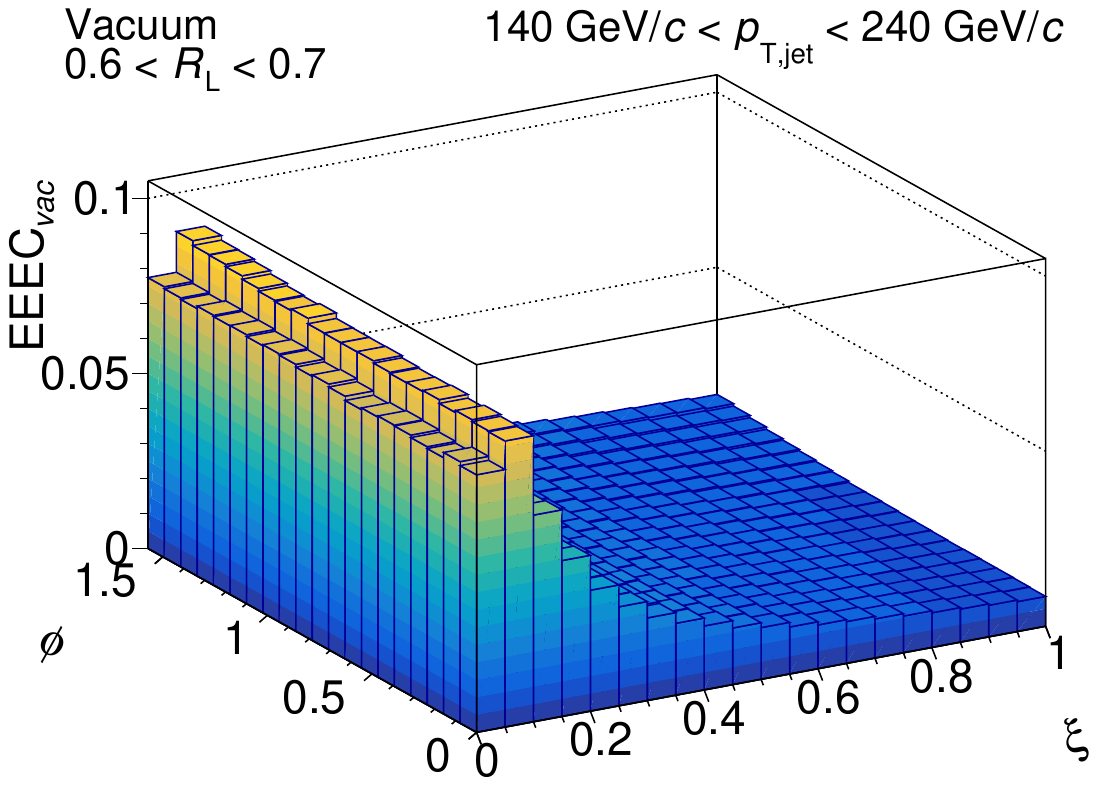}}
\subfloat[]{\includegraphics[width = 0.53\textwidth]
{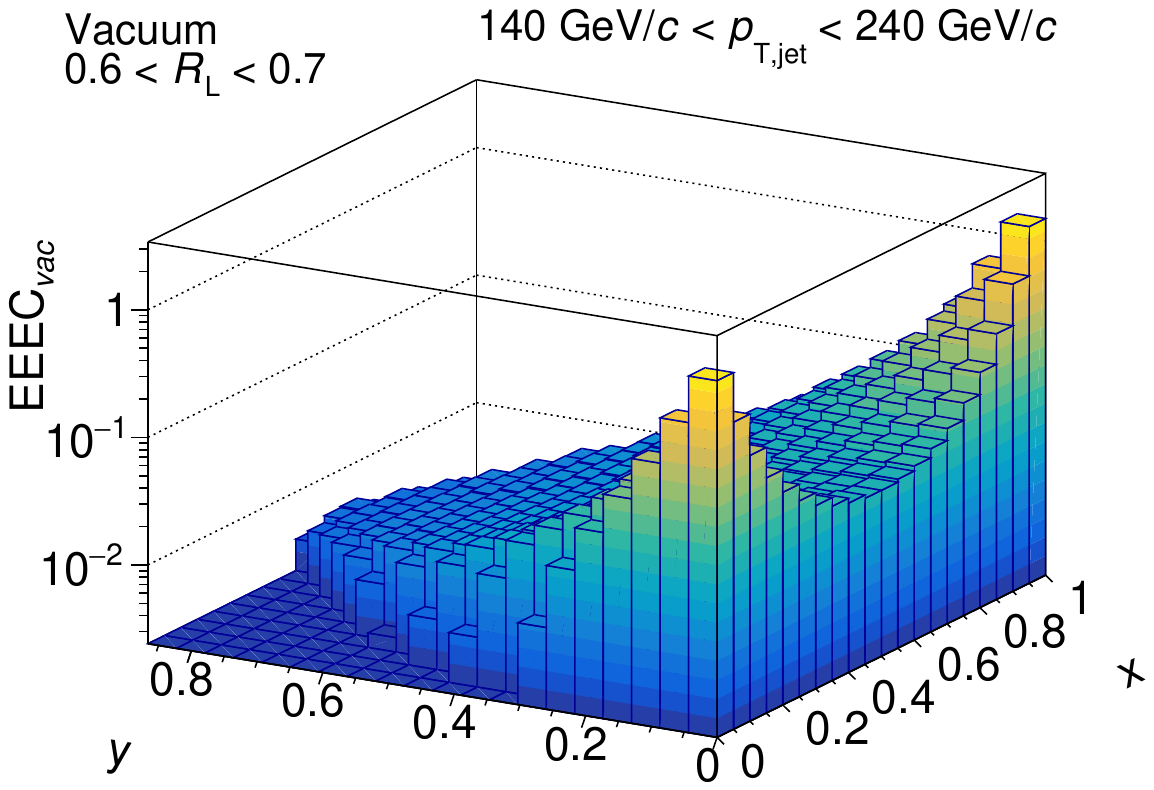}}
\caption{The EEEC for anti-$k_{\rm T}$ jets with $R=0.8$ in vacuum (i.e.~as produced in proton-proton collisions), plotted in $\xi-\phi$ coordinates in (a) and in $x$-$y$ coordinates in (b). The collinear enhancement that originates from the collinear singularity in the splitting function that governs the development of parton showers in vacuum is clearly visible in either the small-$\xi$ region or near the points $(0,0)$ and $(1,0)$ in $(x,y)$ coordinates. The EEEC is dominated by these squeezed triangles; no other shapes exhibit enhancement in vacuum.}
\label{fig:EEEC_vac}
\end{center}
\end{figure}

 Beyond the scaling behavior that is identified by the projected energy correlators, significant additional information is encoded in the shape of higher point correlators, as we shall see. The lowest order correlator with a non-trivial shape dependence 
 is the three-point energy-energy-energy correlator (EEEC), which was first computed in elementary collisions in Ref.~\cite{Chen:2019bpb}. The three-point correlator is a function of three-angles $R_{\rm S}$, $R_{\rm M}$, and $R_{\rm L}$. It is convenient to separate the scaling behavior, which is set by the overall size of the correlator which we shall characterize by $R_{\rm L}$, from the shape of the correlator, specified by the triangle with sides $R_{\rm L}$, $R_{\rm M}$ and $R_{\rm S}$.
 Following Ref.~\cite{Komiske:2022enw}, it is convenient for studies of 
 jets produced in elementary collisions that form entirely in vacuum
 to specify the shape of the three-point correlator via the coordinates $(R_{\rm L}, \xi, \phi)$, where
\begin{align}
\xi\equiv\frac{R_{\rm S}}{R_{\rm M}} \,, \qquad \phi&\equiv\arcsin \sqrt{1 - \frac{(R_{\rm L}-R_{\rm M})^2}{R_{\rm S}^2}}
\,.
\label{eq:transf}
\end{align}
Because of the collinear singularity of the splitting functions that govern
the development of a parton shower in vacuum, the correlator is dominated by small values of $\xi$ in this case.  That is, if one first specifies that two of the three angles in the correlator are separated by $R_{\rm L}$, it is then most likely to find
energy flow in a third direction that is almost collinear with one or other of the first two angles.
We shall refer to triangles with small $\xi$, namely with $R_{\rm S}\ll R_{\rm M}\sim R_{\rm L}$, as squeezed triangles; for jets in vacuum, the EEEC is dominated by such configurations.
The variable $\phi$ can be thought of as an angular variable when $\xi$ is small.
In the case of a squeezed triangle, it specifies the orientation of $R_{\rm S}$ relative to $R_{\rm L}$, and for jets in vacuum the EEEC at small $\xi$ has little dependence on $\phi$. We illustrate the geometry of the triangle that specifies the EEEC in Fig.~\ref{fig:EEEC_shapes}, and plot the full EEEC for jets in proton-proton collisions in Fig.~\ref{fig:EEEC_vac}. The $x$-$y$ coordinates used in these figures are defined in \Sec{sec:coords}.


The EEEC was first measured in high-$p_{\rm T}$ jets in vacuum using CMS open data~\cite{Komiske:2022enw,Chen:2022swd}.
We reproduce this behavior in Fig.~\ref{fig:EEEC_vac} using 
our calculation (described below) of anti-$k_{\rm T}$ jets with radius $R=0.8$ in vacuum using the Hybrid Model (which describes jets in vacuum using a particular tune of PYTHIA 8).
In vacuum, the EEEC 
is dominated by the small-$\xi$ region of squeezed triangles, as expected. Also as expected, there is little dependence on $\phi$.
In this small $\xi$ regime where $R_{\rm S}$ is much smaller than the other two sides, the $\xi$-dependence of the EEEC is proportional to the $R_{\rm S}$-dependence of the two-point correlator evaluated at $R_{\rm S}$, with a proportionality constant that depends on $\alpha_{\rm s}$. The $(\xi,\phi)$ coordinates are designed to illustrate this physics.

The upper-right corner of \Fig{fig:EEEC_vac}(a) ($\xi\sim 1$ with $\phi\sim\pi/2$) describes the equilateral triangle region of the EEEC and the lower-right corner of \Fig{fig:EEEC_vac}(a) ($\xi\sim 0$ with $\phi\sim 0$) describes the flattened triangle region, see Fig.~\ref{fig:EEEC_shapes}. These regions of the EEEC are not prominent in vacuum. 
In this work, we compute the full shape-dependent three-point correlator for heavy-ion collisions for the first time. 
In heavy-ion collisions this function encodes not only the energy loss experienced by the partons in the jet shower as they traverse the droplet of QGP but also the hadrons originating from the jet-induced wake in the medium.
We shall in fact see that the hadrons coming from jet wakes fill in the equilateral region of the EEEC -- a region that is 
not at all prominent for jets in vacuum.

\section{New Coordinates for the Three-Point Correlator}\label{sec:coords}

The $\xi$-$\phi$ coordinates described in the previous Section are well-suited to providing a visual representation of the EEEC for jets in vacuum QCD in that they highlight an important feature of the theory, namely its collinear singularity. 
However, away from the small-$\xi$ region, where the triangle that defines the three-point correlator is not squeezed, the Jacobian for this coordinate system is not flat. In fact, as we show explicitly in Appendix~\ref{sec:coords_art}, the $\xi$-$\phi$ coordinate system has a Jacobian with a sharp peak for precisely equilateral triangles with $\xi=1$ and $\phi=\pi/2$.
This means that if we use this coordinate system to plot the correlator in a case where, unlike for jets in vacuum, the physics populates the equilateral region of the correlator one must always be careful to plot ratios in which the Jacobian of the coordinate system cancels, as otherwise the effects coming from the physics of interest will be obscured by the effects of the Jacobian. Or, we should choose a new coordinate system in which the Jacobian is flat for triangles that are not squeezed, including for triangles that are close to equilateral. Since, as we have foreshadowed in the introduction and as we shall see below, the soft hadrons coming from jet wakes populate the EEEC in the equilateral region, we shall employ new coordinates with a flat Jacobian so as to obtain a faithful visual representation of the physics, even when the Jacobian does not cancel in what we plot.
There are many ways to choose new coordinates with a flat Jacobian; we shall employ one, others should be investigated also.

We provide a complete description of the 
new coordinate system with a flat Jacobian that we shall employ as well as the transformation between these coordinates and $\xi$-$\phi$ coordinates in~\App{sec:coords_art}.
We use an $x$-$y$ plane to characterize the shape of the triangle that defines the EEEC.
As we do throughout, we denote
the largest separation between two of the three angles by $R_{\rm L}$, and use $R_{\rm L}$ to characterize the scale of the EEEC correlation. To define the $x$-$y$ coordinates with which we shall specify the shape of the triangle, we first rescale the triangle by $R_{\rm L}$ so as to set the longest side length of the rescaled triangle to $1$. 
We then place the two vectors 
that span the longest side of the EEEC-triangle at $(0,0)$  and $(1,0)$ in the $x$-$y$ plane. 
The third vector that defines the three-point correlator sits at the point $(x,y)$, and serves to specify the shape of the triangle.
Since flipping the sign of $y$ from positive to negative would lead to the same shape EEEC, we shall always take $y$ to be positive.
We illustrate how triangles of three different shapes (one squeezed, one flattened, and one equilateral) are represented in the $(x,y)$ coordinates in~\Fig{fig:EEEC_shapes}. It should be apparent from this construction that the Jacobian for the (Cartesian) $x$-$y$ coordinate system is flat. 
Note also that
since
we describe the EEEC correlator by triangles whose longest side is 1, the domain of $(x,y)$ is bounded by two arcs as well as by the $y=0$ axis.
There is no particular advantage to using the $x$-$y$ coordinates to describe the EEEC for jets in vacuum, where the equilateral region is hardly populated and the physics of interest is peaked around squeezed triangles.  However, for orientation and because we will refer to it below, 
in \Fig{fig:EEEC_vac}(b) we show the EEEC for jets in vacuum in $(x,y)$ coordinates.  The collinear enhancement that sits at small $\xi$ in \Fig{fig:EEEC_vac}(a) is 
apparent
at (0,0) and (0,1), which are the collinear points in the $x$-$y$ coordinates. One way to say this is that if we require that the jet has particles at (0,0) and (0,1) (which, recall, are separated by an angle $R_{\rm L}$) and then asks where is it most likely to find a third particle, if the jet in question is a jet in vacuum then the answer is that it is by far most likely to find the third particle near one or other of the first two.  We shall see that for jets in heavy-ion collisions the hadrons in the jet that originate from the wake change this dramatically.

\section{The Hybrid Strong/Weak Coupling Model}\label{sec:wake_review}

The hybrid strong/weak coupling model, introduced and extensively described in Refs.~\cite{Casalderrey-Solana:2014bpa,Casalderrey-Solana:2015vaa,Casalderrey-Solana:2016jvj,Hulcher:2017cpt,Casalderrey-Solana:2018wrw,Casalderrey-Solana:2019ubu,Hulcher:2022kmn}, addresses the challenge of describing the interaction of jet parton showers that traverse the droplet of QGP with which they are concurrently produced in heavy-ion collisions, the consequent modification of the parton shower and the droplet, and the consequences for jets as identified, reconstructed and analyzed in experimental data. 
The model is based on the observation that while the initial production of energetic partons and their showering occur at scales for which QCD is weakly coupled, both the physics of QGP and of its interaction with jet constituents involve scales of the order of the QGP temperature, at which QCD is strongly coupled and the QGP that it describes is a strongly coupled liquid.

In vacuum, high-energy partons produced with a large virtuality scale $Q\sim p_{\rm_{T}}^{\rm jet}$ at a hard scattering relax their virtuality down to the QCD confinement scale via successive splittings, as determined by the perturbative, high-$Q^2$, DGLAP evolution equations, resulting in a parton shower.  QGP, however, is a strongly coupled liquid with a lower value of $\eta/s$, the dimensionless measure of internal dissipation as a liquid flows than for any other known liquid.
And, the most probable interactions between a jet parton and the expanding cooling droplet of QGP produced in a heavy-ion collision involve low-momentum transfers of order the QGP temperature $T$, which is comparable to the QCD confinement scale, and thus cannot be weakly coupled. 
This makes
describing the evolution of a parton 
shower within the dynamically evolving droplet of QGP produced in a heavy-ion collision a truly challenging task.
The existence of a separation in scales between those of the hard scattering and the splitting via which the parton shower forms and the lower scales that characterize the jet/medium interplay raises the hope of a future description in terms of an effective field theory at strong coupling that includes the hydrodynamic evolution of a droplet of  QGP and the jet/medium interplay that has been rigorously matched to weakly coupled perturbative QCD at higher scales.
The Hybrid Model is not that but it should be seen in that spirit.

In the hybrid strong/weak coupling model, the development of a jet while interacting with QGP is carried out by blending a holographic formulation of the strongly-coupled interaction of jet constituents with QGP into a perturbative QCD treatment of the production of hard partons and parton showering. 
The 
model adopts a perturbative description for the evolution of energetic partons, governed by the initially high virtuality $Q \gg T$, where $T$ is the QGP temperature. This perturbative evolution is described by the event generator PYTHIA 8~\cite{Sjostrand:2014zea},
including initial state radiation but not multi-parton interactions. To take into account the different production rate of hard processes in PbPb collisions with respect to pp, initial state nuclear effects are accounted for by using modified parton distribution functions following the EPS09 parametrization~\cite{Eskola:2009uj}. 
Next, these hard parton showers 
must be embedded into a space- and time-dependent background consisting of an expanding and cooling droplet of QGP described via a hydrodynamical calculation, averaged over collisions with a given centrality class~\cite{Shen:2014vra}. In order to do this, the parton showers from PYTHIA must be endowed with a spacetime structure, which we do by assigning a lifetime to each individual parton according to $\tau=2 E/Q^2$, where $E$ is the parton's energy and $Q$ its virtuality~\cite{Casalderrey-Solana:2011fza}.
The origin of the hard event is determined 
probabilistically according to the density distribution of the nuclear 
overlap for the same centrality class, and its azimuthal direction is chosen randomly. The  hydrodynamic background then provides local, that is for each spacetime point, QGP temperatures necessary for the computation of the modifications imparted on the
parton showers by the droplet of QGP. 
The separation between
the physics of jet formation at high scales, described as in vacuum, and the interaction between the jet partons and the medium at lower scales that is intrinsic to the Hybrid Model
is reminiscent of many other approaches to the problem in which there is a sizeable phase space for jet formation whose dynamics is largely unaffected by the medium, see for example Ref.~\cite{Caucal:2018dla}. 

Interactions between the energetic partons in the jet shower and the droplet of QGP are dominated by soft momentum exchanges. In the Hybrid Model, these are described
using results for the energy loss of an energetic parton within strongly coupled plasma with a holographic description, derived at infinite coupling and large $N_c$~\cite{Chesler:2014jva,Chesler:2015nqz}.
The energy loss rate $dE/dx$ for an energetic color charge in the fundamental representation 
that has traveled a distance $x$ then takes the form~\cite{Chesler:2014jva,Chesler:2015nqz}
\begin{equation}
\label{eq:elossrate}
   \left. \frac{\rmd E}{\rmd x}\right|_{\rm strongly~coupled}= - \frac{4}{\pi}\, \frac{E_{\rm in}}{x_{\rm stop}}\, \frac{x^2}{x_{\rm stop}^2} \frac{1}{\sqrt{1-(x/x_{\rm stop})^2}} \quad ,
\end{equation}
where $x_{\rm stop}\equiv  E_{\rm in}^{1/3}/(2{T}^{4/3}\kappa_{\rm sc})$ is the maximum distance the energetic excitation with initial energy $E_{\rm in}$ can travel within the strongly coupled medium before thermalizing and stopping. $\kappa_{\rm sc}$ is an $\mathcal{O}(1)$ parameter 
that can be calculated in ${\cal N}=4$ supersymmetric Yang-Mills (SYM) theory, where it is proportional to $\lambda^{1/6}$.
In the Hybrid Model, $\kappa_{\rm sc}$ is chosen by
fitting to high-$p_{\rm T}$ hadron and jet suppression measured 
at the LHC~\cite{Casalderrey-Solana:2018wrw}. 
The fit results in a thermalization distance $x_{\rm stop}$ that is longer by a factor of 3-4 for QGP in QCD than  in
${\cal N}=4$ SYM theory at the same temperature.
In the Hybrid Model,
we take  a spatially and temporally varying
$T$ from the hydrodynamic background and 
apply Eq.~(\ref{eq:elossrate}) to each parton in the jet shower,
with $T$ and hence $x_{\rm stop}$ varying as a function of $x$ along the parton trajectory.
Finally, note that the holographic stopping distance of a gluon is reduced by a factor $(C_A/C_F)^{1/3}$ compared to that of a quark~\cite{Gubser:2008as}, which means that in the Hybrid Model $\kappa_g=(9/4)^{1/3} \, \kappa_{\rm sc}$.

At strong coupling, the energy and momentum lost by the energetic partons in the parton shower hydrodynamizes already at distances $\sim 1/T$, exciting a wake in the droplet of QGP which evolves hydrodynamically as the droplet  expands, flows and cools.
When the droplet, including the wake deposited in it by the passing jet, reaches the freeze-out hypersurface it
falls apart into thousands of soft hadrons, with the hadrons originating from the wake carrying the energy and momentum lost by the parton shower.
        In the Hybrid Model, the momentum distributions of those hadrons (in $p_{\rm T}$, azimuth, and rapidity) is described via a relatively crude approximation to what would be obtained by applying the Cooper-Frye prescription~\cite{PhysRevD.10.186} to the droplet of QGP with
the jet-induced wakes, and subtracting what would be obtained in the absence of any jet.
If one assumes that the jet wakes are a small perturbation to the hydrodynamic flow (a reasonable assumption),
assumes that the perturbation to the spectra of hadrons
produced at freezeout coming from the jet wakes
is small at all momenta (which need not be a good assumption 
at all $p_{\rm T}$), assumes that the flow in the background fluid is dominantly longitudinal and boost invariant (neglecting the interplay between jet wakes and radial flow 
is not a good approximation), and assumes that the wake stays close in rapidity to the jet (which is the case for the moving fluid behind the jet but need not be the case for energy in sound waves),
the perturbation to the momentum distribution of the hadrons at freezeout coming from the wake of 
a jet with azimuthal angle $\phi_j$ and rapidity $y_j$
that has lost momentum and energy $\Delta p_{\rm T}$ and
$\Delta E$ takes the form~\cite{Casalderrey-Solana:2016jvj}
\begin{equation}
\label{eq:onebody}
\begin{split}
E\frac{\rmd\Delta N}{\rmd^3p}=&\frac{1}{32 \pi} \, \frac{m_T}{T^5} \, \textrm{cosh}(y-y_j)  \exp\left[-\frac{m_T}{T}\, \textrm{cosh}(y-y_j)\right] \\
 &\times \Bigg\{ p_{\rm_{T}} \Delta p_{\rm_{T}} \cos (\phi-\phi_j) +\frac{1}{3}m_T \, \Delta M_T \, \textrm{cosh}(y-y_j) \Bigg\} \, ,
\end{split}
\end{equation}
where $p_{\rm_{T}}$, $m_T$, $\phi$ and $y$ are the transverse momentum, transverse mass, azimuthal angle and rapidity of the emitted thermal particles and where $\Delta M_T\equiv \Delta E/\cosh y_j$. The analytical expression (\ref{eq:onebody}) is straightforward to implement in the Hybrid Model~\cite{Casalderrey-Solana:2016jvj}.

The distribution (\ref{eq:onebody}) can become negative
in a region of $\phi$ around the direction opposite to that of the jet. 
This is a consequence of the boost experienced by
a fluid cell at freezeout caused by the wake of a jet.
Before freezeout, most of the momentum lost by the parton shower is carried by QGP fluid behind the jet that is moving in the direction of the jet. The Cooper Frye prescription ensures that by boosting the fluid in its wake in its direction, the jet enhances the production of soft particles in its direction and 
depletes the production of soft particles in the opposite direction, relative to what the Cooper-Frye description would have yielded at freezeout in the absence of any jet wake.
In the Hybrid Model, where the distribution (\ref{eq:onebody}) is positive it is implemented by adding hadrons with this distribution and where
it is negative it is implemented by adding hadrons
with this distribution but with a negative energy.
In any observable (like the jet shape, for example) where what enters is the energy flow at one cell on the
celestial sphere, this subtraction is done correctly simply by
adding the energies of the hadrons in a cell upon treating the energies of the ``negative particles'' to be negative.
In an energy correlator observable, the 
subtraction procedure must be 
handled with care as the energy of a ``negative particle'' 
has to be accounted for by subtracting it from
that of a nearby positive particle before
the energy correlator is computed.
We describe this subtraction procedure in App.~\ref{app:negasub}.

It has been understood since the distribution (\ref{eq:onebody}) was derived in
Ref.~\cite{Casalderrey-Solana:2016jvj}
that, because of the various simplifying assumptions
made in its derivation, this simple analytical expression
describes a distribution of soft hadrons coming from
a jet wake that is too soft.
The comparisons made in 
Ref.~\cite{Casalderrey-Solana:2016jvj} between Hybrid Model calculations with the hadrons originating from jet
wakes described by Eq.~(\ref{eq:onebody})
and experimental measurements of observables including the jet shape and missing-$p_{\rm T}$ observables~\cite{CMS:2015hkr}
indicate that Eq.~(\ref{eq:onebody}) predicts a few too many very soft particles coming from the wake and not  enough such particles in the 2-4 GeV range of $p_{\rm T}$.
The analytical and compact expression Eq.~(\ref{eq:onebody}) does not account for the coupling of the 
evolution of the hydrodynamic wake deposited by a jet
with the transverse hydrodynamic flow (i.e.~radial expansion)
of the droplet of QGP produced in a realistic heavy-ion collision. It has been shown that these dynamics lead, on average, to a narrower in angle, and harder in $p_{\rm_{T}}$ distribution for the wake hadrons~\cite{Yan:2017rku,Tachibana:2020mtb,Casalderrey-Solana:2020rsj}, as the jet wakes are boosted in the radial direction by the radial hydrodynamic flow before they subsequently freeze out. 
An efficient, computationally feasible description of a realistic wake that accounts for this and other effects is in the works~\cite{wakeprep}, but in this paper we shall use Eq.~(\ref{eq:onebody}). 
Consequently, the predictions presented in the present paper are not expected to agree quantitatively with future measurements. 
However, using Eq.~(\ref{eq:onebody}) should suffice for the purpose of understanding where in the space of two- and three-point energy correlators the effects of jet wakes are to be found. And, since the spectrum of the wake hadrons
that it yields is too soft, it likely underestimates the magnitude of their contribution to energy-energy and energy-energy-energy correlators.

Note that the Hybrid Model does not include any further modification of the parton shower or of the hadrons from the wake after freezeout. At freezeout, 
the remaining partons in the parton shower are hadronized via the Lund string model implemented in PYTHIA8. (Some of the softer partons in the shower lose all of their energy to the wake. The more energetic partons from the shower lose only some of their energy to the wake; after this energy loss, they are the shower partons that are then hadronized via PYTHIA8.)
The hadrons that come from the fragmentation of the partons in the modified parton shower, together with the hadrons that result from the freezeout of the wake which are 
described by Eq.~(\ref{eq:onebody}), ensure energy-momentum conservation, event-by-event. These hadrons, together, are then what we analyze, first reconstructing jets and then computing energy correlators.

Finally, we note that various previous works have added elements to the Hybrid Model to investigate physical effects that we shall not investigate in this paper.
Medium-induced Gaussian-distributed relatively soft transverse kicks felt by the shower partons~\cite{Casalderrey-Solana:2016jvj} and harder, larger angle, elastic (or Moli\`ere) scattering of shower partons off
medium partons~\cite{Hulcher:2022kmn} have both been investigated, as has introducing the QGP resolution length, namely the length scale such that when two jet partons are closer than this they lose energy as if they were a single parton~\cite{Hulcher:2017cpt}.
All are turned off in this work.
We leave the investigation of how each of these effects influences energy correlator observables to future work.

\subsection{Simulation Details}\label{sec:setup}

The studies in this paper are performed using a variety of samples (inclusive jets or jets produced in association with a hard photon; jets from proton-proton collisions or PbPb collisions, in the latter case either with the hadrons coming from the wake or with the wake turned off) generated using the hybrid strong/weak coupling model framework as described above. Each sample contains $\sim $ 1 million events, all for collisions with a center of mass energy of $\sqrt{s} = 5.02$ TeV per nucleon-nucleon collision. The heavy-ion events simulated correspond to the 0--5\% most central collisions. 
In the inclusive jet samples, we set
$\widehat{p}_{\rm T, min}=100$ GeV/$c$ in PYTHIA; this corresponds to the minimum $p_{\rm T}$ of the initial hard scattering from which the parton showers that PYTHIA then describes originate. We subsequently reconstruct inclusive jets with $p_{\rm T,{\rm jet}}>140$~GeV$/c$.
For the $\gamma$-jet samples, we selected only those events containing a photon with $p_{{\rm T}}^{\gamma} > 140$ GeV/$c$ (considering both photons from the hard scattering and those produced during the parton shower evolution). In these events, we reconstructed and studied jets with $p_{\rm T,{\rm jet}} > 40$ GeV/$c$. The PYTHIA version used is 8.244, with NNPDF2.3 parton distribution functions (PDFs) for proton-proton collisions.
For PbPb collisions, the PDFs are modified
according to the EPS09LO~\cite{Eskola:2009uj} nuclear PDFs. While initial state radiation is on, multi-parton interactions are off. The default value of \texttt{TimeShower:pTmin} has been modified to 1 GeV/$c$.

In this study, full hadron-level jets are reconstructed using \texttt{FastJet 3.4.2}~\cite{Cacciari:2011ma} with the anti-$k_{\rm T}$ algorithm~\cite{Cacciari:2008gp} with a specified resolution parameter $R$ and an acceptance of $|\eta| \leq 2.0$. In an experimental context, it may be beneficial to use track-based jets due to the excellent angular resolution of tracking detectors. This can be incorporated into calculations of the energy correlators using the track function formalism \cite{Chang:2013rca,Chang:2013iba,Jaarsma:2023ell,Chen:2022pdu,Chen:2022muj,Jaarsma:2022kdd,Li:2021zcf}, and has been found to have a minimal affect on the behavior of the correlator observables~\cite{Jaarsma:2023ell}. Unless otherwise specified (see \Sec{sec:exp}), no cuts on the $p_{\rm T}$ of constituents that enter the jet were applied. In order to identify a $\gamma$-jet event, a number of cuts were applied in order to closely mimic the selections performed in experimental measurements~\cite{ALICE:2020atx}. First, the photon must have $\Delta \phi > 2\pi /3$ in relation to the jet. In this case, the photon recoiling from the jet must also be ``isolated" which we define as having less than 5 GeV/$c$ of transverse energy in a cone of $R$ = 0.4 around the photon. We emphasize that for $\gamma$-tagged jets, we use the $p_{\rm T}$ of the photon as a proxy for the jet $p_{\rm T}$ when constructing the energy correlators. 

To examine the impact of the hadrons originating from jet wakes on energy correlator observables in either inclusive jets or $\gamma$-jet events, we shall make comparisons among three different samples of simulated events:
\begin{itemize}
    \item {\textit{Vacuum}: Simulations of proton-proton collisions, in which the jets shower and develop entirely in vacuum. Note that Hybrid Model vacuum samples are nothing more than PYTHIA simulations, with the various settings described above. These samples serve as references. Differences between observables in a vacuum sample and observables in either of the Hybrid Model samples can be attributed to the consequences of the parton shower having plowed through and interacted with a droplet of QGP.}
 \item {\textit{Medium with wake}: Hybrid Model simulations of PbPb collisions, in which the parton showers interact with, and lose energy and momentum to, the droplet of QGP produced in the heavy-ion collision as described around Eq.~(\ref{eq:elossrate}). The energy and momentum gained by the droplet takes the form of a wake in the hydrodynamic QGP liquid that, after freezeout, becomes soft hadrons as described around Eq.~(\ref{eq:onebody}).
 This case includes all the contributions to the final hadron distributions for jets in PbPb collisions, as simulated by the Hybrid Model.}    
 \item {\textit{Medium without wake}: Simulations of PbPb collisions as above, except that the hadrons originating from the wake are {\it not} included. Jets in this sample only include hadrons originating from the medium-modified parton shower, as simulated in the Hybrid Model. This is an unphysical scenario -- turning the wake off violates momentum and energy conservation. However, looking at observables in this sample is particularly illustrative and informative, as the comparison with observables in the ``medium with wake'' sample isolates the impact of the wake on any chosen observable.}
\end{itemize}
Throughout this  paper, we will use the terms vacuum, medium with wake, and medium without wake to refer to these three different samples. 
Note that for all results presented in this work, we normalize by the number of jets in the selected jet $p_{\rm T}$ range for each sample.

\section{The Angular Scale of the Wake from Projected Energy Correlators}\label{sec:proj}

In this Section, we begin by investigating the two- and three-point projected correlators, as well as their ratios. While the primary focus of this paper is on the shape dependence of the three-point correlator, the study of the two- and three-point projected correlators in the Hybrid Model is interesting in its own right. This is especially true for comparing how the effects of the wake in the two-point correlator compare with previous studies of the two-point correlator in  QGP. Studies of the two-point correlator in  QGP \cite{Andres:2022ovj,Andres:2023xwr,Andres:2023ymw,Barata:2023vnl,Barata:2023zqg,Yang:2023dwc,Barata:2023bhh,Barata:2024nqo,Andres:2024ksi} have primarily used perturbative approaches, and have not focused on medium response. The principal exception to this is
the calculation of \cite{Yang:2023dwc} using the coupled linear Boltzmann transport (CoLBT) \cite{Li:2010ts,He:2015pra,Cao:2016gvr,Luo:2023nsi,Chen:2017zte,Chen:2020tbl,Zhao:2021vmu} approach. 
Studying the two-point correlator on a strongly coupled wake is interesting for exploring the hydrodynamics of  QGP and, more generally, the hydrodynamics of a strongly-coupled gauge theory. Although we will use a complete simulation, it would be interesting to perform analytic studies of the energy correlators for simple sources coupled to a strongly coupled $\mathcal{N}=4$ medium. While many such calculations exist for one-point functions, e.g.~\cite{Chesler:2007an}, to our knowledge they do not exist in the literature for two-point functions. Furthermore, improving the understanding of the structure of the correlator using a hydrodynamic background is interesting for looking for modifications from rare hard scattering events
\cite{DEramo:2012uzl,DEramo:2018eoy,Hulcher:2022kmn}, including those that are sensitive to the color coherence of dipoles within the jets~\cite{Pablos:2024muu}. We hope that experimental measurements of the two-point correlator in heavy-ion collisions can motivate such calculations.

\begin{figure}[t]
    \begin{center}
    \subfloat[]{\includegraphics[width = 0.53\textwidth]{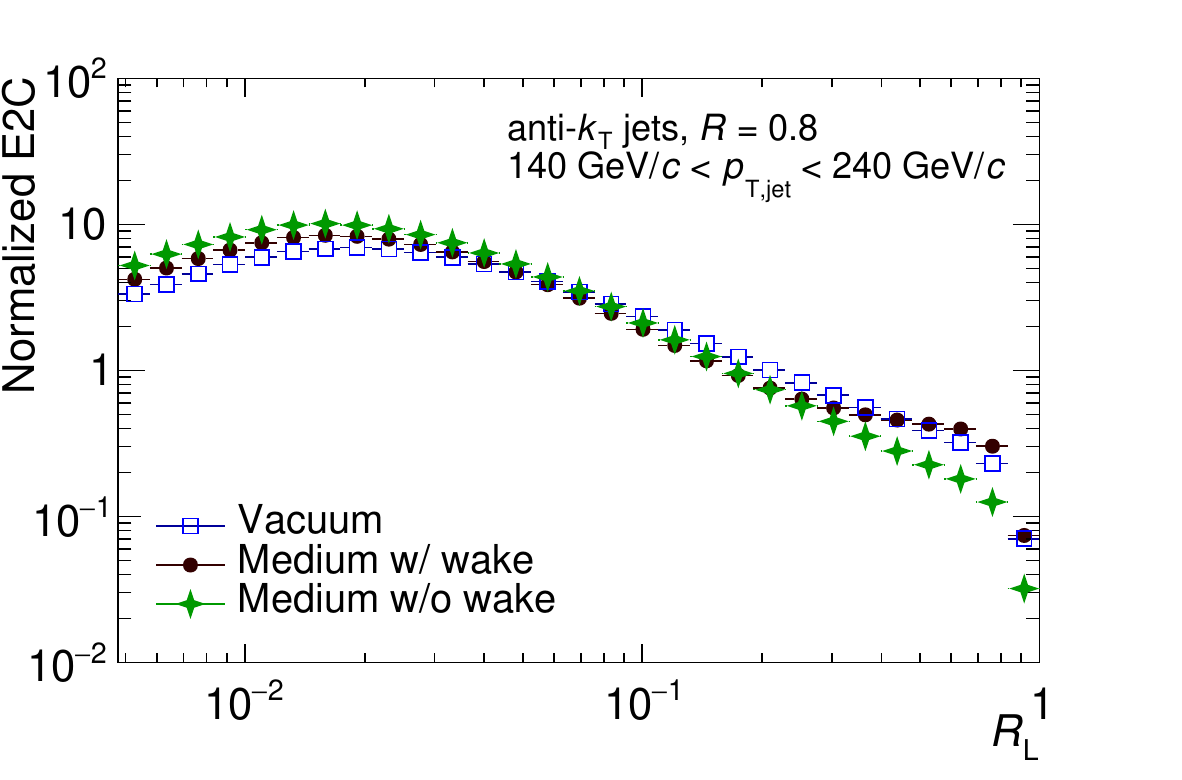}}
    \subfloat[]{\includegraphics[width = 0.53\textwidth]{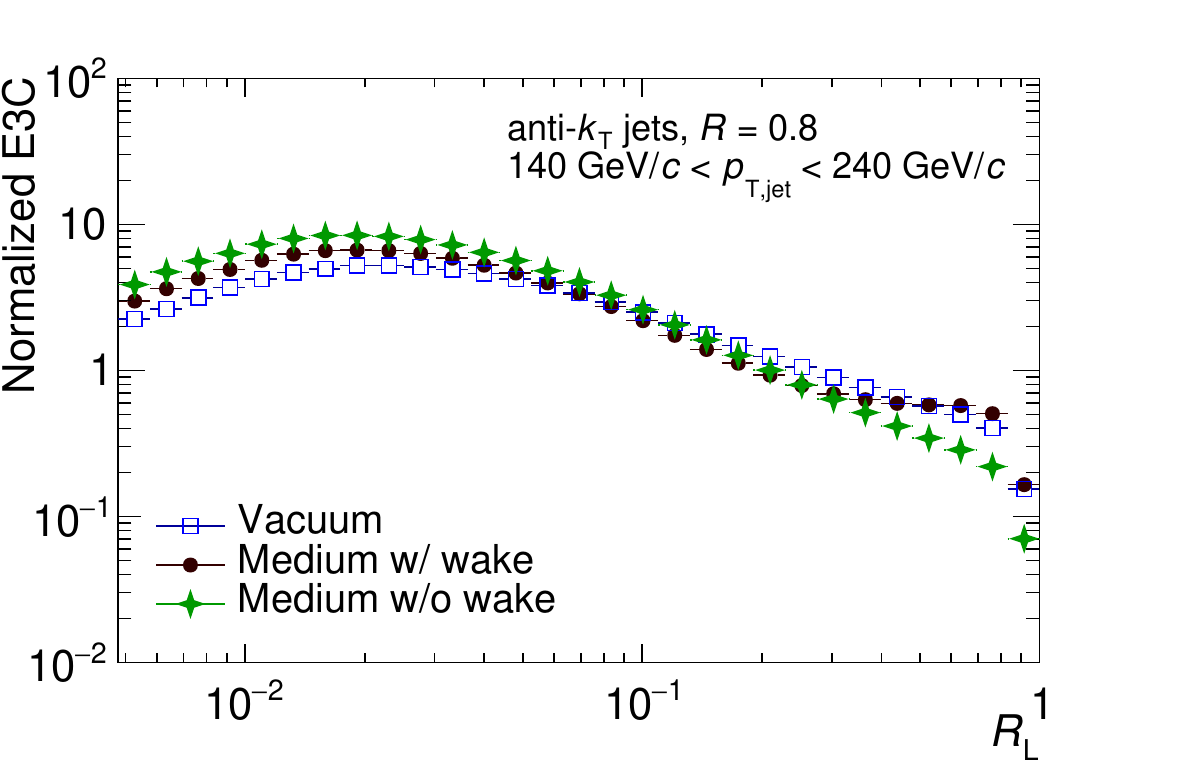}}
    \caption{The E2C two-point correlator (a), and E3C projected three-point correlator (b), as a function of $R_{\rm L}$ for inclusive jets with $R=0.8$ and $140 < p_{\rm T, jet} < 240 $ GeV/$c$
    in vacuum, medium with wake, and medium without wake. These distributions are per-jet and bin-width normalized. The wake imprints itself as a modification in the behavior of both the E2C and the E3C at large angles, meaning that neither scales in the way that they do in vacuum.}
    \label{fig:projectedE2CE3C}
    \end{center}
\end{figure}

As we have discussed in \Sec{sec:intro}, we shall characterize the overall angular scale of each $N$-point 
correlator by $R_{\rm L}$, the largest distance between the $N$ points ($N$ particles) that define the correlation.
The projected $N$-point correlator, namely the ENC defined in Eq.~(\ref{eq:projection}),
is specified by $R_{\rm L}$, upon integrating over all shapes of the $N$-points that have a given $R_{\rm L}$.
For the EEC this is trivial, as the only angle it depends on is $R_{\rm L}$ meaning that the E2C is the same as the EEC.  
In this Section we shall look at the E2C and the first nontrivial example, the E3C, for inclusive jets. The E3C is a function of $R_{\rm L}$, upon integrating over triangles with all possible $R_{\rm M}$ and $R_{\rm S}$ for a given $R_{\rm L}$ or, equivalently, with all possible $(\xi,\phi)$ or all possible $(x,y)$ for a given $R_{\rm L}$, see \Sec{sec:coords}.

The E2C for inclusive jets in vacuum, in medium with wake, and in medium without wake are plotted in the left panel of \Fig{fig:projectedE2CE3C}. When comparing the in-medium cases to the  in-vacuum case, there are several notable features. First, in both the medium with wake and medium without wake, the peak of the E2C shifts to the left, to 
a smaller angle than in vacuum.
For jets in vacuum, the peak in the E2C 
marks the transition between the small angle regime where the E2C takes on a scaling form governed by hadronization and the large angle regime where it has a different scaling form governed by the dynamics of the parton shower. The parametric dependence of the location of this peak, in vacuum, is
$R_{\rm L}\sim \Lambda_{\text{QCD}}/p_{\rm{T,jet}}$ ~\cite{Komiske:2022enw} which means that we can immediately interpret its shift to the left in heavy-ion collisions as arising from the 
energy loss experienced by parton showers as they traverse droplets of QGP. The inclusive jets that
we have selected in the heavy ion samples with $p_{\rm T,jet}>140$~GeV$/c$ have this $p_{\rm T}$ after they have lost energy; if they had developed in vacuum instead, they would have had a higher $p_{\rm T}$. Remarkably, this is encoded in the location in $R_{\rm L}$ of the peak of the E2C: the percentage via which this peak is shifted to the left in PbPb collisions relative to pp collisions is a reflection of the average percentage energy lost by the parton showers in medium. 
Second,
in particular in the case of the medium without wake we see that there is a modest modification to the scaling of the E2C to the right of the peak, in the region where the E2C is controlled by the dynamics of the parton shower, and a modest increase in the height of the peak.  These effects are due to parton energy loss in medium, both via the resulting modification to the parton showers and via the bias toward selecting an inclusive jet sample in
which the jets are narrower. (This bias
is also a consequence of energy 
loss, which biases inclusive jet selection toward those jets with a given $p_{\rm T}$ which lose the least energy~\cite{dEnterria:2009xfs,Renk:2012ve,Spousta:2015fca,Milhano:2015mng,Casalderrey-Solana:2016jvj,Rajagopal:2016uip,Brewer:2017fqy,Brewer:2018mpk,Casalderrey-Solana:2019ubu,Caucal:2020xad,Du:2020pmp,Brewer:2021hmh,Caucal:2021cfb,Pablos:2022mrx,Kang:2023ycg,CMS:2024zjn}.) When soft hadrons from the wake are included, this effect remains visible but it is partially obscured by a third feature$\ldots$

The third feature that we see in the the left panel of \Fig{fig:projectedE2CE3C}, at large angles $R_{\rm L}$, is the most interesting one for our discussion in this paper. 
If we first compare the E2C at large angles for the medium without wake to that in vacuum, we see that the modification of the parton shower changes the scaling of the E2C at large angles, which makes sense as that is where it is governed by the dynamics of the parton shower.
But when we compare the E2C at large angles for the medium with wake to that without the wake, we see a significant
additional enhancement at large values of $R_{\rm L}$, such that the E2C no longer scales at large angles as it does in vacuum.
By comparing the E2C for PbPb collisions with and without jet wakes -- something that we can do in our model study but that is impossible to do in the analysis of experimental data -- we see that this observed 
enhancement at large values of $R_{\rm L}$ is due to the incorporation of hadrons coming from the wake in the jets. 
Although in the Hybrid Model this enhancement is unambiguously due to to the presence of the wake, all other implementations of jet-medium interactions that have been investigated to date show substantial modifications to the E2C at similar angular scales~\cite{Andres:2022ovj,Andres:2023xwr,Andres:2023ymw, Barata:2023vnl,Barata:2023zqg,Yang:2023dwc,Barata:2023bhh,Barata:2024nqo,Andres:2024ksi} that also break the scaling of the E2C seen at large angles for jets in vacuum. These analyses all involve more significant modification to the parton showers than in the Hybrid Model -- it is a feature of the strong coupling dynamics built into the Hybrid Model that the only modification to the parton shower arises from energy loss. We leave to future work the investigation of the E2C and higher-point energy correlators in the Hybrid Model augmented by the addition of elastic scattering between jet partons and weakly coupled medium partons resolved at high momentum transfer~\cite{Hulcher:2022kmn}, which further modifies the parton showers.

Our goal in this work is to search for observables that may offer a 
more unique signature of the hadrons coming from jet wakes, in particular looking for observables constructed from higher-point correlators in which modifications of the parton shower are separated in correlator-space from modifications to the correlator caused by the hadrons 
from jet wakes.  In the next Section, we shall show that the full shape-dependent three-point energy-energy-energy correlator offers the opportunity to do exactly this.  In the remainder of this Section, we begin our analysis of three-point correlators by looking at the projected E3C.

\begin{figure}[t]
    \begin{center}
    \includegraphics[scale=0.6]{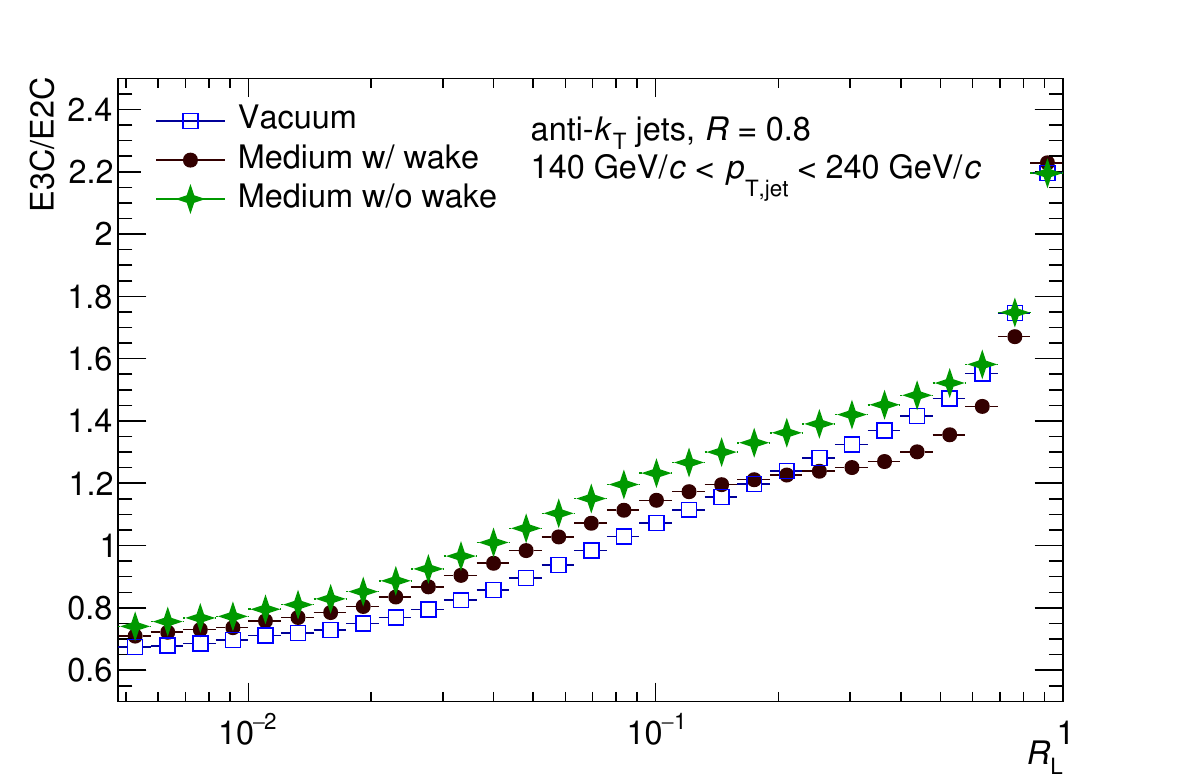}
    \end{center}
    \caption{The ratio of the two- and three-point projected correlators (E3C/E2C) in the vacuum, medium without wake, and medium with wake inclusive jet samples. Comparing medium with wake to vacuum, we see that the medium imprints itself as a modification of the E3C/E2C ratio above a characteristic angular scale, with the vacuum scaling behavior seen only at smaller angles. By comparing medium with wake to medium without wake, we see that the hadrons coming from the wake make a bigger contribution to modifying the E3C/E2C ratio than does the modification of the parton shower.}
    \label{fig:E3CEEC_ratio}
\end{figure}

In the right panel of \Fig{fig:projectedE2CE3C}, we repeat the analysis that we have just discussed, this time for the E3C.  Every aspect of the discussion of the E2C in the  paragraphs above can be repeated here for the E3C. We see the same three physical effects again, encoded in the E3C in much the same way as in the E2C.
Now, though, we can consider the ratio of the three-point projected correlator to the two-point projected correlator (E3C/E2C), plotted in \Fig{fig:E3CEEC_ratio}. This observable was originally proposed for vacuum measurements in Ref.~\cite{Chen:2020vvp} with the motivation that the ratio cancels leading non-perturbative power corrections, and that in the regime where the correlators are controlled by parton shower dynamics the ratio of the scaling form of the E3C to the scaling form of the E2C
is directly proportional to the value of the QCD coupling constant $\alpha_{\rm s}$. For jets in vacuum, the E3C/E2C ratio an be calculated with control, and it has recently been measured in experimental 
data~\cite{Fan:2023,CMS:2024mlf}. 
The E3C/E2C ratio for jets in vacuum can be found in \Fig{fig:E3CEEC_ratio}. For jets in vacuum, the observable exhibits a characteristic power law scaling, governed by the twist-2 anomalous dimensions~\cite{Chen:2020vvp}. 
\Fig{fig:E3CEEC_ratio} also shows the same ratio for jets
in PbPb collisions with and without the hadrons coming from jet wakes.  Since the hadrons from the wake break the vacuum scaling for both the E2C and the E3C separately, it is no surprise that their ratio also does not scale as in vacuum.
At very small angles, though, the ratio flattens similarly in all cases, since both the E3C and E2C exhibit 
vacuum-like hadronization-governed scaling in this regime in PbPb collisions and since the wake does not contribute to the correlators at these very small angles.

Let us focus further on the behavior of the E3C/E2C ratio in medium. First, in the case without wake, we observe a mild deviation in the scaling at large angles. We believe that this is due to the modification of the parton shower
by parton energy loss in the Hybrid Model.
With the wake turned on, the ratio no longer scales in the way that it does in vacuum.
It is interesting that we observe an almost flat region 
in the dependence of the E3C/E2C ratio as a function of $R_{\rm L}$ over a range of relatively large angles.
This may indicate that the wake contributes a distribution of hadrons that is nearly uniform as a function of angular scale, since if this is the case and if the hadrons from the wake dominate then this ratio would be flat.
Note that this does not imply that the wake has a uniform overall shape, just that the angular distribution of the hadrons coming from the wake is ``boring'' within whatever the shape of its overall envelope may be, with no spikiness or substructure at smaller angles. 
See \App{sec:coords_art} for further discussion, but it is certainly the case that this observation 
motivates further modeling and analysis of the E3C and of
of higher point correlators, which enable us to probe the shape and structure of the wake and not just its scaling.
It will also be very interesting to study the behavior of this ratio in other simulations of jets in QGP, or in analytic calculations incorporating medium effects. In particular, it would be interesting to see how different models of parton energy loss, medium-induced radiation, and medium response modify the E3C/E2C ratio.

We believe that our initial investigation motivates the experimental measurement of the E3C/E2C ratio in heavy-ion collisions, as an observable that is sensitive to the modification of jet properties in medium originating from both modifications of the parton shower and the wake in the droplet of QGP. Furthermore, this observable should have nice experimental properties, since many experimental uncertainties should cancel in the ratio.

\section{Shape-Dependent Energy-Energy-Energy Correlators for Jets in QGP}\label{sec:shape}

In \Sec{sec:proj}, we have seen how the dynamics of jet wakes are imprinted on the projected correlators at large angles, breaking the way that they scale in vacuum.
We have also seen, however, that other aspects of the in-medium dynamics of jets, namely the medium-induced modifications of parton showers, also influence the projected correlators at comparable large values
of the angle $R_{\rm L}$. We emphasize that in the case of the Hybrid Model where parton showers are modified only via energy loss this yields smaller modifications to the 
projected correlators than do the hadrons from jet wakes.
Motivated by this, we now look beyond just the projected three-point correlator so as to study the shape-dependence of the EEEC and not just the dependence on its overall angular scale.
The EEEC has been calculated in vacuum QCD in Refs.~\cite{Chen:2019bpb} and studied further in Refs.~\cite{Chen:2022jhb,Chang:2022ryc,Chen:2022swd, Komiske:2022enw}. The vacuum EEEC is shown in \Fig{fig:EEEC_vac} for $0.6 < R_{\rm L} < 0.7$. In both the $\xi$-$\phi$ coordinate system and the $x$-$y$ coordinate system we see that the EEEC is enhanced in the squeezed triangle region due to the collinear singularity of the vacuum splitting functions, so much so that all other regions in the EEEC phase space, namely regions described by triangles that are not squeezed, are relatively unpopulated.

\begin{figure}[t]
    \begin{center}
    \subfloat[]{\includegraphics[width = 0.53\textwidth]{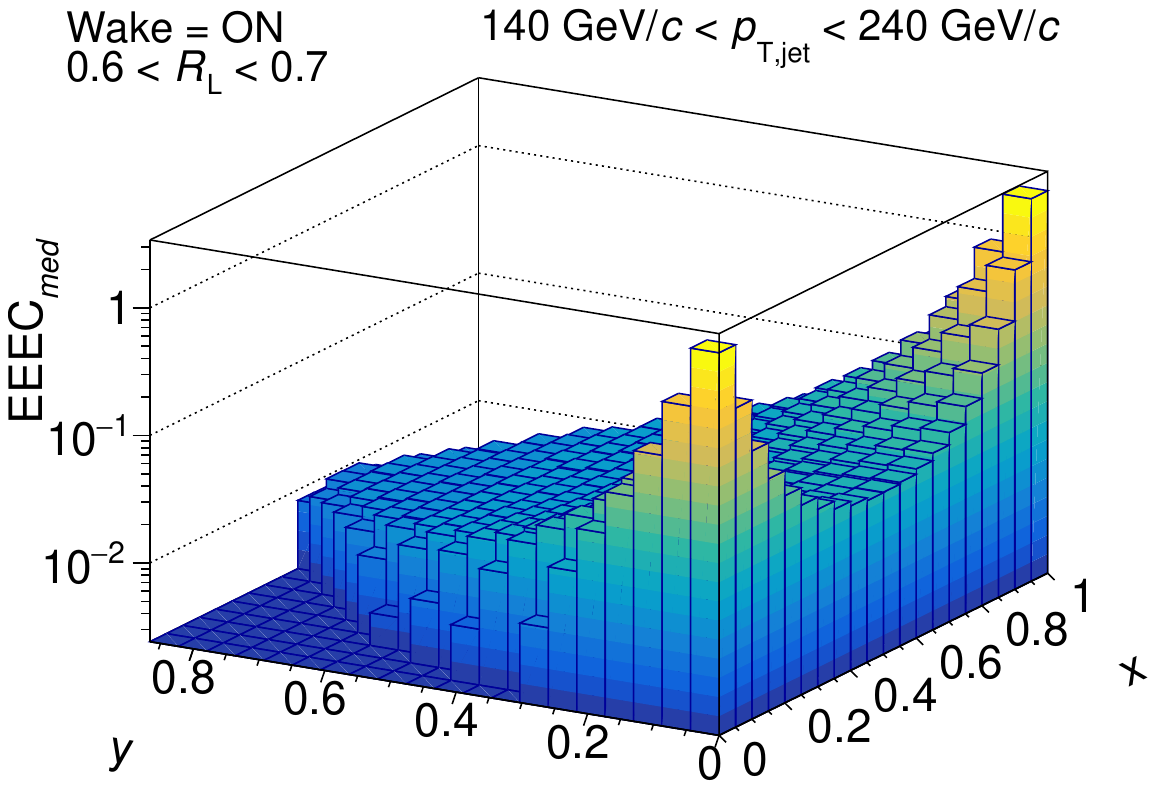}}
    \subfloat[]{\includegraphics[width = 0.53\textwidth]{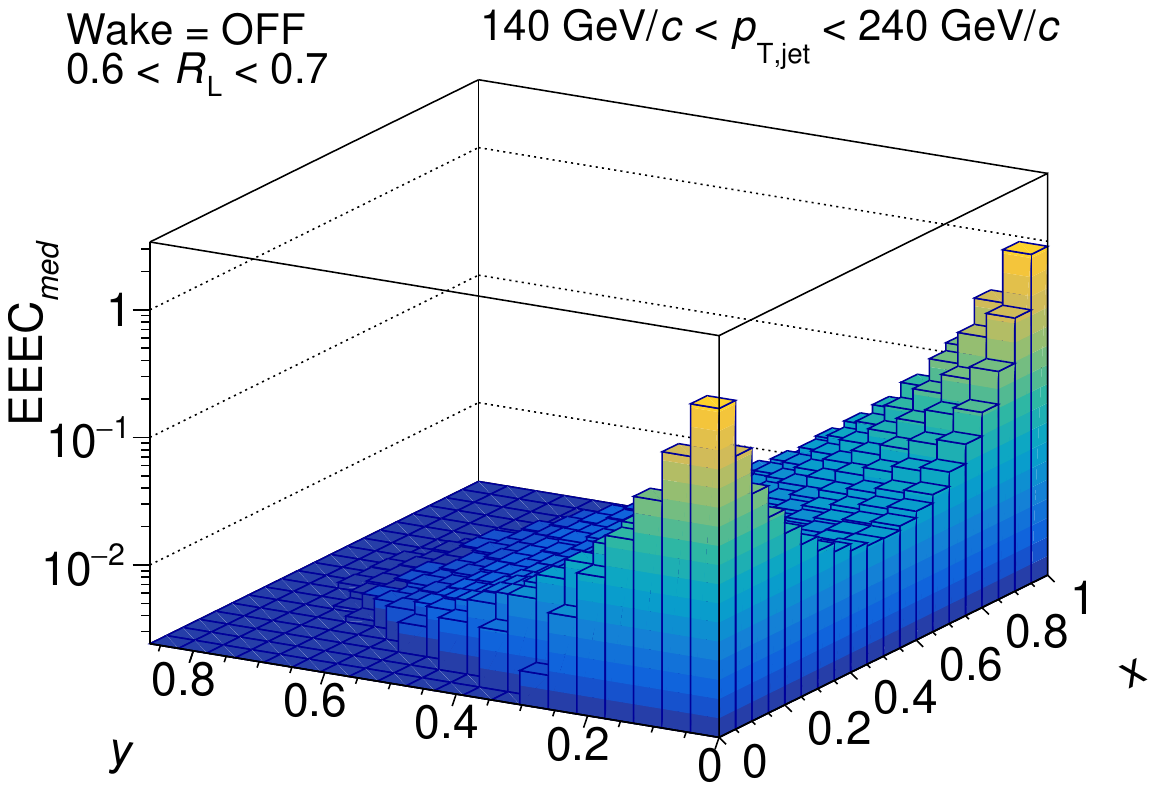}}
    \end{center}
    \caption{The in-medium shape-dependent EEEC both with the wake (a) and without the wake (b). The wake manifests as a clear enhancement in the equilateral region of the EEEC, where hadrons originating from the wake dominate the correlator. These EEECs should be compared to the EEEC for jets in vacuum, plotted in the right panel of \Fig{fig:EEEC_vac}.}
    \label{fig:wake}
\end{figure}

Since the projected E3C correlator shows the dependence of physical effects on the overall angular scale of the correlation, we can use the right panel of \Fig{fig:projectedE2CE3C} to identify the angular scales
at which the effects of hadrons from the wake become 
significant. We see an enhancement in the E3C 
from the wake at values of $R_{\rm L}$ between 
$\sim$ 0.3-0.4 and 0.8, with the fall-off beyond 0.8 arising because
we are looking at $R=0.8$ jets.
Guided by this, to investigate the effect of the 
hadrons from the wake on the full shape-dependent EEEC
we shall focus on the EEEC for relatively large triangles with
$0.6 < R_{\rm L} < 0.7$, for the $R = 0.8$ jets that we 
are using in our analysis. This is in fact the reason why we picked this $R_{\rm L}$-range for our depiction of the EEEC in vacuum in \Fig{fig:EEEC_vac}.
We show the EEEC in this $R_{\rm L}$-range for PbPb collisions with and without hadrons from the wake in the left and right panels of \Fig{fig:wake}.

Before we compare the two panels in \Fig{fig:wake}, it is instructive first to compare the right panel here (jets in-medium, without wake) to the EEEC for jets in vacuum in the right panel of \Fig{fig:EEEC_vac}.  We see that the modification of the parton shower in the Hybrid Model, which arises via every parton in the shower losing momentum and energy to the
medium, depresses the EEEC everywhere, including but not limited to in the
collinear squeezed-triangle regions (around $(x,y) = (0,0)$ and $(1,0)$) which dominate the 
vacuum EEEC.
To see the effect of the hadrons coming from the wake, we now compare the left panel in \Fig{fig:wake} to the right panel. We see that hadrons coming from the wake enhance the EEEC everywhere, for triangles of every shape.
In the squeezed-triangle regions, the diminution from energy loss and the enhancement from the wake 
seem to roughly cancel -- which is similar to what we see in the E3C for $0.6 < R_{\rm L} < 0.7$ in the right panel of \Fig{fig:projectedE2CE3C}.
Other mechanisms that yield medium-induced modifications of the parton shower not seen in the Hybrid Model would further modify the height and shape of the two peaks in the EEEC in \Fig{fig:wake}.  Focusing only on the collinear regions of the EEEC would make it challenging to separate different physical effects. 
Instead, we can focus on the equilateral region of the EEEC, namely the triangles with larger values of $y$, where we
see by comparing the left and right panels of \Fig{fig:wake} that hadrons originating from the wakes of jets completely dominate the EEEC correlator in PbPb collisions!
Comparing to \Fig{fig:EEEC_vac}, we see that in PbPb collisions the hadrons from jet wakes  populate the equilateral region of the EEEC much more than for jets in vacuum. 
When the wake is turned on, as shown in the left panel of \Fig{fig:wake}, one can see immediately that the hadrons from the wake very effectively ``fill in" the region of the EEEC that is otherwise less populated, which is to say the equilateral region.

\begin{figure}
    \begin{center}
    \subfloat[]{\includegraphics[width = 0.53\textwidth]{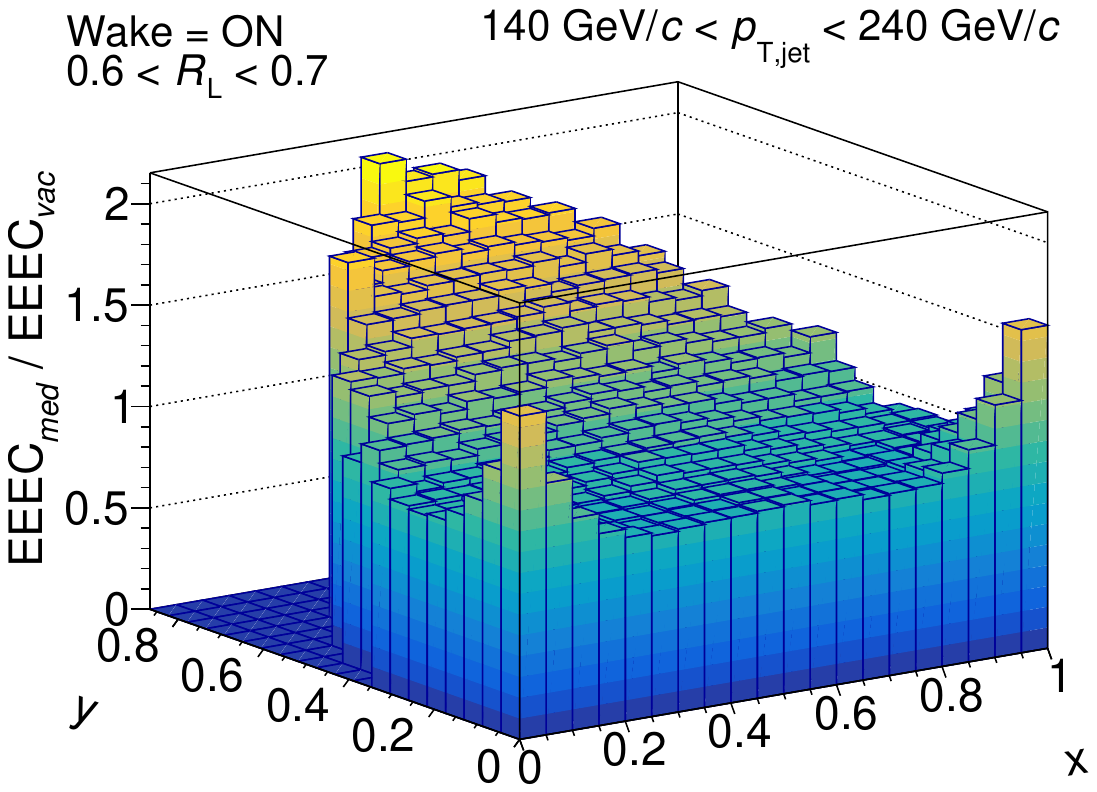}}
    \subfloat[]{\includegraphics[width = 0.53\textwidth]{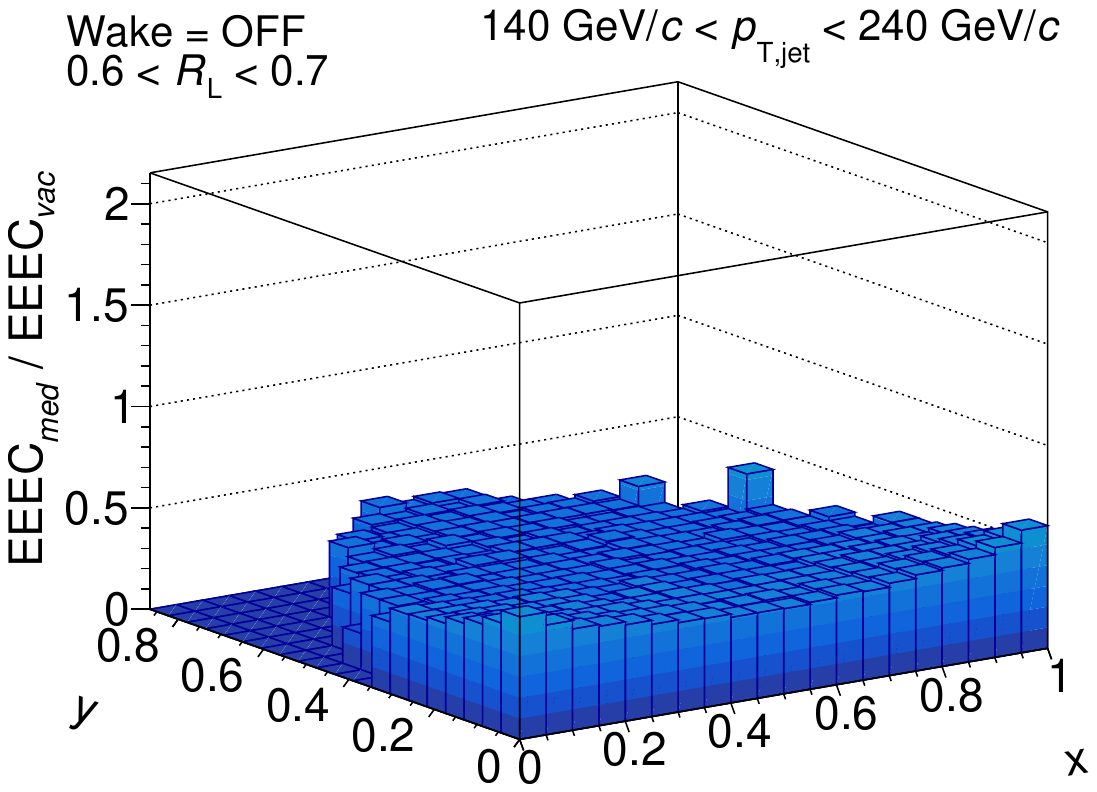}}
    \end{center}
    \caption{The ratio of the shape-dependent EEEC in-medium with wake (a) and without wake (b) to the EEEC in vacuum.
     One can see the enhancement in the equilateral region in the left plot. In comparison, this behavior does not manifest in the case where we remove the wake. Each sample is normalized by the number of jets before taking the ratios.}
    \label{fig:wake_ratio}
\end{figure}

In order to better visualize the effects that we have just discussed, it is useful to look at the ratios of the EEEC in medium with or without the wake  to the EEEC in vacuum, plotted in \Fig{fig:wake_ratio} in $x$-$y$ coordinates.
When hadrons from the wake are turned off, in the right panel of \Fig{fig:wake_ratio}, one can see that the ratio is relatively flat and below 1 over the whole phase space.
This reflects the physics of parton energy loss in the Hybrid Model.  (In fact, if one looks carefully one can see that the suppression of the EEEC is least in the collinear regions, where the most energetic particles are found. This 
reflects the fact that for energy loss in a strongly coupled medium described by Eq.~(\ref{eq:elossrate}), partons with the highest energies lose a smaller fraction of their energy.)
The contrast with the left panel of \Fig{fig:wake_ratio}
is dramatic.
When the hadrons from the wake are included, one can see that there is a substantial enhancement of the EEEC PbPb/vacuum ratio in the equilateral region,
with the ratio of the correlator in-medium to in vacuum reaching values of 2. The comparison between the
two panels of \Fig{fig:wake_ratio} provides striking evidence that the EEEC in PbPb collisions is dominated
by hadrons coming from jet wakes in the equilateral region of the correlator, namely the region
of the EEEC that describes the 
correlation between the energy flow in three well-separated directions.

\begin{figure}[t]
    \begin{center}
    \subfloat[]{\includegraphics[width = 0.53\textwidth]{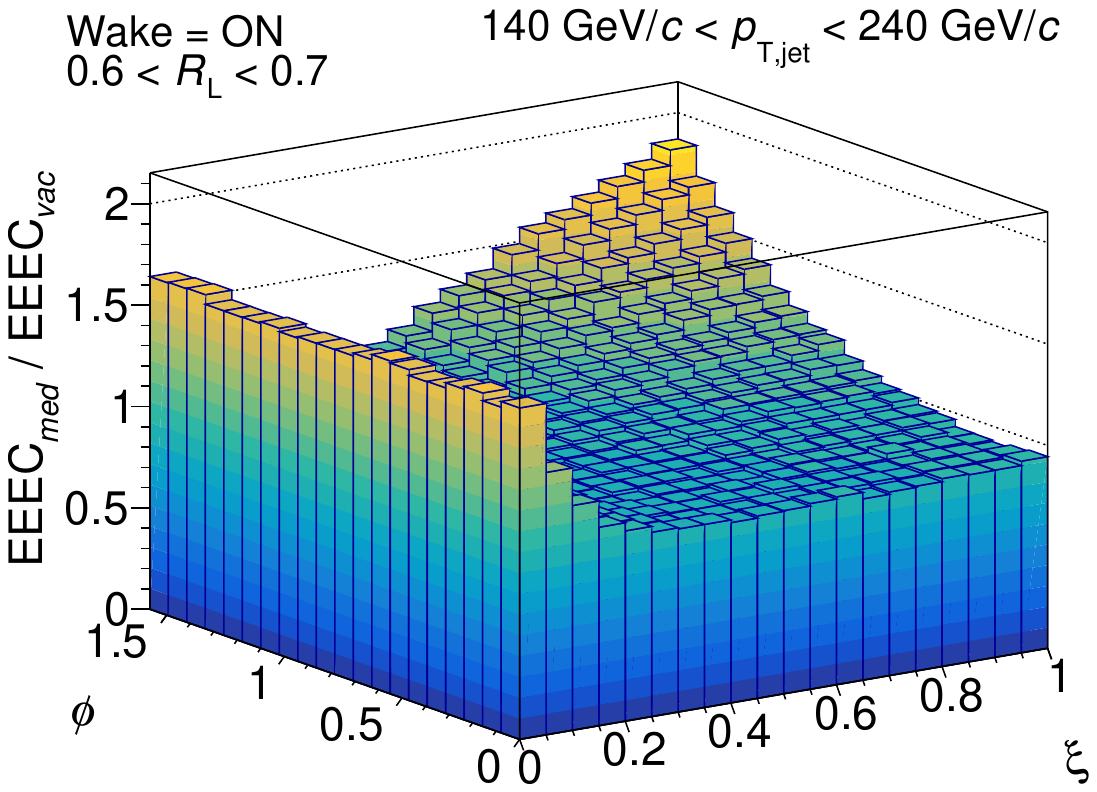}}
    \subfloat[]{\includegraphics[width = 0.53\textwidth]{figures/Latest_Jun/EEEC_xy/WakeOnOverVac/wovervac_final.pdf}}
    \end{center}
    \caption{The ratio of the shape dependent EEEC in medium with wake to vacuum in $\xi$-$\phi$ coordinates (a) and $x$-$y$ coordinates (b). (Panel (b) here is the same as panel (a) in \Fig{fig:wake_ratio}.) The wake shows up as an enhancement in the equilateral region irrespective of the choice of coordinates.}
    \label{fig:compare_coordinates}
\end{figure} 

As we have discussed in \Sec{sec:coords} and describe in detail in \App{sec:coords_art}, plotting 
the EEEC for PbPb collisions from \Fig{fig:wake} in $\xi$-$\phi$ coordinates would introduce coordinate artifacts in the visualization, given that the Jacobian for the $\xi$-$\phi$ coordinates is far from flat in the equilateral region.  The coordinate Jacobian cancels
in the ratio of the EEEC for jets in PbPb collisions to the EEEC for jets in vacuum, however, so in \Fig{fig:compare_coordinates}
we compare this ratio in $\xi$-$\phi$ coordinates to 
that in $x$-$y$ coordinates.
Recall that in $\xi$-$\phi$ coordinates the collinear squeezed-triangle region is at small-$\xi$ and the 
equilateral triangle region is at $\xi=1$, $\phi=\pi/2$.
So, in \Fig{fig:compare_coordinates} we see the substantial enhancement in the EEEC for jets in PbPb collisions relative to that in vacuum in the equilateral region
in both coordinate systems.

It is also instructive to separate the EEEC  into correlations only of hadrons from the parton shower (loosely speaking, correlations of jet particles), correlations only of
hadrons from the wake, and jet-jet-wake and jet-wake-wake correlations. We defer this to \App{sec:coords_art},
but note here that the enhancement originating from the wake in the collinear regions can be attributed to jet-jet-wake correlations 
while the enhancement from the wake in the equilateral region is dominantly from jet-wake-wake correlations.
This separation, which we can do in this study but which cannot be done in experimental data, further confirms that the dominant correlations that fill in the equilateral region of the EEEC are correlations of the jet with wake particles.  In the remainder of this Section we focus instead on qualitative checks that confirm this picture
that could also be performed using experimental data.

\begin{figure}[t]
    \begin{center}
    \subfloat[$n=0.5$]{\includegraphics[width = 0.53\textwidth]{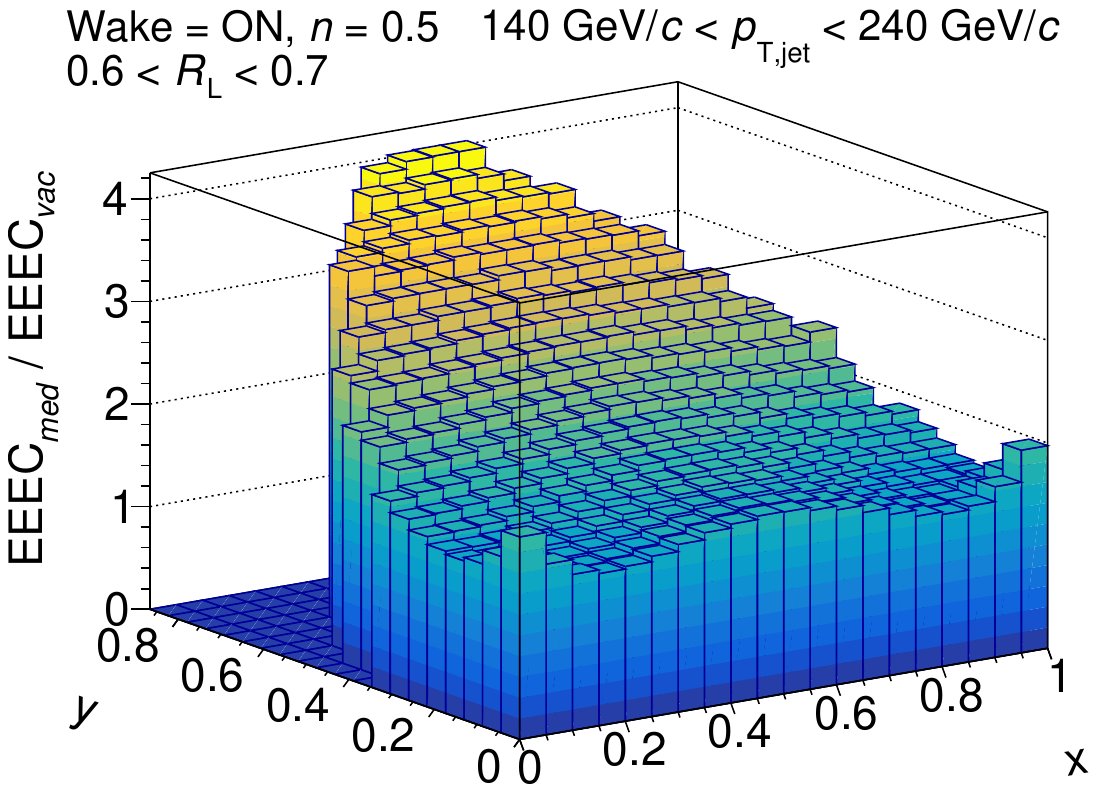}}
    \subfloat[$n=1$]{\includegraphics[width = 0.53\textwidth]{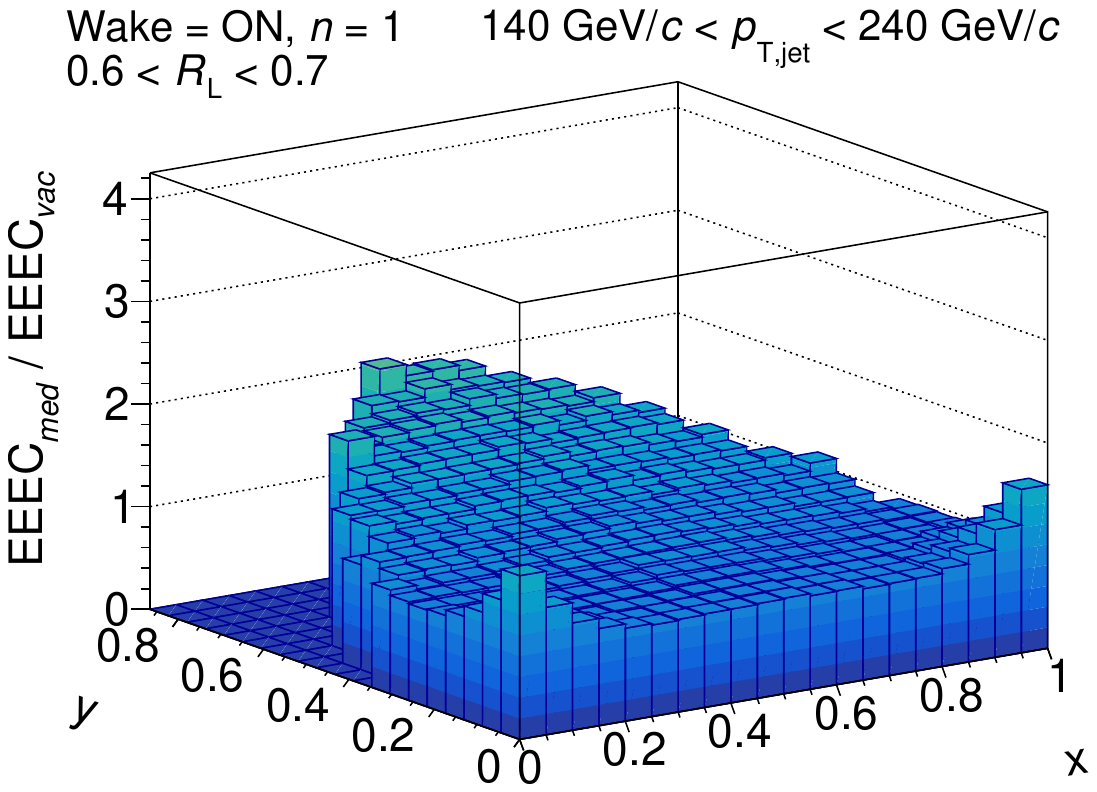}}\\
    \subfloat[$n=1.5$]{\includegraphics[width = 0.53\textwidth]{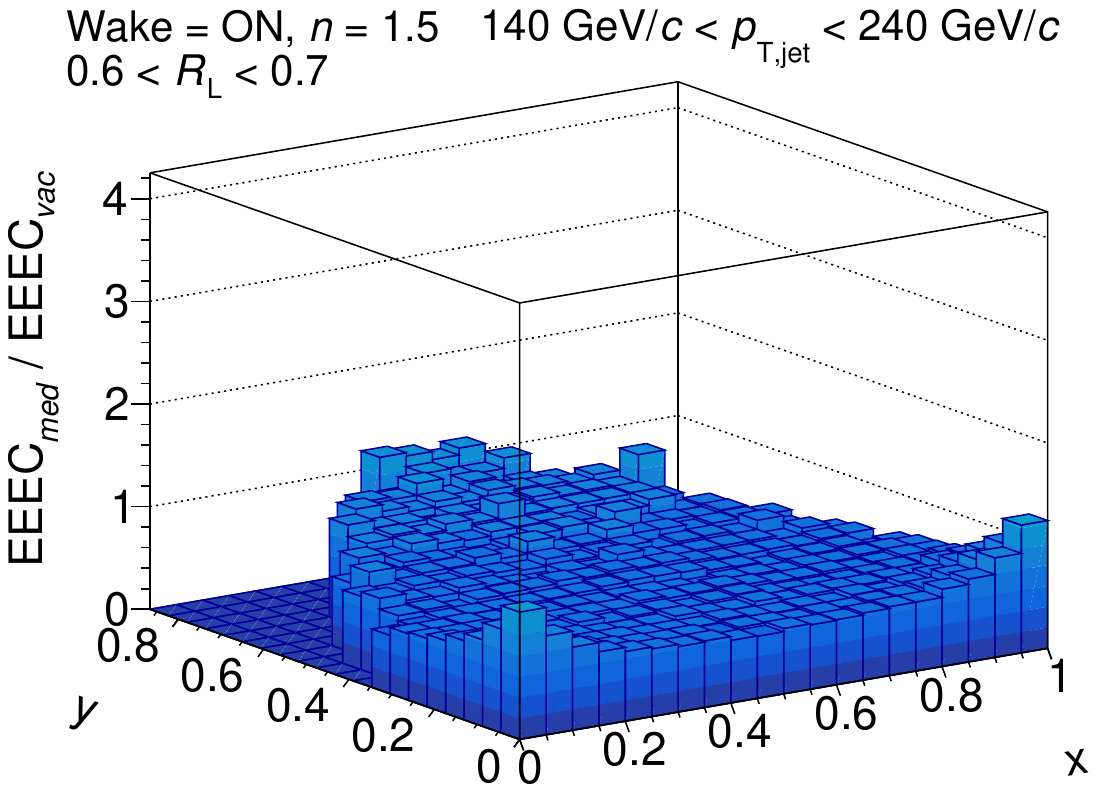}}
    \caption{Ratio of the shape dependent EEEC in PbPb collisions including hadrons from the wake to the EEEC for jets in vacuum, with three different choices of the energy weighting $n$. Panel (b), with $n=1$, is the same as panel (b) in \Fig{fig:wake_ratio} but with a different choice of vertical axis. We see from panels (a) and (c) that when the soft contribution is enhanced or suppressed by correlating ${\cal E}^{0.5}$ or ${\cal E}^{1.5}$, this enhances or suppresses the  contribution of hadrons from the wake in the equilateral region.}
    \label{fig:EnergyWeights}
    \end{center}
\end{figure}

Changing the energy weighting in the correlator
on the right-hand side of
\Eq{eq:projection} so as to look at the three-point correlator of ${\cal E}^n$ for values of $n$ other than 1
can provide another confirmation that the features we have identified in the equilateral region originate from jet
wakes.
Since the wake contributes soft hadrons to the jet, and hence to the EEEC, we can make the EEEC 
less sensitive to contributions from the wake by choosing $n>1$, for example $n=1.5$, or more sensitive to wake contributions by choosing $n<1$, for example $n=0.5$.
In \Fig{fig:EnergyWeights} we show the ratio of the shape-dependent EEEC in PbPb collisions to that for jets in vacuum for three choices of $n$.
When one scans the value of $n$ we can see that the behavior in the equilateral region changes dramatically. For the case of $n = 0.5$, the enhancement of the correlation in the equilateral region relative to that in
vacuum becomes even greater than with $n=1$.
Increasing the value of $n$ to 1.5 reduces the enhancement of the equilateral region significantly.
The fact that the enhancement of the EEEC in the equilateral region can be tuned by enhancing or suppressing the 
contribution of soft hadrons confirms what we have already seen in other ways, namely that the observed enhancement is due to the presence of the wake.
More generally, we believe that using different energy weights $n$ 
will be particularly useful for mapping out the shape in correlator-space of jets in medium in
different models, since it allows us to map out the shape \emph{as a function of energy}.  We have used the angular degrees of freedom that specify the shape of the triangle that defines the EEEC to find separate regions dominated by hadrons from the wake versus hadrons from the parton shower. For our Hybrid Model study, this suffices. But as this study is repeated in models in which the parton shower is more modified than for jets in a strongly coupled liquid as in the Hybrid Model, the ability to vary $n$ as well will become more valuable.

\begin{figure}
    \begin{center}
    \subfloat[$0.4< R_{\rm L} <0.5$]{\includegraphics[width = 0.53\textwidth]{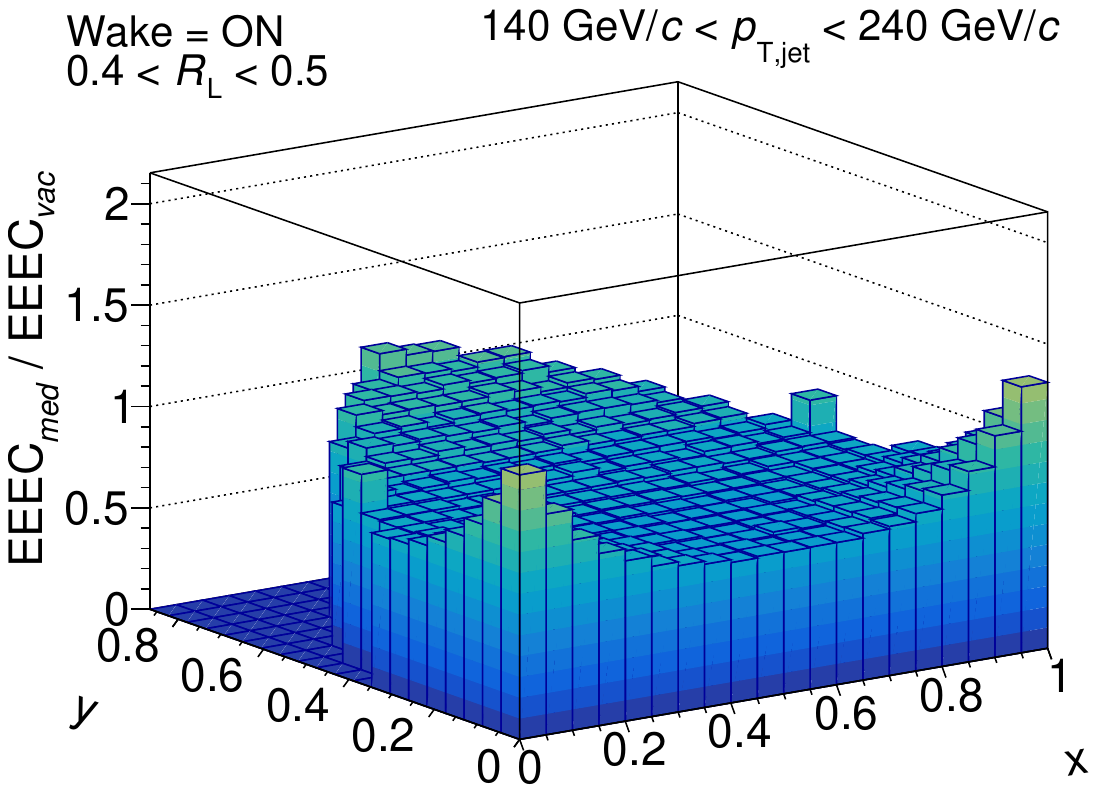}}
    \subfloat[$0.5< R_{\rm L} <0.6$]{\includegraphics[width = 0.53\textwidth]{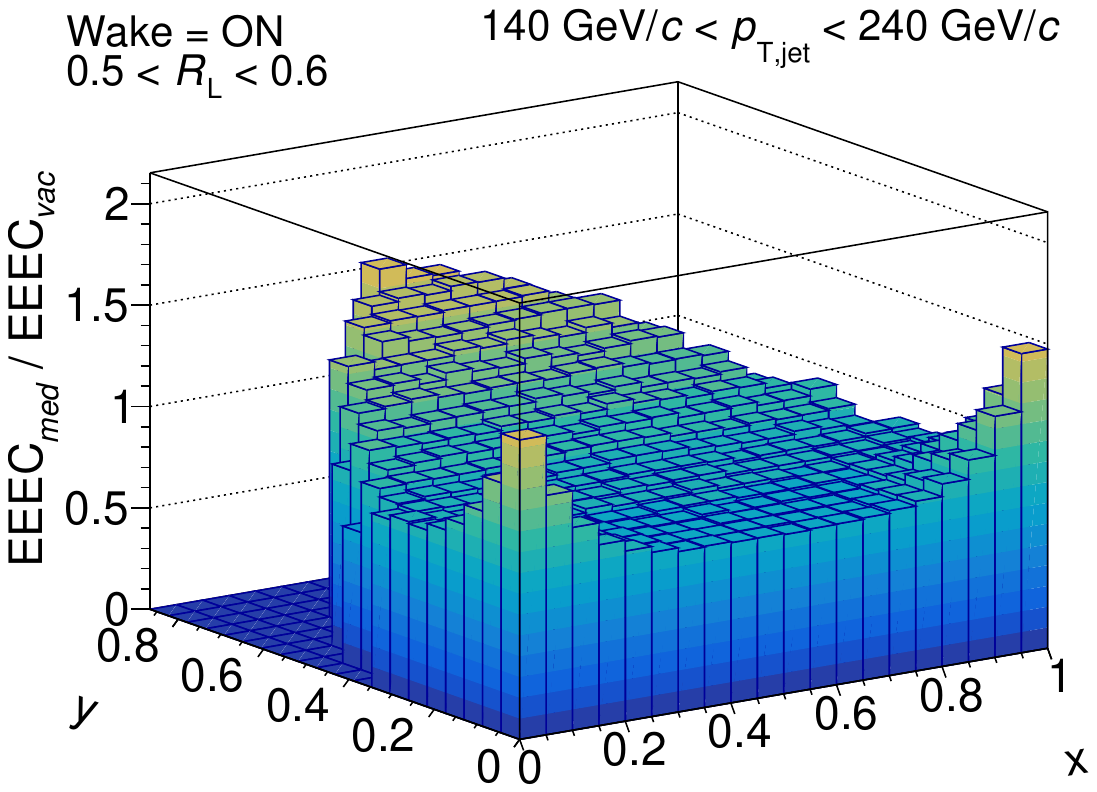}}\\
    \subfloat[$0.6<  R_{\rm L} <0.7$]{\includegraphics[width = 0.53\textwidth]{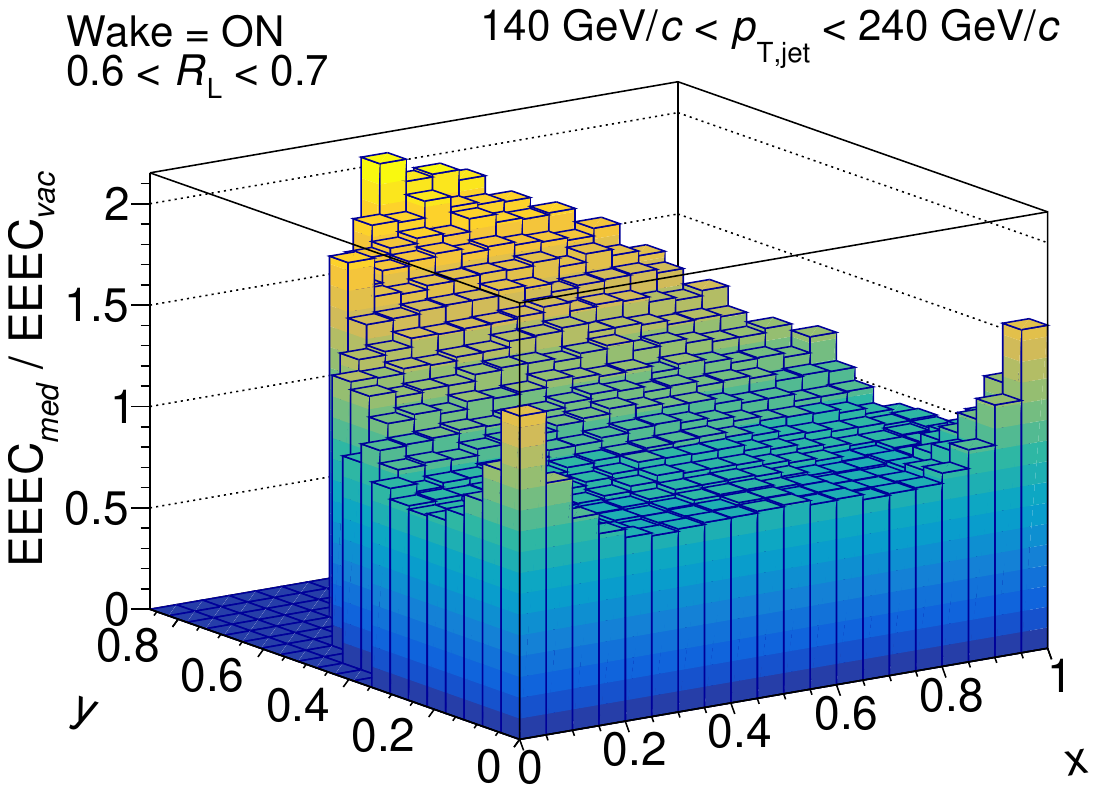}}
    \caption{Ratio of the shape dependent EEEC in PbPb collisions including hadrons from the wake to the EEEC for jets in vacuum, for triangles with three different choices for the range of $R_{\rm L}$, namely three different choices of the overall angular scale. The enhancement of the EEEC in the equilateral region is much larger at the larger values of $R_{\rm L}$ where, as we saw in  \Fig{fig:projectedE2CE3C}(b), the projected E3C is sensitive to the medium. This provides a consistency check of our understanding of the shape-dependent EEEC.}
    \label{fig:wake_diffRslice}
    \end{center}
\end{figure}

Another way to confirm our interpretation of
the equilateral enhancement is to demonstrate that this effect disappears as one moves 
to a region of (equilateral) triangles with a smaller
overall scale $R_{\rm L}$ where, as 
seen in the E3C in \Fig{fig:projectedE2CE3C}(b),
the contribution of hadrons from the wake should be small.
In \Fig{fig:wake_diffRslice}, the EEEC PbPb/vacuum ratio is shown for three different $R_{\rm L}$ slices. As expected, 
the enhancement in the equilateral region is reduced when 
the overall angular scale of the correlator is reduced.

The combination of each of these checks provides confidence that the enhancement of the equilateral region in the ratio of the EEEC in the medium as compared to vacuum is imaging both the energy flow and shape dependence of the wake. The additional information encoded in the shape dependence should lead to an improved discriminating power between various medium response mechanisms, as it has allowed us
to separate, and thus separately visualize, the imprint in the EEEC of the medium-induced modification of parton showers and of the hadrons coming from jet wakes.
It will be very interesting, therefore, to 
study the EEEC in models that implement different
medium-induced modifications of parton showers and
that have different implementations of the response of the medium to a parton shower and the resulting modification of jets as reconstructed from experimental data.
We leave this to future work.

\begin{figure}
    \begin{center}
    \subfloat[]{\includegraphics[width = 0.53\textwidth]{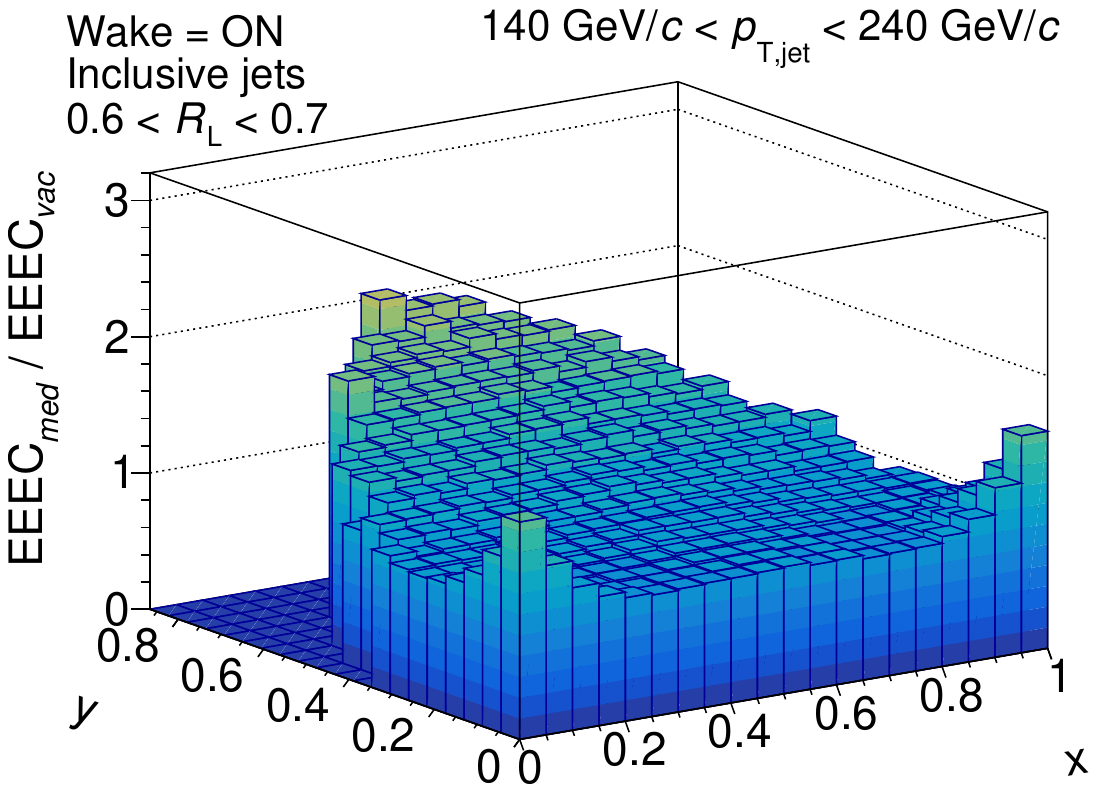}}
    \subfloat[]{\includegraphics[width = 0.53\textwidth]{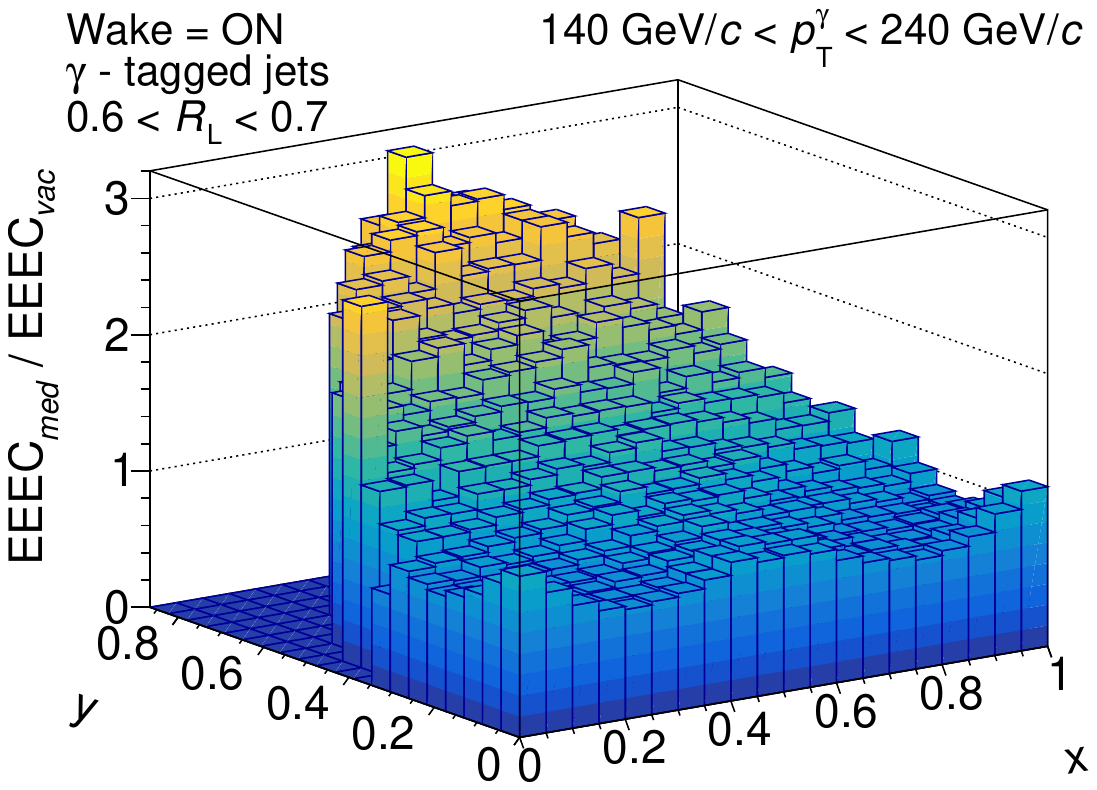}}
    \end{center}
    \caption{The ratio of the shape-dependent EEEC in medium with wake to the EEEC in vacuum, for inclusive jets (a) and for $\gamma$-tagged jets (b). Panel (a) is the same as panel (a) in \Fig{fig:wake_ratio}, but with a different vertical scale. The contribution to the EEEC of hadrons from the wake  is enhanced for $\gamma$-tagged jets. We shall see in \App{app:negasub} that in reality the comparison here reflects a reduction in the EEECs for inclusive jets in medium caused by overlapping negative wakes of away-side jets in the same event.}
    \label{fig:gamVSinc}
\end{figure} 

Finally, in \Fig{fig:gamVSinc} we compare the EEEC for $\gamma$-tagged jets (jets produced in association with a photon with $p_{\rm T}^\gamma>140$~GeV$/c$ and $<240$~GeV$/c$) and inclusive jets (with $p_{\rm T,jet}>140$~GeV$/c$ and $<240$~GeV$/c$). 
We described our procedure for selecting $\gamma$-tagged jets in \Sec{sec:setup}.
We see that, in comparison to inclusive jets, $\gamma$-tagged jets exhibit a larger enhancement in the equilateral region. There are two main factors that contribute to this observation. First, using $\gamma$-tagged jets offers us better access to the unquenched jet $p_{\rm T}$. Second, there is little of the contamination from the superposition of wakes that exists in the inclusive case. As we discuss in detail in \App{app:negasub}, in an inclusive jet sample there is likely to be at least one jet in a direction in the transverse plane that is approximately opposite to the jet selected for inclusion in the sample, and if this is so then the ``negative wake'' from that away-side jet will serve to reduce the magnitude of the wake from the selected jet.  We defer the full discussion to \App{app:negasub}; it suffices here to
say that what we see in \Fig{fig:gamVSinc} provides 
further confirmation, albeit somewhat indirect in this case, that the enhancement of the EEEC in the equilateral 
region originates from jet wakes.

We have now confirmed in multiple ways that
the dramatic enhancement of the EEEC that we have 
found for jets in PbPb collisions relative to jets
in vacuum in the equilateral region, where the EEEC describes the correlation of energy flow in three well-separated directions, is due to the incorporation of
soft hadrons originating from jet wakes in the jets that
are reconstructed from experimental data.
In essence, we have shown that looking at the EEEC for
triangles of different shapes with the same overall angular scale allows us to separate, and separately visualize,
the hadrons in jets that come from jet wakes  in the droplet of QGP  and the hadrons in jets that come from the medium-modified parton shower. Further investigations, using other models and exploring other energy correlator observables, are obviously warranted. 
In addition, and more important in the long run, what we have seen provides strong motivation for the measurement of the full shape-dependent EEEC in experimental heavy-ion collision data. To this we now turn.

\section{Towards Experimental Measurements}\label{sec:exp}

In recent years there has been significant experimental progress in measuring multi-dimensional jet substructure observables in heavy-ion collisions across a variety of kinematic regimes. (For a recent review, see Ref.~\cite{Cunqueiro:2021wls}). These experimental developments have enabled the first measurements of the energy correlators in heavy-ion collisions, with a preliminary measurement recently being released by the CMS collaboration~\cite{CMS-PAS-HIN-23-004}. In the previous Section of this paper, an idealized case was presented in order to explore the way the jet wake is imprinted in the full shape-dependent energy-energy-energy correlator. In this Section, experimental considerations relevant to this observable will be discussed, and results with more realistic experimental parameters will be shown. 

There are a number of key challenges in measuring the wake of jets in experiment.
The first is the challenge that we introduced in \Sec{sec:intro} and have discussed throughout: the modification of jets arises both from the response of the medium to the passage of the jet and from the medium-induced modification of the parton showers, making it
challenging to distinguish the contributions of jet wakes to observables, leading to challenges in the interpretability of measurements.
We have shown that measurements of the shape dependence of the EEEC will advance our understanding, as we have found regions in the space of shapes of the correlator-triangles (squeezed region) where the medium-modified parton showers dominate the correlator and a region (equilateral region) where hadrons from the wake dominate. Further investigations using other models for jets in QGP will surely yield further insights into how best to use higher-point energy correlator observables toward these goals. In this Section we turn to the experimental challenges, and make a start on optimizing the EEEC observable to meet them.
The two key experimental challenges are related to each other and are both consequences of the fact that the hadrons from the wake are very soft.  One aspect of this challenge is that a fraction, and possibly a significant fraction, of these hadrons will be
below the minimum $p_{\rm T}$ threshold imposed in experimental measurements due to worsening detector resolution at low $p_{\rm T}$. We have investigated the effects of introducing a minimum $p_{\rm T}$ threshold in our analysis, as we discuss below.
The other, related, challenge is background subtraction, which is made more challenging by the softness of the effects we are investigating, by the largeness of the 
anti-$k_{\rm T}$ radius $R$ used in reconstructing jets, and (in the analysis of inclusive jets) by the complications arising from the overlap between the wake of the jet we select and the antipodal negative wakes of any other jets in the event. Due to the higher-dimensional nature of the observable, unfolding (described in Refs.~\cite{DAgostini:1994fjx, Hocker:1995kb, Andreassen:2019cjw}), which typically is applied in addition to the background subtraction in order to correct for smearing effects, will be made more difficult. In this Section we discuss the choices that can be made so as to assess and ameliorate these experimental challenges.

As described in \Sec{sec:setup}, throughout most of this paper we have analyzed samples of inclusive jets with an anti-$k_{\rm T}$ radius of $R = 0.8$. (The suppression of jets with this $R$ has measured at high $p_{\rm T}$ by CMS~\cite{CMS:2021vui}.) We have looked at jets with transverse momenta within the range $ 140 < p_{\rm T, jet} < 240 \text{ GeV}/c$, pseudorapidities $|\eta| \leq 2.0$, and have made no kinematic cuts on the jet constituents. In most of our plots we have employed the energy weighting 
$n = 1.0$ in the correlators that we have computed.

For measurements focused upon the hadrons coming from
jet wakes, or indeed from medium response regardless of how it is treated, it is advantageous to utilize jets with
larger 
radii: the larger the anti-$k_{\rm T}$ radius $R$, the more hadrons from the jet wake will be included in the reconstructed jets.
However, due to the fact that an experimental jet must be fully contained within the fiducial acceptance of the detectors and also on account of the challenges associated with background subtraction, measurements with large radius jets can be more difficult. Even in proton-proton collisions, large radius jets are more susceptible to contamination from the underlying event, which goes as $R^{2}$~\cite{Dasgupta:2007wa,CMS:2020caw}. 
In planning an experimental measurement of the EEEC, therefore, it will be natural to assess how well the measurement can be performed using jets reconstructed with values of $R$ that are smaller than 0.8.  With this in mind, in this Section we shall employ $R=0.6$ and, given this choice, we shall look at the EEEC with $0.4<R_{\rm L}<0.5$.
Furthermore, any experimental analysis will only employ jet constituent hadrons with  $p_{\rm T}$ above some threshold. In this Section, we shall introduce a constituent hadron $p_{\rm T}$ threshold $p_{\rm T}^{\rm const}>0.7$~GeV$/c$,
which is within the reach of experiments at RHIC and the LHC~\cite{CMS:2011iwn,CMS:2015hkr,ALICE:2022wpn, STAR:2020xiv,PHENIX:2020alr}. Reducing $R$ and introducing a constituent $p_{\rm T}$ threshold in order to make the analysis more experimentally realizable will each reduce the number of hadrons in a reconstructed jet that come from its wake, and will thus diminish the enhancement of the EEEC in the equilateral region.  The Hybrid Model in fact overestimates this diminishment because, as discussed in \Sec{sec:wake_review} the crude treatment of the freezeout of the wake that yields Eq.~(\ref{eq:onebody}), results in a wake that is too soft and too wide in angle~\cite{Casalderrey-Solana:2016jvj}.

There are also (at least) two choices that we can make that will increase the signal of jet wakes in the equilateral region expected in an experimental measurement of the EEEC.
The first is to use $\gamma$-tagged jets instead of using inclusive jets.
Electromagnetic probes have a (much) larger mean free path than the size of the droplet of QGP produced in a heavy-ion collision, meaning that the photon has negligible interactions with the medium. 
For this reason, $\gamma$-tagged jets offer a way to select on the unmodified hard scale of the jet, which is useful for removing effects arising from a selection bias~\cite{Brewer:2021hmh}. Numerous experimental measurements have been made of $\gamma$-tagged jets and their substructure over a wide kinematic range~\cite{CMS:2024zjn, ATLAS:2023iad,He:2024roj, STAR:2023ksv}. For this analysis, $\gamma$-tagged jets additionally offer a cleaner signal of wake-like effects as the confounding effect of the superposition of wakes (from the selected jet and the other from the away-side jet) is removed and, as we have seen in \Fig{fig:gamVSinc}, this superposition reduces the impact of jet wakes on the EEEC in the equilateral region in inclusive jet samples relative to what is seen in $\gamma$-jet samples. In this Section, therefore, we shall analyze $\gamma$-jet samples.
The second knob that we can turn
in order to increase the sensitivity to soft particles like those coming from jet wakes, and in so doing enhance the signal of jet wakes in the EEEC, 
is to reduce the energy weight $n$ in the correlator on the right-hand side of 
\Eq{eq:projection}. When $n$ is less than 1, the sensitivity to soft particles is enhanced, whereas when it is greater than 1, the sensitivity to soft particles is suppressed. Indeed, we have seen in \Fig{fig:EnergyWeights}
that choosing $n=0.5$, as we shall do in this Section,
enhances the signature of jet wakes seen in the equilateral region of the EEEC.
One should, however, keep in mind that dialing $n$ down may present both theoretical and experimental challenges. Theoretical challenges will emerge due to enhanced sensitivity to non-perturbative corrections in the pQCD calculation of the EEEC in vacuum, whereas experimental challenges will emerge in terms of performing background/detector corrections in experimental data (for example, via an unfolding procedure). 
We are hopeful that our new $x$-$y$ coordinates are beneficial for background subtraction and subsequent unfolding, since a uniform background will be flat in this coordinate system. Another possible avenue to avoid the kinematic limitations arising from the jet radius would be to construct the energy correlator using all particles in an event as opposed to only those within a reconstructed jet as was originally proposed in  Ref.~\cite{Basham:1978bw} (and subsequently measured in Ref.~\cite{OPAL:1993pnw}).

\begin{figure}
    \begin{center}
        \includegraphics[scale=0.6]{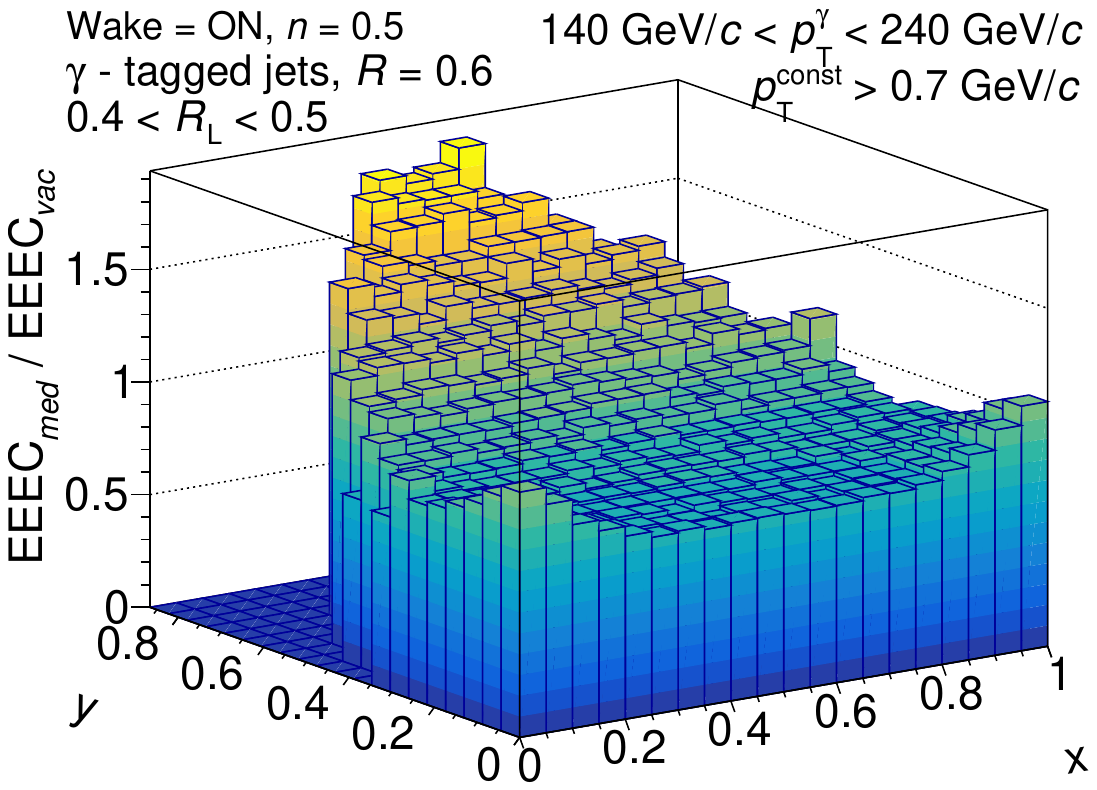}
        \caption{The ratio of the shape-dependent EEEC in medium with wake to the EEEC in vacuum, for anti-$k_{\rm T}$ jets with $R= 0.6$ found in association with a selected $\gamma$ whose $p_{\rm T}$ lies between 140 and 240 GeV$/c$.  We analyze the EEEC for triangles with $0.4<R_{\rm L}<0.5$ and with an energy weight $n=0.5$ for these $\gamma$-tagged jets using only their constituent hadrons whose $p_{\rm T}$ is above a 0.7 GeV/$c$ threshold. With these choices, we see that the shape of the wake manifests itself very clearly in the equilateral region of the EEEC.}
        \label{fig:EEEC_experiment}
    \end{center}
\end{figure}

In \Fig{fig:EEEC_experiment}, we plot the ratio of the EEEC for $\gamma$-tagged jets in medium, including hadrons from the wake, to the EEEC for jets in vacuum, reconstructing the jets with anti-$k_{\rm T}$ $R = 0.6$, as has been done in experimental analyses~\cite{CMS:2021vui,ALICE:2023waz}, and employing a constituent $p_{\rm_T}$ threshold of 0.7 GeV/$c$. We analyze the EEEC with an energy weighting of $n=0.5$, meaning that we are looking at the correlation of ${\cal E}^{0.5}$ in three directions. With these choices,
choices that we expect are along the lines of the choices that will need to be made in an analysis of experimental data, we can see that the soft hadrons coming from jet wakes continue to be the dominant contribution to the EEEC in the 
equilateral region. We are optimistic that measurements with parameters along these lines are possible, and that 
the results of our investigations that we have reported here provide strong motivations for making such measurements as well as for further more detailed studies, both in models and in experimental data, of the shape dependent three-point correlator in heavy-ion collisions.

\section{Conclusions and a Look Ahead}\label{sec:conc}

It has been a long-held vision that the substructure of jets reconstructed in heavy-ion collisions has in
some way encoded in it, and thus experimental
measurements thereof can in some
way reveal, the dynamics of how a shower
of high energy partons is modified as it propagates through
a droplet of QGP and the dynamics of how the hydrodynamic droplet
responds to the passage of a jet through it.
Understanding the dynamics of a parton shower in a strongly coupled medium and understanding the dynamics of a wake that has been excited in a droplet of QGP are
both outstanding and difficult challenges in QCD.
Measurements of jet substructure in heavy-ion collisions
present us with the opportunity to learn about both from the same events, but there is a risk that each may obscure our vision of the other.
We have shown in this paper that
if we extend the measurement and analysis of 
energy correlator observables to three-point 
correlators, we can begin to separate the regions in
the space of correlators where the dynamics of 
parton showers and the dynamics of jet wakes
are each imprinted. This improves the prospects for  experimental
measurements to image, and teach us about, both.

There has been significant recent progress in the theoretical understanding and experimental measurement of energy correlator observables. In the case of jets in QGP, this has primarily focused on the simplest two-point correlator. This observable has now been studied both analytically and in simulations, for a variety of different models of jets in QGP, and has very recently been measured by the CMS experiment. 
While the quantitative predictions differ from model to model, medium modifications to the two-point correlator are qualitatively similar. This is ultimately due to the fact that the qualitative features of the two-point correlator 
are to a significant degree only sensitive to the
overall angular scale or regime of scales that characterize the relevant phenomena, without providing a more complete view into the underlying dynamics.
This motivates the exploration of higher point correlation functions, which enable us to begin to map out the dynamics of jets in QGP and of QGP probed by jets. 
In this paper, we have taken steps in this direction by performing the first study of three-point energy correlators in heavy-ion collisions, using the Hybrid Model to simulate the interaction of 
high-$p_{\rm T}$ parton showers with the droplet of QGP and
the soft hadrons originating from the wakes that they
excite in the hydrodynamic medium. 

Starting with the projected energy correlators, we first showed that the wake imprints itself in the projected two- and three-point correlators, E2C and E3C, and in the ratio E3C/E2C. In all these cases, it does so in a way that breaks the scaling behavior that characterizes these correlators for parton showers in vacuum. 
In fact, because we can turn jet wakes off in the Hybrid Model, we can also see smaller effects in the E2C, E3C and E3C/E2C ratio that are caused by the modification of the parton shower. These effects are found across a similar range of angular scales as those where the hadrons originating from jet wakes make their mark.
In the Hybrid Model, as in a strongly coupled liquid,
parton showers are modified only by parton energy loss.
In models for the jet-medium interaction that are based upon weakly coupled dynamics, the more substantial medium-induced modifications to the parton shower translate into larger modifications to the projected correlators.  In any case, though, the imprints of modifications to the parton showers and of 
the hadrons coming from jet wakes span a comparable range of angular scales, meaning that the projected correlators by themselves are not well-suited to separating these effects.
Measuring the E3C/E2C ratio is nevertheless of interest for at least two reasons. First, it is
advantageous from an experimental perspective as it allows for the cancellation of some uncertainties. And, second,  we find that in the range of angular scales where the contributions from the wake dominate in the Hybrid Model, the E3C/E2C ratio is close to flat.  This makes sense as a characteristic of energy correlators dominated by hadrons from the wake, as they are expected to be randomly distributed within some overall envelope as a function of angle, with no substructure at smaller angles than that.

We then performed the first study of the shape dependence of the three-point correlator in heavy-ion collisions. While the shape of the three-point correlator for jets in vacuum can be, and has been, studied using perturbative QCD calculations, heavy-ion collisions present a particularly appealing use case by virtue of the goal of separating and studying medium-induced modifications to the parton shower and jet-induced modifications to the droplet of QGP.
This is somewhat analogous to one of the
goals of analyzing the three-point correlator
in cosmology, where one aims to isolate and extract information about the dynamics of inflation from the
shape-dependence of the correlator.
Here, we hope to unravel the dynamics of jets in QGP
from the dynamics of QGP probed by jets, with the goal
of imaging both (in correlator space) and improving our understanding as a result.

For this study we have introduced 
a new set of coordinates -- with a flat Jacobian -- that are suitable for the study of the three-point energy-energy-energy correlator (EEEC) in heavy-ion collisions where we do not focus our attention solely on the collinear regime.
We look at correlations between the energy flow in three directions pointing in a triangle whose longest side is $R_{\rm L}$; the collinear regime corresponds to squeezed triangles in which the third vector is very close to one of the other two.
This regime is of interest, as this is where modifications to the parton showers make their mark in correlator space.
But this is not the only regime of interest.
The hadrons coming from jet wakes fill in the regions
of correlator space that are relatively unpopulated by
hadrons from parton showers, as the hadrons coming
from jet wakes introduce EEE correlations that form triangles of all shapes.
The least collinear triangle is an equilateral triangle, and indeed we have found that in the equilateral region of the EEEC, farthest from the region populated by the parton shower, the correlation is dominated by hadrons coming from jet wakes.
The consequence is that in this region the EEEC is substantially larger for jets in PbPb collisions than for jets in vacuum. We have confirmed via several different systematic checks (including varying $R_{\rm L}$, varying the weight $n$ of the three ${\cal E}^n$'s that we correlate, separately correlating hadrons from wakes and hadrons from the parton shower, and comparing inclusive jets to $\gamma$-jets) that the large enhancement of the EEEC in the equilateral region in PbPb collisions originates from
jet wakes.
Since the correlations among 
hadrons originating from the wake
dominates all across the region in which the three vectors that define the correlator are comparably far from each other, in the coordinates that we have introduced the shape-dependent EEEC  gives us an image of the wake of a jet in correlator space. 

The conclusions from our study provide added motivation
for the effort (that is already underway) to improve the theoretical description of the soft hadrons spread over a large angle produced from jet wakes as they freeze out, as the simplified description employed in the Hybrid Model of today is known to yield a distribution that is too soft and distributed over too large an angle.  We very much look forward to confronting 
predictions for the image of the wake of a jet in 
EEEC correlator space from such an improved treatment to
experimental measurements.

The conclusions from our study motivate
the experimental measurement of the projected three-point correlator and the E3C/E2C ratio in heavy-ion collisions, and provide
very strong motivation for the experimental 
effort that will be needed in order to measure
the shape-dependent EEEC. We closed our investigation
by providing an example of the kinds of choices (of $R$, $R_{\rm L}$, constituent $p_{\rm T}$ threshold, and $n$,
as well as choosing to use $\gamma$-jets) that will need to be made in order to optimize a feasible experimental analysis, showing that our conclusions remain clear and strong.
Pursuing such measurements will require the development of sophisticated background subtraction and unfolding techniques. Our results highlight the importance of such an experimental program.  

Once such a program is well underway and analysis tools have been honed on experimental data, an exciting longer term prospect to contemplate is comparisons between
the measured shapes (in correlator space) of jet wakes in central PbPb collisions, in more peripheral PbPb collisions as a function of the angle between the jet direction and the reaction plane, and in collisions of smaller ions.
Perhaps one day by imaging the wakes of jets in smaller collision systems in which the wakes have had less time to hydrodynamize, we can 
image their hydrodynamization as it happens -- something that is today only a dream.

In addition to providing strong motivation for an experimental campaign to measure the shape-dependent EEEC, our work opens the way to many further theoretical investigations with considerable promise of their own: 

\begin{itemize}

\item
An obvious next step is to repeat our study using a model that includes either
weakly coupled elastic scattering of 
jet partons off medium partons
or medium-induced radiation or both, as either should introduce more substantial modifications to the parton showers than in our Hybrid Model study, meaning more substantial modifications to the EEEC in the collinear region where parton shower dynamics dominates.

\item 
We have focused on isolating and imaging the contribution to the EEEC from jet wakes in correlator space, which has led us to focus on the equilateral region. The Hybrid Model is well-suited for this investigation. 
Repeating our study using different models with varying medium-induced modifications to the parton shower, as above,
will motivate careful analysis of the shape of the EEEC around the collinear region, and that this in turn will
increase the motivation for
analytic calculations that build upon the early
work in Ref.~\cite{Fickinger:2013xwa}, where the 
three-point splitting function was computed in medium.

\item
It will also be interesting to investigate higher-than-three-point correlators.  It will take a study of the four-point correlator, which is specified by four vectors that make a tetrahedron whose scale can again be
set by its longest side but whose shape needs five vectors to specify rather than two, 
to ascertain whether this increase in complication yields bewilderment or clarity.  Perhaps it will make it possible to find regions of the five-dimensional shape space where different types of modifications to the parton shower each dominate, or perhaps not. 
Upon developing an intuition for this that we do not
yet have, perhaps we may also find well-motivated ways of projecting the five-dimensional shape space down to two or three dimensions. And, perhaps we can learn from progress that has been made along analogous lines in cosmology.

\item 
We can also speculate about what may be learned from characterizing the shape (in correlator space) of the wake-dominated regime of the EEEC 
using a celestial block decomposition of N-point energy correlators. In Refs.~\cite{Chen:2021gdk,Chen:2022jhb}, this was performed for the three-point correlator computed at both weak coupling~\cite{Chen:2019bpb}, and at strong coupling~\cite{Hofman:2008ar}. 
It would be interesting to study such a decomposition for the three-point correlator computed 
for hadrons originating from the freezeout of a hydrodynamic fluid in a more general context than just the analysis of the wakes of jets.
In the present context, it is also worth investigating whether, and if so how, this decomposition could serve as another means by which to separately characterize effects
originating from jet wakes from effects originating 
from medium-induced modifications to the parton showers.

\item 
Although we focused on the three-point correlator of the energy flow operator, in the context of heavy-ion collisions, it might be natural to consider the fluctuation operator $\delta {\cal E}(\vec{n})\equiv (\mathcal{E}(\vec{n})-\langle \mathcal{E}(\vec{n}) \rangle)/\langle \mathcal{E}(\vec{n}) \rangle$, and its multi-point correlators $\langle \delta {\cal E}(\vec{n}_1) \delta {\cal E}(\vec{n}_2) \cdots \delta{\cal E}(\vec{n}_k) \rangle$ \cite{Hofman:2008ar}.  At weak coupling, the additional terms in correlators of the fluctuation operator are less singular in collinear limits, and therefore do not modify the scaling behavior as compared to the standard energy correlators making them of less interest. However, it would be interesting to investigate correlators of the fluctuation operators on the strongly coupled wake, and in heavy-ion collisions more generally.

\item 
It will also be very interesting to study three-point correlators that involve a suitably defined jet axis. Although other choices merit investigation, using the winner-take-all (WTA) jet axis~\cite{Larkoski:2014uqa} seems particularly promising as this ensures that there is a high-$p_{\rm T}$ hadron along the jet axis. 
Correlators of this type could be natural for studying cylindrically symmetric features of the wake, which are not easily manifest in the $N$-point correlators. 
For example, a traditional jet shape observable 
can be thought of as a two-point correlator between the WTA jet axis and one other point or as a one-point energy correlator with respect to the WTA axis, as in Ref.~\cite{Neill:2018wtk}. 
The three-point correlator between the WTA jet axis and two other points has the potential to yield results that are
complementary to what we have found but that may turn out to be easier to implement in an experimental analysis.  
Furthermore, one can imagine engineering such correlators with the goals of imaging the dynamics in different ways.
Correlating the WTA axis, ${\cal E}^{0.5}(\vec{n}_1)$ and ${\cal E}^{0.5}(\vec{n}_2)$ would yield a study complementary to our own whereas correlating 
the WTA axis, ${\cal E}^{2}(\vec{n}_1)$ and ${\cal E}^{0.5}(\vec{n}_2)$ would yield an observable that is analogous to the jet shape but for a jet with substructure.
If one in addition specifies the angle between the WTA jet axis and the reaction plane in the definition of such correlation observables, studies along the lines
pioneered recently in Ref.~\cite{Xiao:2024ffk} may also be possible.

\end{itemize}
We have only just scratched the surface. We expect that the physics of shape-dependent EEECs has only just begun to stimulate the imagination of theoretical and experimental 
physicists. The interplay between new ideas, new observables, and new experimental data
will lead to 
new ways to image and understand the dynamics of jets in QGP and of QGP probed by jets and new discoveries.

\acknowledgments
We thank Carlota Andr\'es, Jo\~{a}o Barata,
Helen Caines, Kyle Devereaux, Fabio Dom\'inguez, Wenqing Fan, Laura Havener, Jack Holguin, Kyle Lee, Yen-Jie Lee, Yacine Mehtar-Tani, Guilherme Milhano, Rachel Steinhorst, Andrew Tamis, Jesse Thaler, Jussi Viinikainen, Xin-Nian Wang, and HuaXing Zhu for useful discussions. Research supported by the U.S. Department of Energy, Office of Science, Office of Nuclear Physics under grant Contract Numbers DE-SC0004168, DE-SC0011088 and DE-SC0011090. IM is supported by startup funds from Yale University. DP is funded by the European Union's Horizon 2020 research and innovation program under the Marie Sk\l odowska-Curie grant agreement No 101155036 (AntScat), by the European Research Council project ERC-2018-ADG-835105 YoctoLHC, by the Spanish Research State Agency under project 
PID2020-119632GB-I00, by Xunta de Galicia (CIGUS Network of Research Centres) and the European Union, and by Unidad de Excelencia Mar\'ia de Maetzu under project CEX2023-001318-M. ASK was supported by a Euretta J. Kellett Fellowship, awarded by Columbia University. KR is grateful to the CERN Theory Department for hospitality and support. The authors would like to express special thanks to the Mainz Institute for Theoretical Physics (MITP) of the Cluster of Excellence PRISMA$^+$ (Project ID 390831469), for its hospitality and support as this work was completed.

\appendix
\section{Coordinate Choices and Artifacts} \label{sec:coords_art}

\begin{figure}[t]
    \begin{center}
    \subfloat[]{\includegraphics[width=0.53\textwidth]{figures/Latest_Jun/EEEC_nom/WakeOnOverVac/wovervac_final_nom.pdf}}
    \subfloat[]{\includegraphics[width=0.53\textwidth]{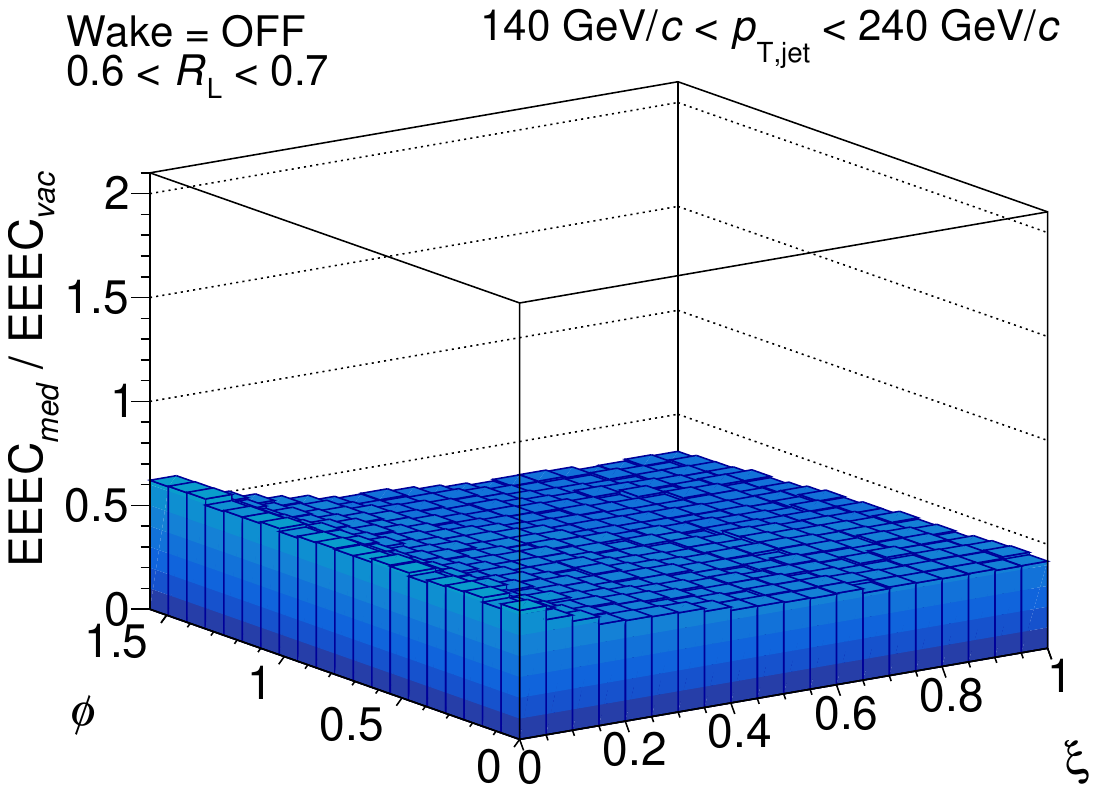}}
    \end{center}
    \caption{The ratio of the shape-dependent EEEC in medium with wake (a) and without wake (b) to the EEEC vacuum, in $\xi$-$\phi$ coordinates. This Figure should be compared to Fig.~\ref{fig:wake_ratio}, where the same ratios are plotted in $x$-$y$ coordinates. Noting that here the equilateral region is $(\xi,\phi)\sim (1,\pi/2)$, we see that here as in Fig.~\ref{fig:wake_ratio} the wake shows up in the left plot as an enhancement in the equilateral region.}
    \label{fig:wakeoffovervacuum_nom}
\end{figure}

\begin{figure}[t]
\begin{center}
\subfloat[Wake-Wake-Wake]{\includegraphics[width = 0.53\textwidth]{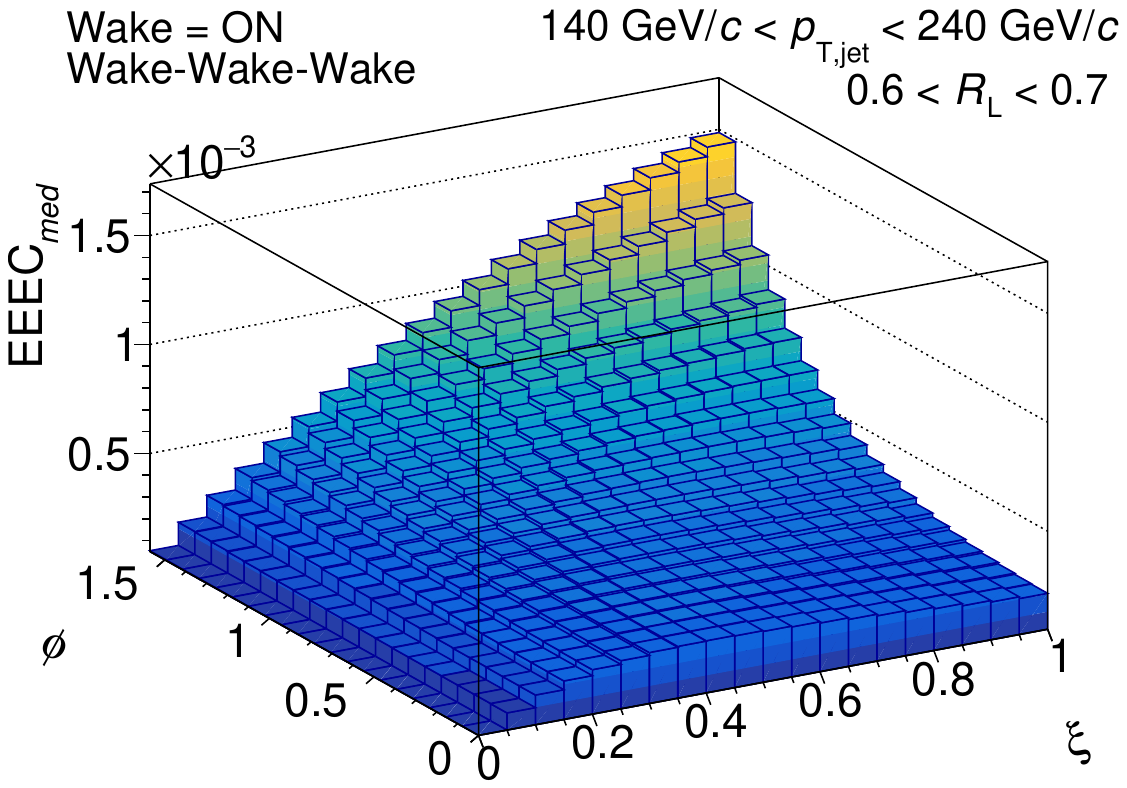}}
\subfloat[Jet-Wake-Wake]{\includegraphics[width = 0.53\textwidth]{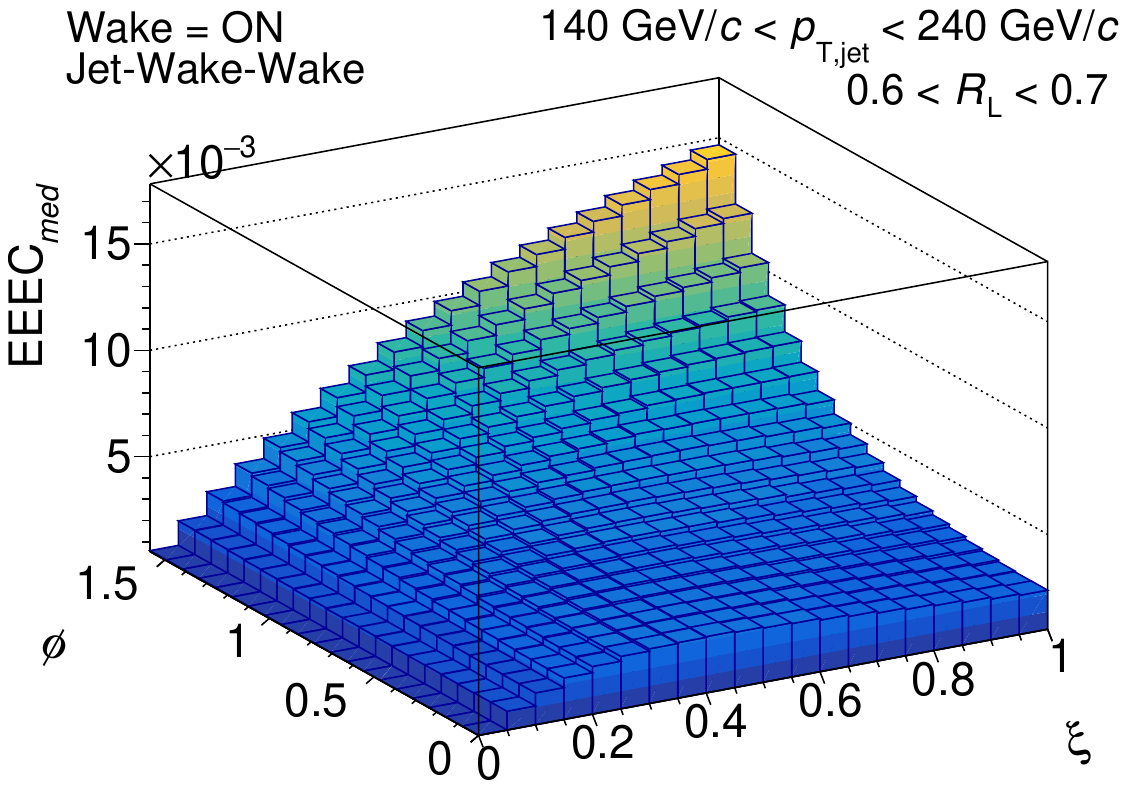}}\\
\subfloat[Jet-Jet-Wake]{\includegraphics[width = 0.53\textwidth]{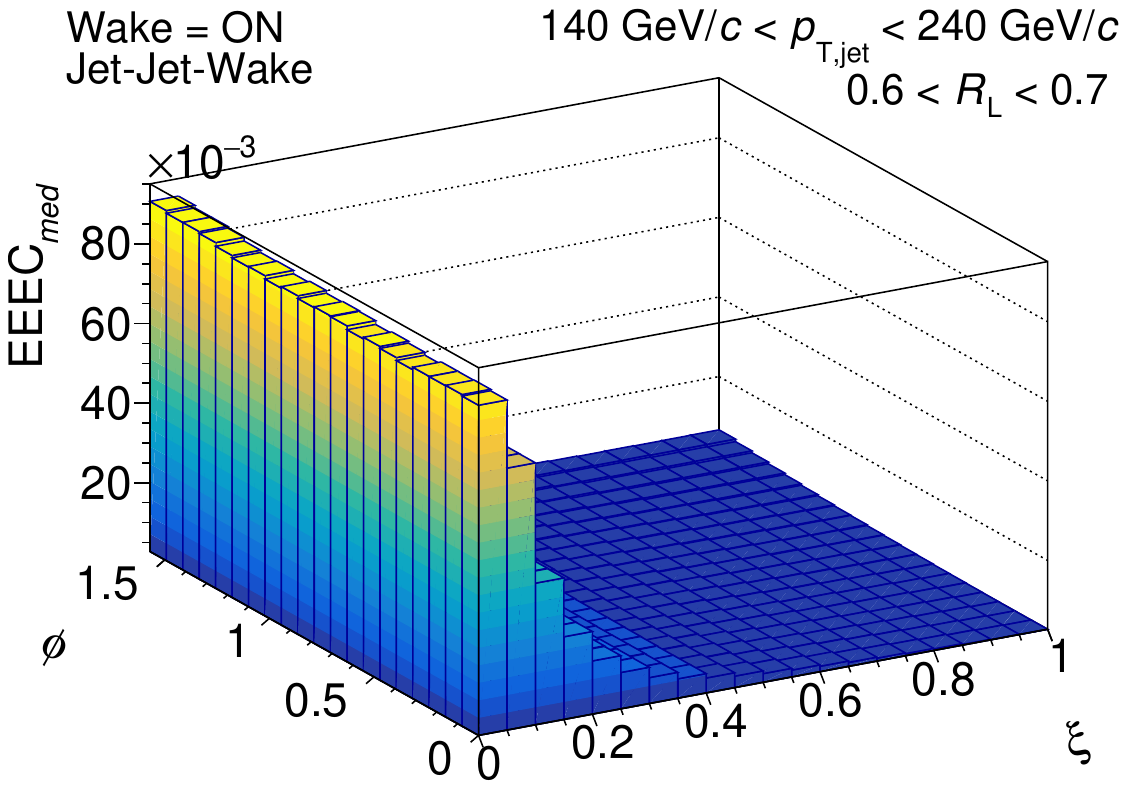}}
\subfloat[Jet-Jet-Jet]{\includegraphics[width = 0.53\textwidth]{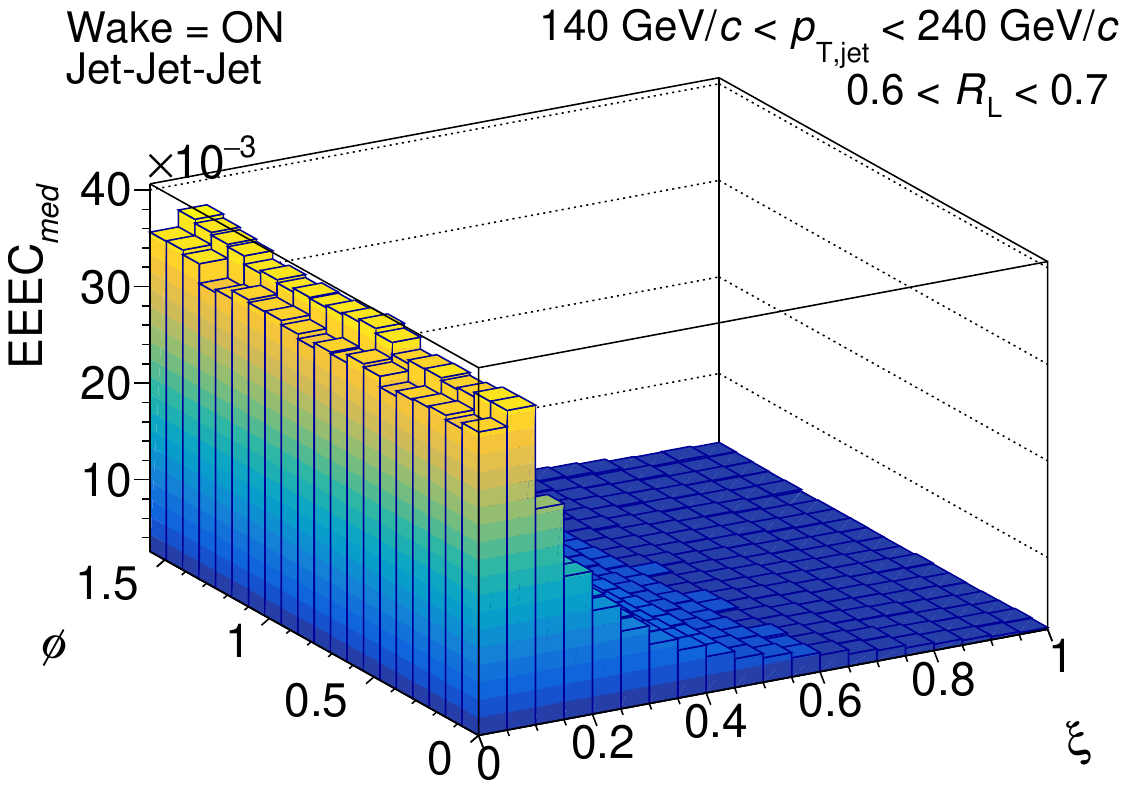}}
\caption{Various contributions to the EEEC for jets in heavy-ion collisions in the ($\xi, \phi$) coordinates. We see that the correlation in the equilateral region is dominated by correlations involving two or more hadrons originating from jet wakes. It also appears as if there is a sharp preference for correlations among hadrons from the wake in directions that form a precisely equilateral triangle; this is a coordinate artifact.}
\label{fig:wake_Contributions_nom}
\end{center}
\end{figure}

In \Sec{sec:coords}, we introduced new coordinates with a flat Jacobian so as to eliminate the effects on the EEEC observable coming from coordinate artifacts, specifically the effects of
a coordinate Jacobian that is not flat. 
Although the $\xi$-$\phi$ coordinate system in the existing literature is useful for studying the features of the three-point correlator in the collinear region which is dominant in vacuum, here we outline the motivation for switching to $x$-$y$ coordinates for visualizing EEECs in the context of heavy-ion collisions where it is important to have a faithful representation of the physics in the equilateral region, where the correlator is dominated by hadrons from jet wakes.

We begin, though, by confirming in \Fig{fig:wakeoffovervacuum_nom} 
that the enhancement of the EEEC in the equilateral region can be seen in either coordinate system.
\Fig{fig:wakeoffovervacuum_nom}
shows the ratio of the EEEC for jets in medium with wake (left panel) and without wake (right panel) to that for jets in vacuum, in the $\xi$-$\phi$ coordinate system. Just as we have seen in the $x$-$y$ coordinate system in \Fig{fig:wake_ratio}, we see that when hadrons from the wake are turned off the ratio is flat and below unity due to parton energy loss. 
And, again as in \Fig{fig:wake_ratio},
when the hadrons from the wake are included we see that this leads to a dramatic enhancement in the EEEC in the equilateral region. This confirms that the choice of coordinates does not impact the central physics conclusions that we draw from ratios of the EEEC for jets in medium to that for jets in vacuum, as the coordinate Jacobian cancels in such ratios. 
We have introduced the $x$-$y$ coordinate system in Section~\ref{sec:coords} because of the advantages that coordinates with a flat Jacobian afford when we look at the EEEC for either jets in medium or jets in vacuum, without taking their ratio. We elaborate upon these advantages in this Appendix.

\Fig{fig:wake_Contributions_nom} shows contributions from different particle-triplet configurations to the EEEC for in-medium jets, in the $\xi$-$\phi$ coordinate system. 
Here, by ``Jet'' particles we mean hadrons originating from the medium-modified parton shower, whereas ``Wake'' particles are hadrons reconstructed as components of the jet that originate from its wake.
One sees that contributions to the EEEC coming from correlations among two or more wake particles are dominant in the equilateral region $(\xi,\phi)\sim (1,\pi/2)$,
as we have already noted  in \Sec{sec:shape}.
However, with this choice of coordinates it also
appears that the correlations among wake particles is
strongly peaked for triangles that are precisely equilateral.
This does not make sense because the hadrons
from the jet wake are approximately uniformly distributed over a broad angular range which should result in correlations for triangles of all shapes including 
equilateral, but not a preference for 
precisely equilateral triangles.
In fact, this perceived sharp preference for equilateral structures is an artifact of the geometry of the $\xi$-$\phi$ coordinate system: it originates from the Jacobian of the coordinate system, not from the physics of jet wakes.

\begin{figure}[t]
\begin{center}
{\includegraphics[width = 0.65\textwidth]{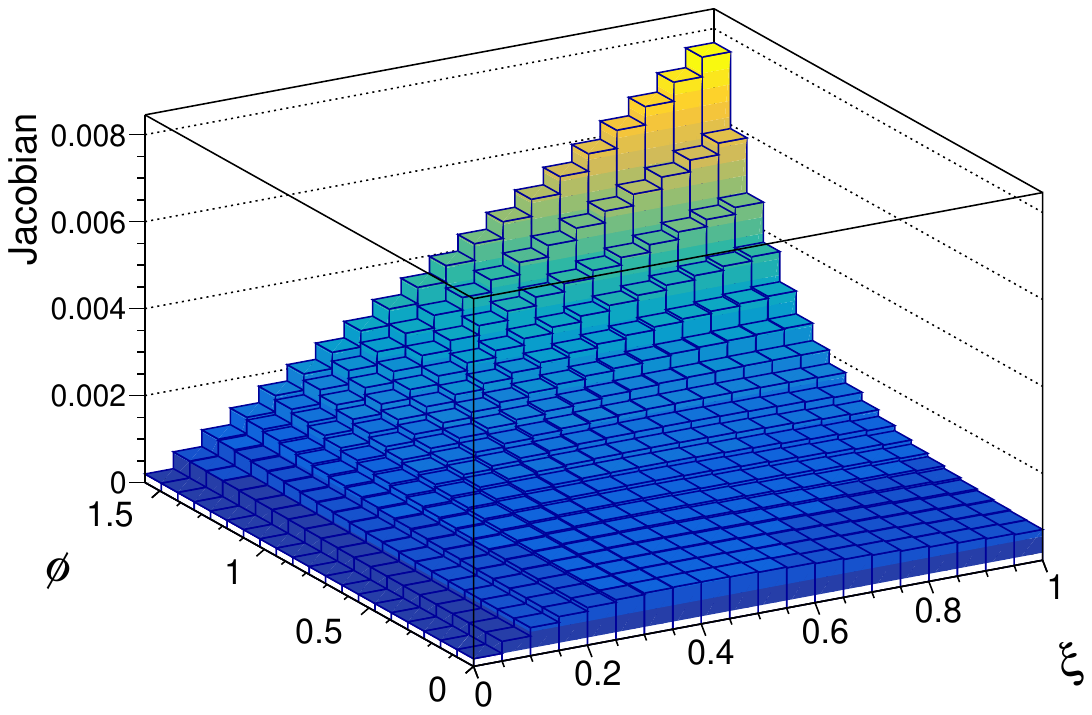}}
\caption{The Jacobian in $\xi$-$\phi$ coordinates. It is far from flat, and has a sharp peak at the equilateral triangle point $(\xi,\phi)= (1,\pi/2)$.}
\label{fig:jacobian_nom}
\end{center}
\end{figure}

In Section \ref{sec:coords}, we defined the $x$-$y$ coordinate system such that for any triplet of particles, the two particles that characterize $R_{\rm L}$ sit 
at $(0, 0)$ and $(1, 0)$, and the third particle sits at some position $(x, y)$  in the plane. Now, imagine that we have a uniform distribution, in $(x, y)$, of all such third particles in the triplets. We can then count, equivalently integrate over, the number of these particles that lie in each $(\xi , \phi)$ bin. This number corresponds to the area of a given $(\xi, \phi)$ bin in the $x$-$y$ coordinate plane. \Fig{fig:jacobian_nom} shows a histogram of these areas; this histogram represents the Jacobian of the $\xi$-$\phi$ coordinate system. One sees that the Jacobian is highly peaked for precisely equilateral triangles and looks qualitatively similar to the EEEC contributions from jet-wake-wake and wake-wake-wake configurations in $\xi$-$\phi$ space in \Fig{fig:wake_Contributions_nom}. This suggests that the visual representation of
the various contributions to the EEEC in \Fig{fig:wake_Contributions_nom}, in particular in the 
equilateral region, is dominated by the coordinate Jacobian, rather than being a faithful visual representation of the physics of the wake.

\begin{figure}[t]
\begin{center}
\subfloat[Wake-Wake-Wake]{\includegraphics[width = 0.53\textwidth]{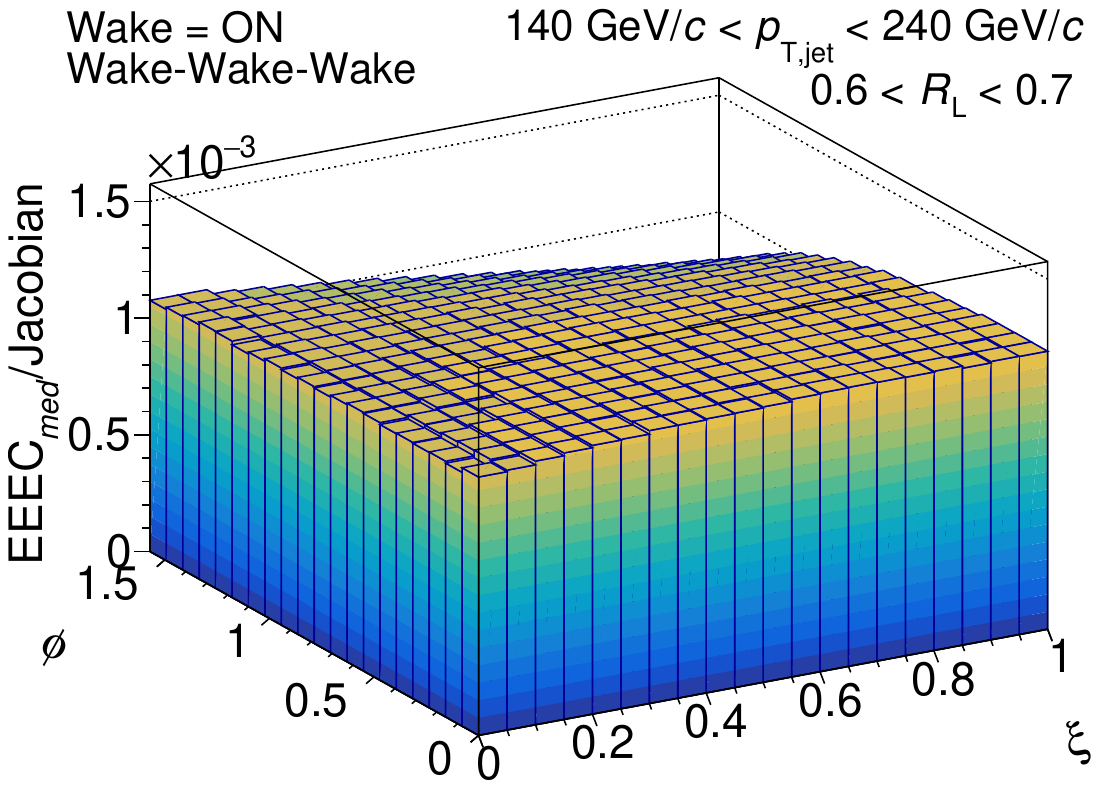}}
\subfloat[Jet-Wake-Wake]{\includegraphics[width = 0.53\textwidth]{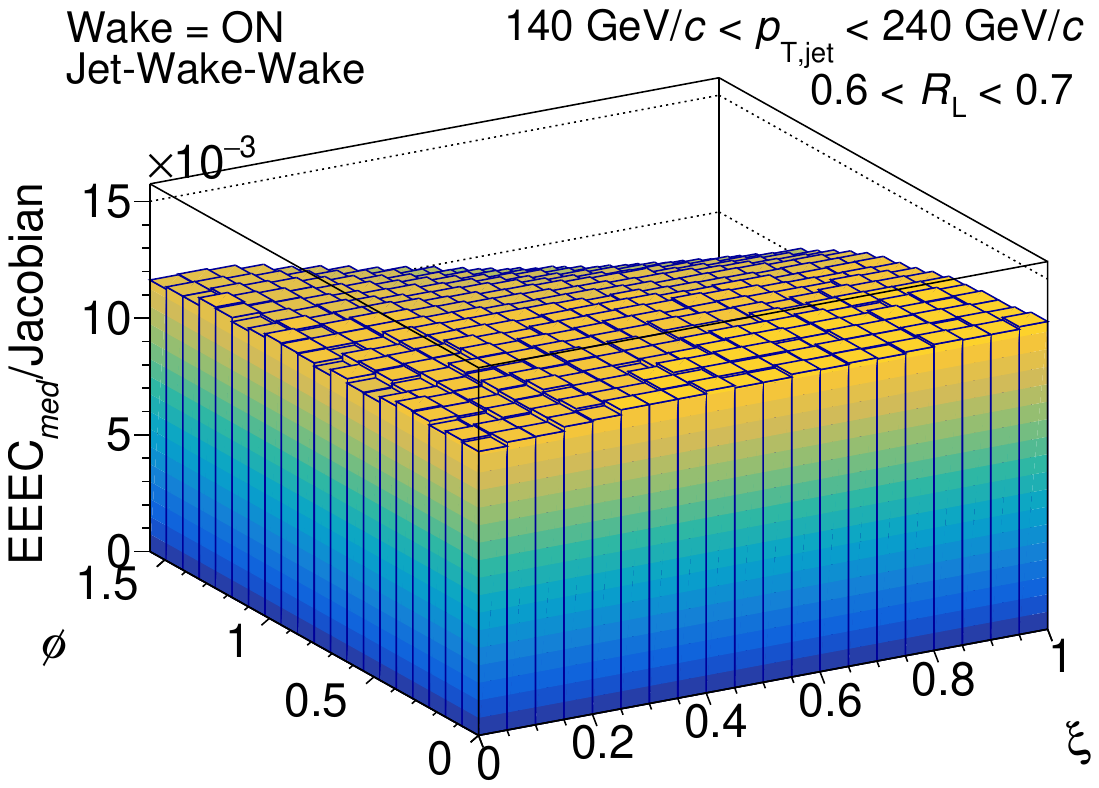}}\\
\subfloat[Jet-Jet-Wake]{\includegraphics[width = 0.53\textwidth]{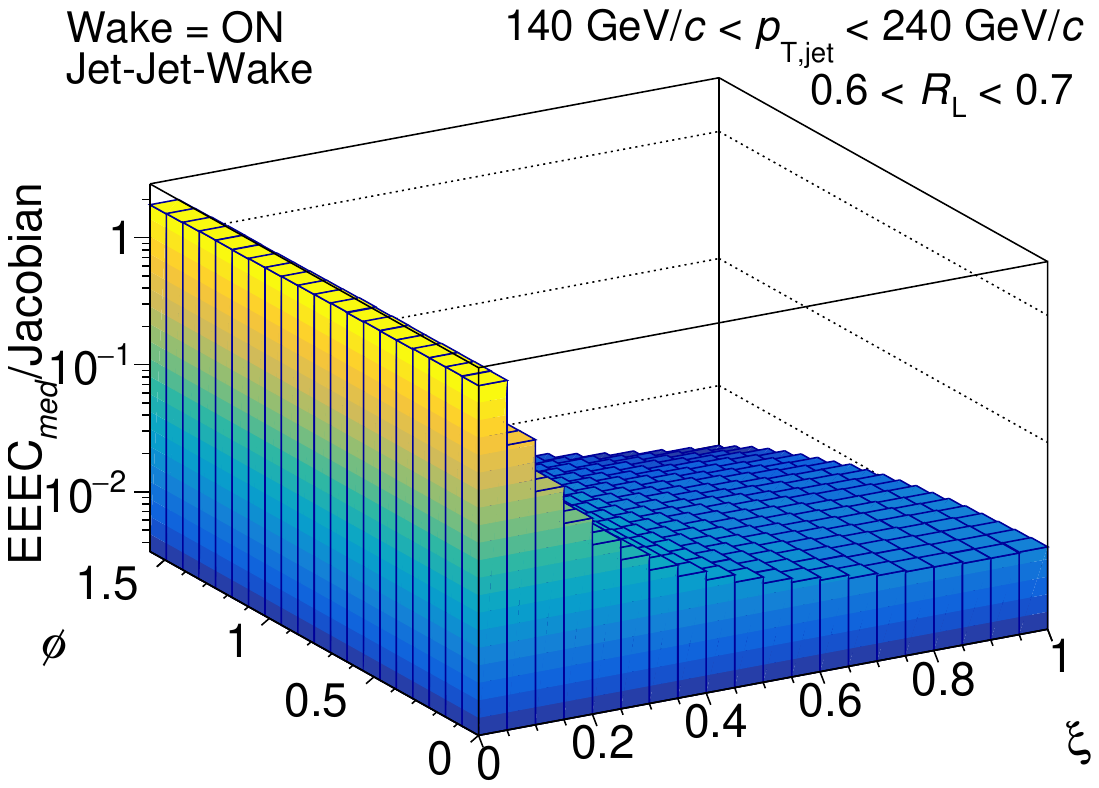}}
\subfloat[Jet-Jet-Jet]{\includegraphics[width = 0.53\textwidth]{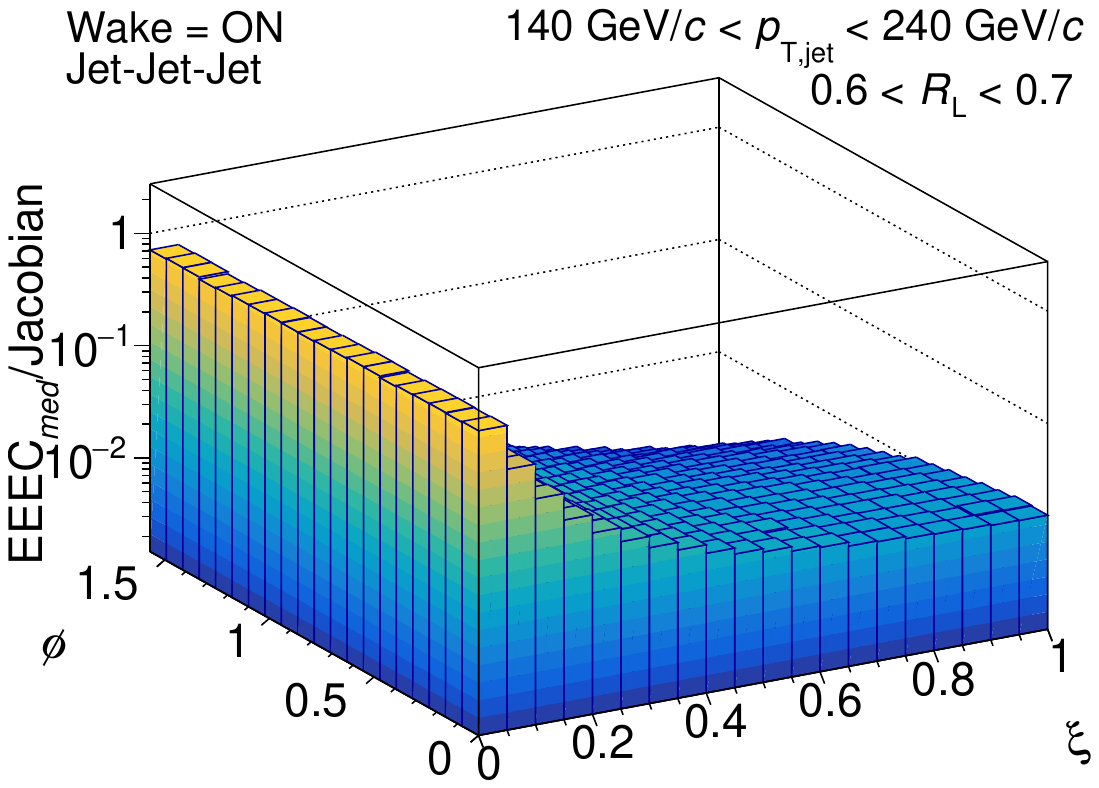}}
\caption{Various contributions to the EEEC for jets in heavy-ion collisions in the ($\xi, \phi$) coordinates, \textit{normalized by the Jacobian}. We see that wake correlations arise for EEECs with triangles of any shape, and are actually 
somewhat depleted in the equilateral region, when normalized appropriately. The enhancement of the correlations in the equilateral region relative to what is seen in vacuum 
reflects how little strength the EEEC has in this region for jets in vacuum. (The different panels have different vertical axes: different linear axes for the top two, log axes for the bottom two, in which the collinear enhancement coming from the parton shower is apparent.)}
\label{fig:JacNorm_wake_Contributions_nom}
\end{center}
\end{figure}

\begin{figure}[t]
\begin{center}
\subfloat[Wake-Wake-Wake]{\includegraphics[width = 0.53\textwidth]{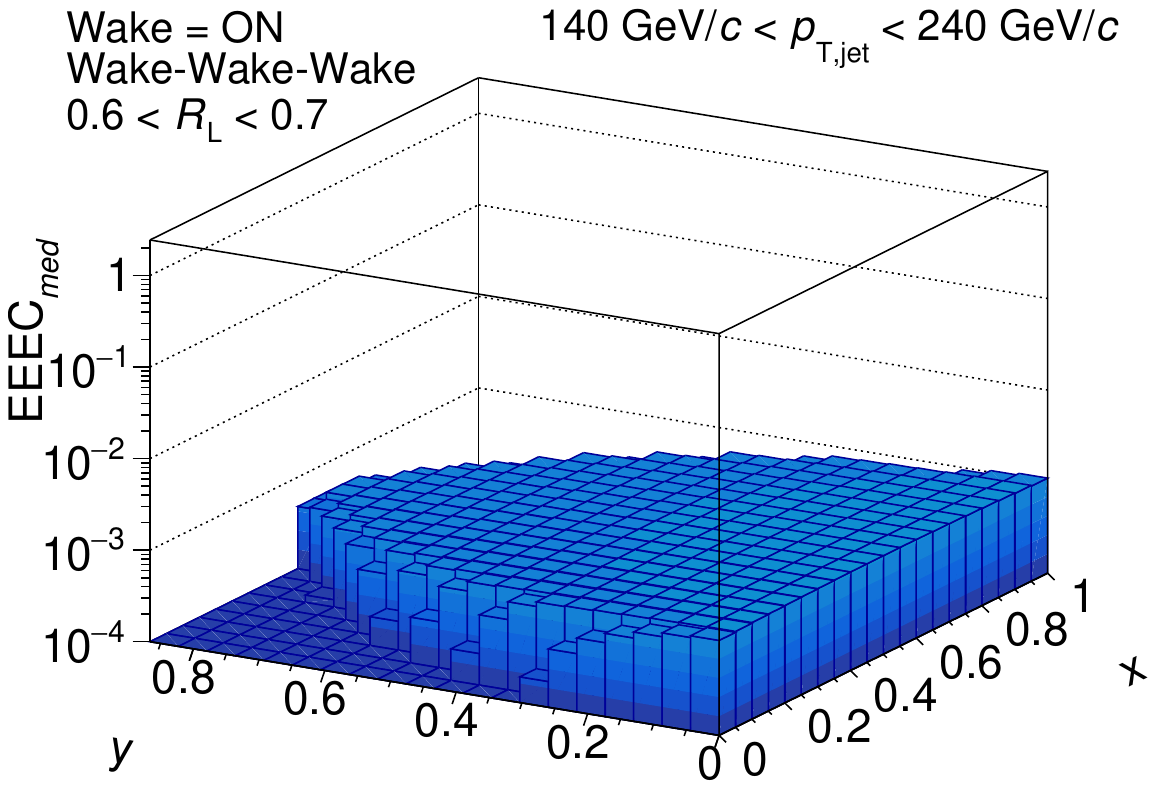}}
\subfloat[Jet-Wake-Wake]{\includegraphics[width = 0.53\textwidth]{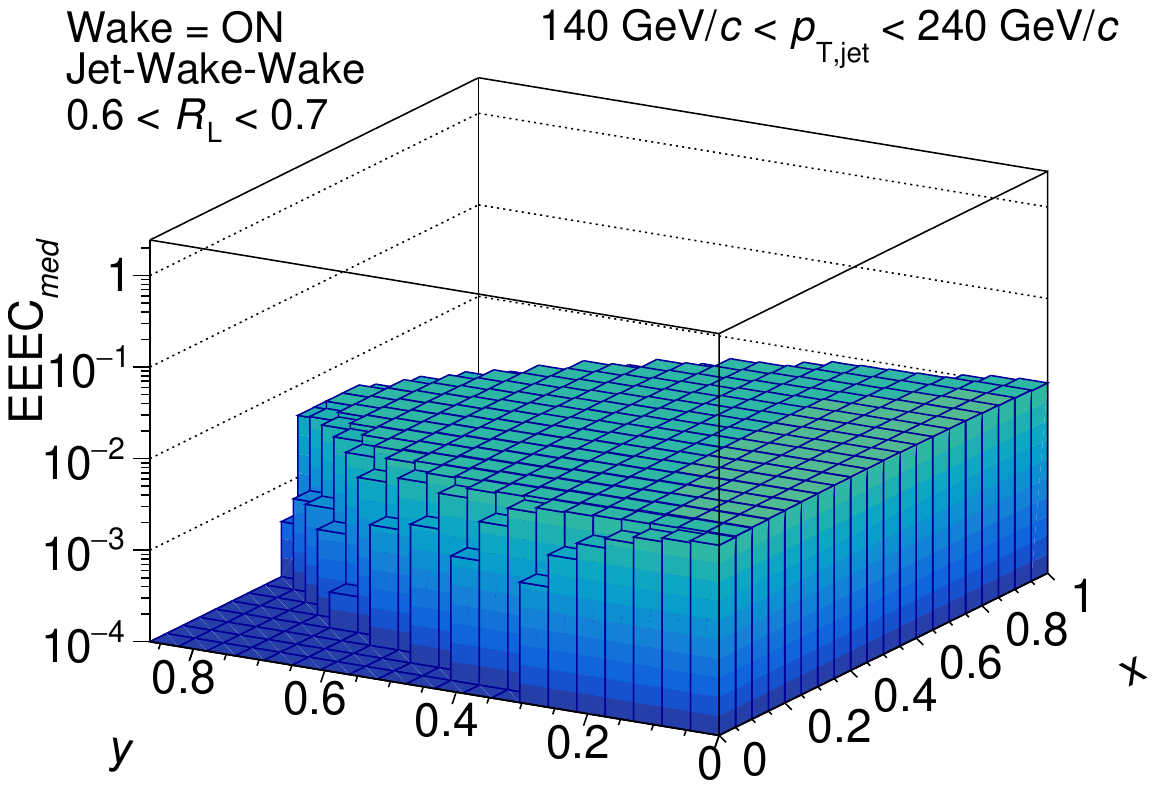}}\\
\subfloat[Jet-Jet-Wake]{\includegraphics[width = 0.53\textwidth]{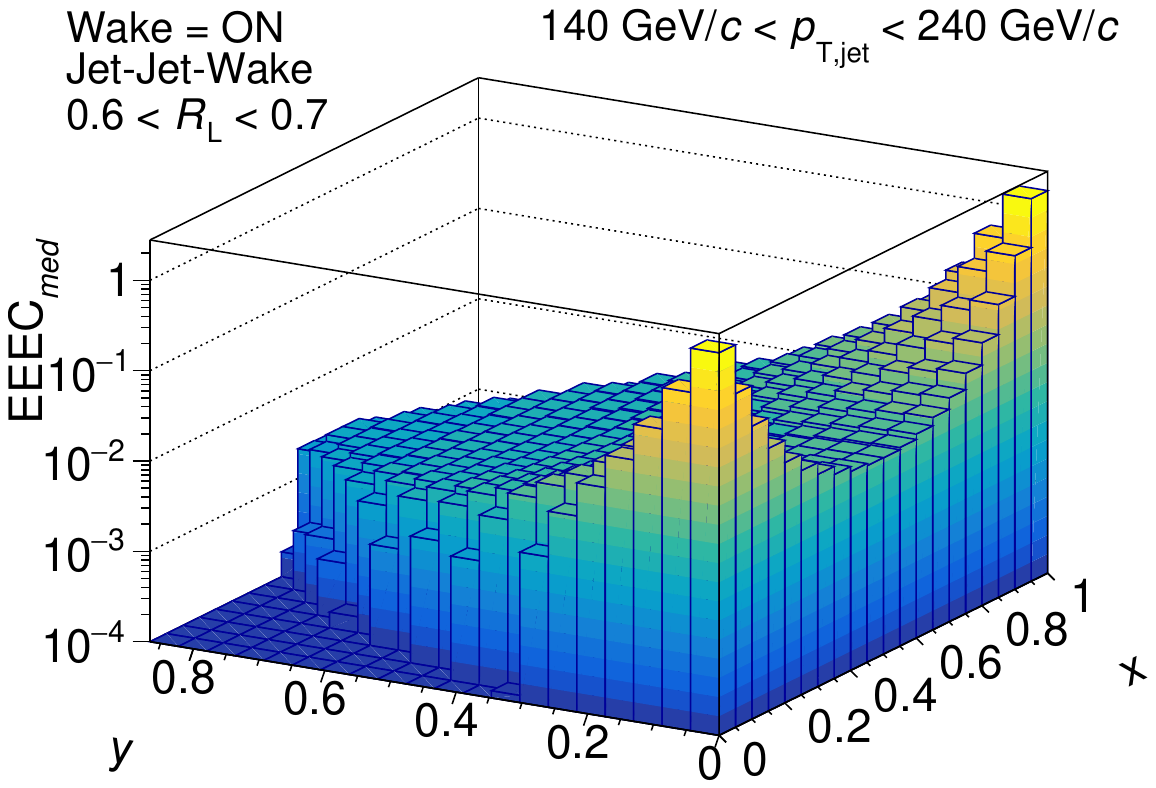}}
\subfloat[Jet-Jet-Jet]{\includegraphics[width = 0.53\textwidth]{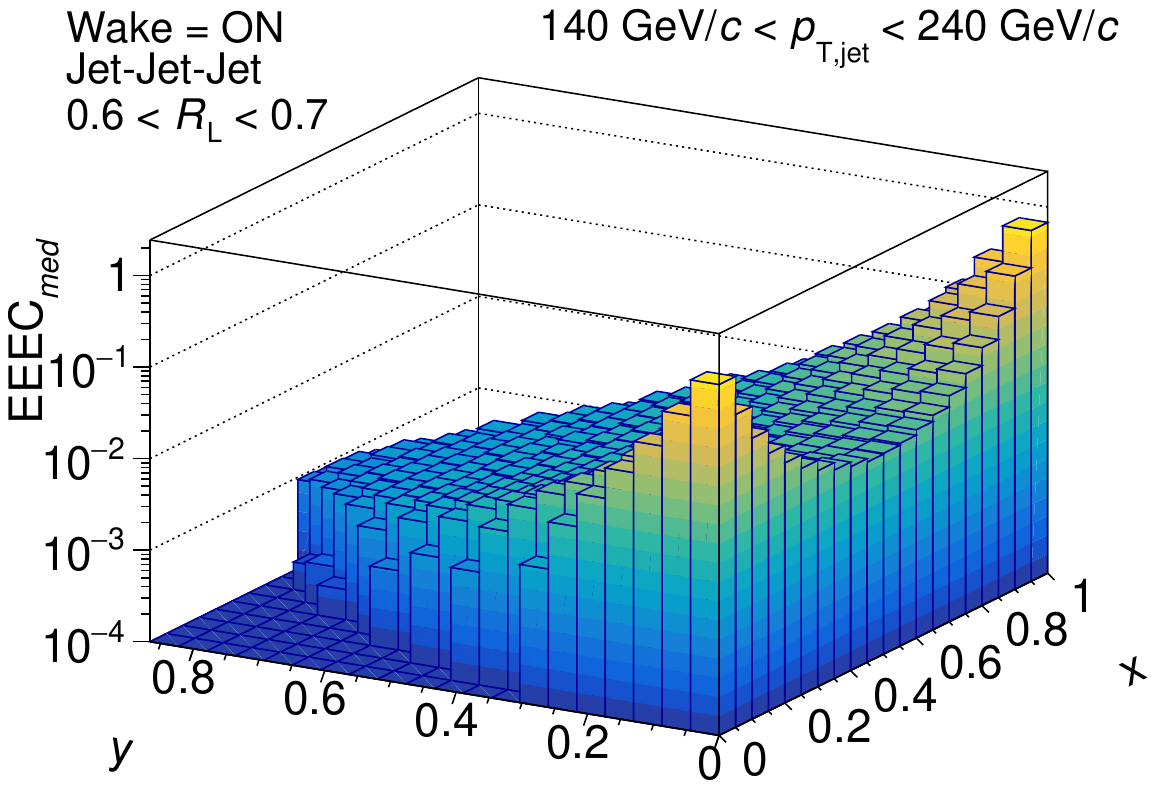}}
\caption{Various contributions to the EEEC for jets in heavy-ion collisions in the $x$-$y$ coordinates. We see that the jet-wake-wake and wake-wake-wake correlations fill out all of the regions of the EEEC that the 
parton shower leaves underpopulated --- compare to the medium-modified jet-jet-jet correlations here or the vacuum EEEC in
\Fig{fig:EEEC_vac}.  The hadrons from jet wakes yield wake-wake-wake and jet-wake-wake correlations with triangles of all shapes, with the equilateral region less prominent than other regions but still populated enough relative to what comes from the parton shower that the wake dominates this region of the EEEC. (The mild shape-dependence of the wake-wake-wake and jet-wake-wake correlations seen in \Fig{fig:JacNorm_wake_Contributions_nom} is harder to see here because we have used the same log scale for the vertical axes in all panels, to make their relative magnitudes apparent.)}
\label{fig:wake_Contributions}
\end{center}
\end{figure}

We can translate the full shape-dependent three-point correlator from the $\xi$-$\phi$ coordinate space to the $x$-$y$ coordinate space by first dividing the EEECs by the Jacobian. We plot these Jacobian-normalized EEECs 
in $\xi$-$\phi$ coordinates in \Fig{fig:JacNorm_wake_Contributions_nom}. The Jacobian-normalized plots now show a modest diminution in the equilateral region, indicating that the perceived prominence of precisely equilateral correlations was 
due to the Jacobian. 
In fact, the diminution in \Fig{fig:JacNorm_wake_Contributions_nom} demonstrates
that the peaks for precisely equilateral triangles
in the jet-wake-wake and wake-wake-wake 
correlations in $\xi$-$\phi$ coordinates depicted in \Fig{fig:wake_Contributions_nom}
are not quite as high as the peak in the coordinate Jacobian depicted in \Fig{fig:jacobian_nom}.
Note that if the correlations among hadrons
from the wake were distributed uniformly in the $x$-$y$ plane, then the Jacobian-normalized wake-wake-wake EEEC in $\xi$-$\phi$ coordinates would be flat.
We see in  \Fig{fig:JacNorm_wake_Contributions_nom}
that it is not actually flat, meaning that the Jacobian-normalized
EEEC shows us the actual shape of the 
three-point correlations 
coming from the wake. We now see that, contrary to the impression given by  \Fig{fig:wake_Contributions_nom}, the correlations of triplets of hadrons where at least two hadrons are from the wake are not sharply peaked when the three vectors form a precisely
equilateral triangle.  Once we have normalized them appropriately by dividing by the Jacobian, we see in \Fig{fig:JacNorm_wake_Contributions_nom} that these correlations
are populated across the whole correlator space of possible triangle shapes by hadrons from the wake.

In \Fig{fig:wake_Contributions} we confirm what we have seen in  \Fig{fig:JacNorm_wake_Contributions_nom} by
plotting the contributions to the EEEC in $(x,y)$ coordinates -- coordinates whose Jacobian is flat.
We have used these coordinates throughout \Sec{sec:shape}
because they immediately provide a faithful
visual representation of the shape of the three-point
correlations coming from the wake without having to divide by the Jacobian.
\Fig{fig:wake_Contributions} shows that correlations of triplets of hadrons that involve at least two hadrons from the wake have the same order-of-magnitude strength in the equilateral region as in the
collinear, squeezed triangle, regions.
Since the equilateral region
is left so unpopulated by the hadrons from the parton shower, the correlations among hadrons from the wake are the dominant
contribution to the EEEC in this region.

\begin{figure}[t!]
\begin{center}
\subfloat[Wake-Wake-Wake]{\includegraphics[width = 0.53\textwidth]{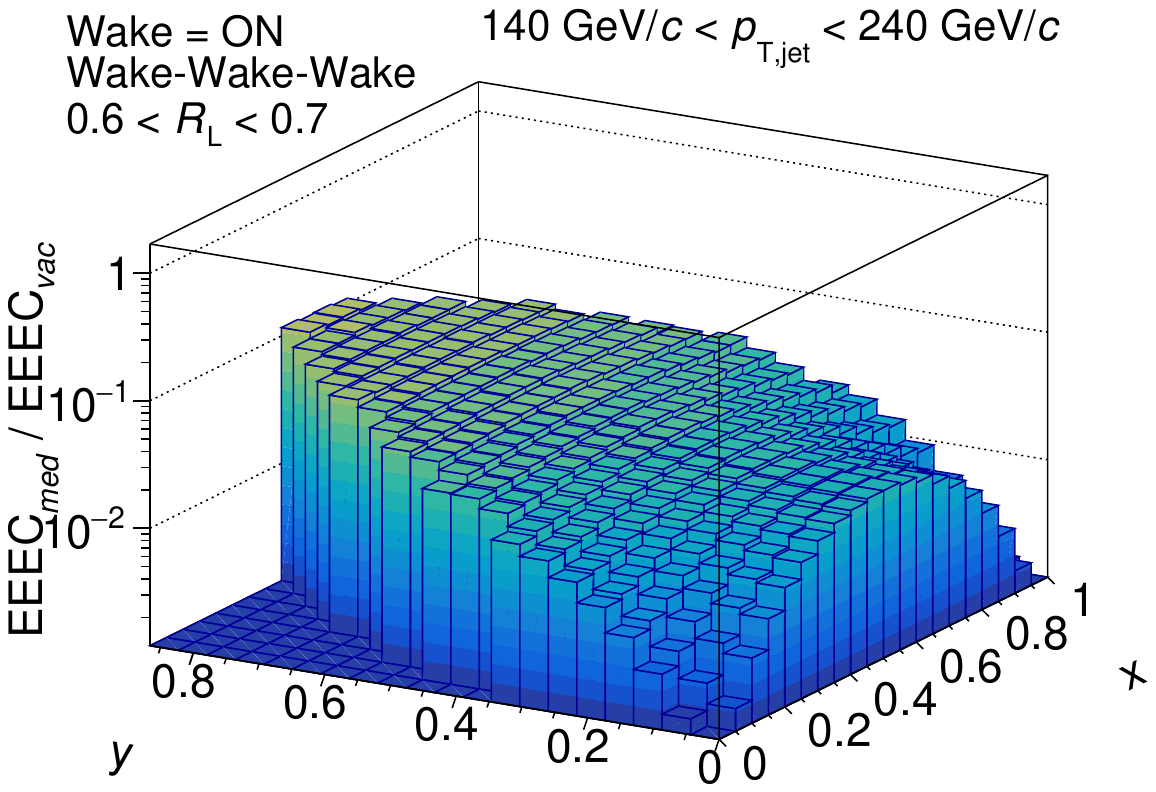}}
\subfloat[Jet-Wake-Wake]{\includegraphics[width = 0.53\textwidth]{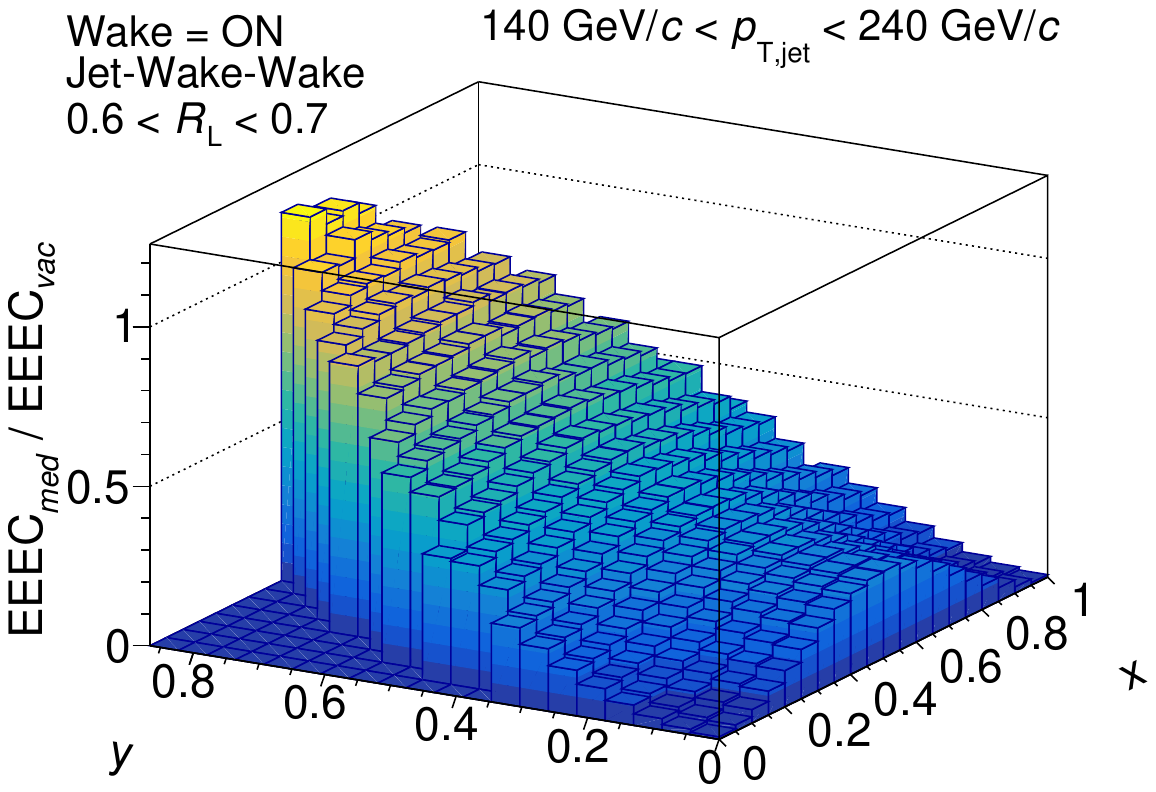}}\\
\subfloat[Jet-Jet-Wake]{\includegraphics[width = 0.53\textwidth]{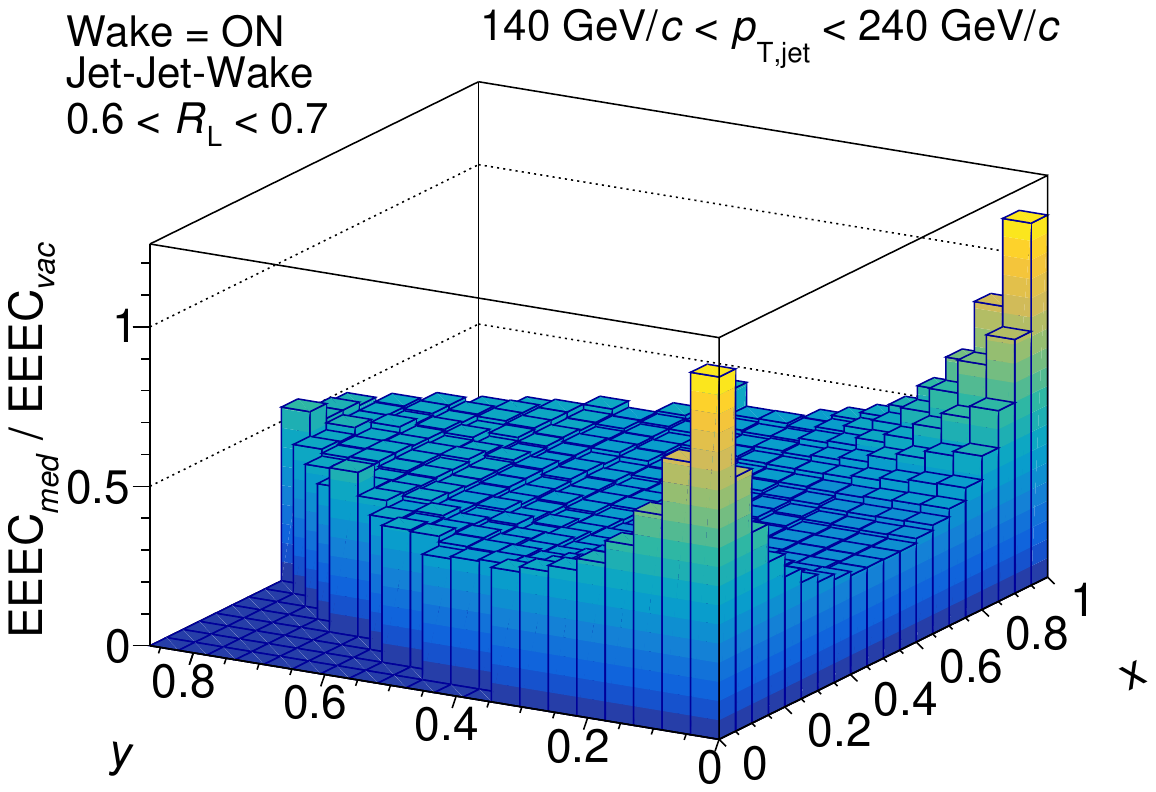}}
\subfloat[Jet-Jet-Jet]{\includegraphics[width = 0.53\textwidth]{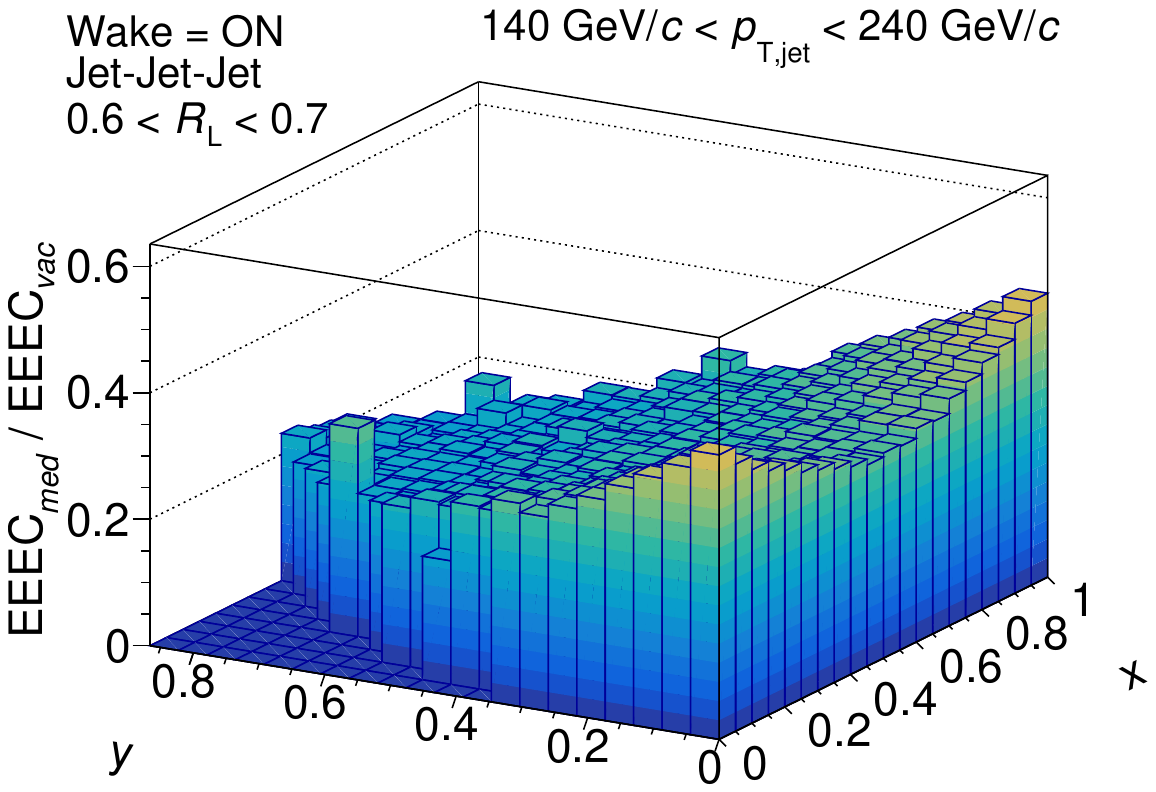}}
\caption{Ratios of various contributions to the EEEC
for jets in heavy-ion collisions with wakes to the EEEC for jets in vacuum,
in $x$-$y$ coordinates. Note that the vertical axis scales are different in each plot in order to make the key features of the plot visible. 
}
\label{fig:wakeovervac_Contributions}
\end{center}
\end{figure}

When we take the ratio of the EEECs in-medium with wake to EEECs in vacuum, the equilateral enhancement coming from wake correlations will be manifest in any choice of coordinates, since any coordinate Jacobian 
is present in both the in-medium EEEC and the vacuum EEEC
and
cancels in the PbPb/vacuum ratio. This is seen clearly in \Fig{fig:wakeovervac_Contributions}. 
Therefore, the equilateral enhancement observed in the in-medium to vacuum EEEC ratio unambiguously encodes the physics of the wake, and not the artifacts of coordinate geometry.  

\section{Negative Wake Subtraction}
\label{app:negasub}

As jets in heavy-ion collisions traverse the droplet of QGP produced in the same collision, they are quenched and deposit momentum and energy into the droplet. By conservation, the momentum and energy lost by jets induces perturbations of the stress-energy tensor of the strongly coupled liquid QGP, inducing jet wakes that carry the ``lost'' momentum and energy. Although some of the energy goes into 
compression (i.e.~sound waves) most of the momentum
is carried by a region of QGP behind the jet that has
been pulled by the jet in its direction. Therefore, when one compares the freezeout of a droplet of QGP without any wake to that of a droplet containing one jet wake, there will be a relative excess of soft particles along the direction of the boost experienced by the fluid cells, accompanied by a relative depletion in the opposite direction. We call this excess of soft particles the ``positive wake" and the depletion of soft particles in the opposite direction the ``negative wake". In the Hybrid Model, final state hadrons either originate from the fragmentation of the partons from the perturbative shower which have not lost all their energy after traversing the medium 
or are soft hadrons coming from the wake as described by Eq.~(\ref{eq:onebody}).
At the orientations where the latter expression is positive, it describes the soft hadrons representing the positive wake; where it is negative,
it describes the soft hadrons representing the negative wake~\cite{Casalderrey-Solana:2016jvj}.

Suppose we have a heavy-ion collision event where we select a jet that is roughly back-to-back in the transverse plane with another jet, called an ``away-side jet". During jet reconstruction, the negative wake particles from the away-side jet will superpose with, and be clustered into, our selected jet (provided that they lie at similar rapidities~\cite{Pablos:2019ngg}). The presence of negative wake particles in the jet poses a problem when calculating energy correlators.  In particular, it is unclear what it means to correlate two or more regions of energy density depletion in the droplet of QGP. When calculating observables that are linear in particle energy, like jet-$p_{\rm T}$ or jet shape, we can simply assign a negative energy weighting to the negative wake particles. However, a two-point energy correlation of two negative wake particles will then yield a positive value. This positive contribution from two correlations of energy depletion in the plasma is difficult to interpret. We avoid such suspicious contributions to the energy correlators by subtracting the energy of the negative wake particles from the energy of nearby positive energy particles in each event, before any observables are calculated.

\subsection{Subtraction Procedure}
The following procedure for negative wake subtraction is based on a similar procedure in Ref.~\cite{Milhano:2022kzx} for subtracting thermal momenta (representing background) from final-state particles after a heavy-ion collision in JEWEL~\cite{Zapp:2012ak,Zapp:2013vla}, and is similar in spirit to a procedure employed in Ref.~\cite{Casalderrey-Solana:2016jvj} 
(for observables where such a procedure is not as important as it is for energy correlators). The idea behind this procedure is to deplete the energy of a negative wake particle from nearby positive energy particles (either from the wake or from the fragmentation of the parton shower). The four-momentum $p^\mu = (E, p_x, p_y, p_z)$ of a particle may be characterized by four parameters: the particle's rapidity $y$, azimuthal angle $\phi$, transverse momentum $p_{\rm_{T}}$, and mass $m_\delta \equiv \sqrt{m^2 + p_{\rm_{T}}^2} - p_{\rm_{T}}$, where $m_\delta$ 
is the difference between the particle's transverse mass and transverse momentum. 
We may then write each particle's 4-momentum as 
\begin{equation}
    p^\mu = \left( (p_{\rm_{T}} + m_\delta) \cosh(y), \, p_{\rm_{T}} \cos(\phi), \, p_{\rm_{T}} \sin(\phi), \, (p_{\rm_{T}} + m_\delta) \sinh(y) \right).
\end{equation}

We then carry out the following steps.

\begin{enumerate}
    \item Create a list of all possible pairs consisting of a negative wake particle $k$ and a positive energy particle $i$ in the event. A positive energy particle is either a  particle originating from the fragmentation of the parton shower or a particle from the positive wake. Order the list in increasing distance $\Delta R = \sqrt{\Delta y^2 + \Delta \phi^2}$.

    \item Remove all pairings with $\Delta R > R_\text{sub}$, where $R_\text{sub}$ is a chosen ``subtraction radius". This step ensures that we only subtract the energy of negative wake particles from other particles locally in the next step. In our analysis, we chose $R_\text{sub} = 0.5$. The dependence of the results on this choice will be explored in Appendix~\ref{app:subCheck}.

    \item Beginning with the first pair $(k, i)$ in the ordered-list, subtract the lower transverse momentum from the higher, and the lower mass from the higher, i.e.,
          \[
          \begin{aligned}
          p_{\rm T}^{(i)} &\geq p_{\rm T}^{(k)} &\Rightarrow \quad p_{\rm T}^{(i)} &\to p_{\rm T}^{(i)} - p_{\rm T}^{(k)} \quad \text{and} \quad p_{\rm T}^{(k)} &\to 0, \\
          p_{\rm T}^{(i)} &< p_{\rm T}^{(k)} &\Rightarrow \quad p_{\rm T}^{(k)} &\to p_{\rm T}^{(k)} - p_{\rm T}^{(i)} \quad \text{and} \quad p_{\rm T}^{(i)} &\to 0,
          \end{aligned}
          \]
          and
          \[
          \begin{aligned}
          m_{\delta}^{(i)} &\geq m_{\delta}^{(k)} &\Rightarrow \quad m_{\delta}^{(i)} &\to m_{\delta}^{(i)} - m_{\delta}^{(k)} \quad \text{and} \quad m_{\delta}^{(k)} &\to 0, \\
          m_{\delta}^{(i)} &< m_{\delta}^{(k)} &\Rightarrow \quad m_{\delta}^{(k)} &\to m_{\delta}^{(k)} - m_{\delta}^{(i)} \quad \text{and} \quad m_{\delta}^{(i)} &\to 0.
          \end{aligned}
          \]
          Continue doing this until the end of the list is reached. Note that the list of pairs must be dynamically updated after the subtraction is done on each pair. So, the subtraction done on one particle applies to all instances of that particle in the list.
    
    \item  Finally, remove all particles with $p_{\rm T} = 0$ from the event. The final list of particles with nonzero $p_{\rm T}$ is the subtracted ensemble.
\end{enumerate}

\begin{figure}[t]
\begin{center}
\subfloat[$R_\text{sub}=0.5$]{\includegraphics[width = 0.53\textwidth]{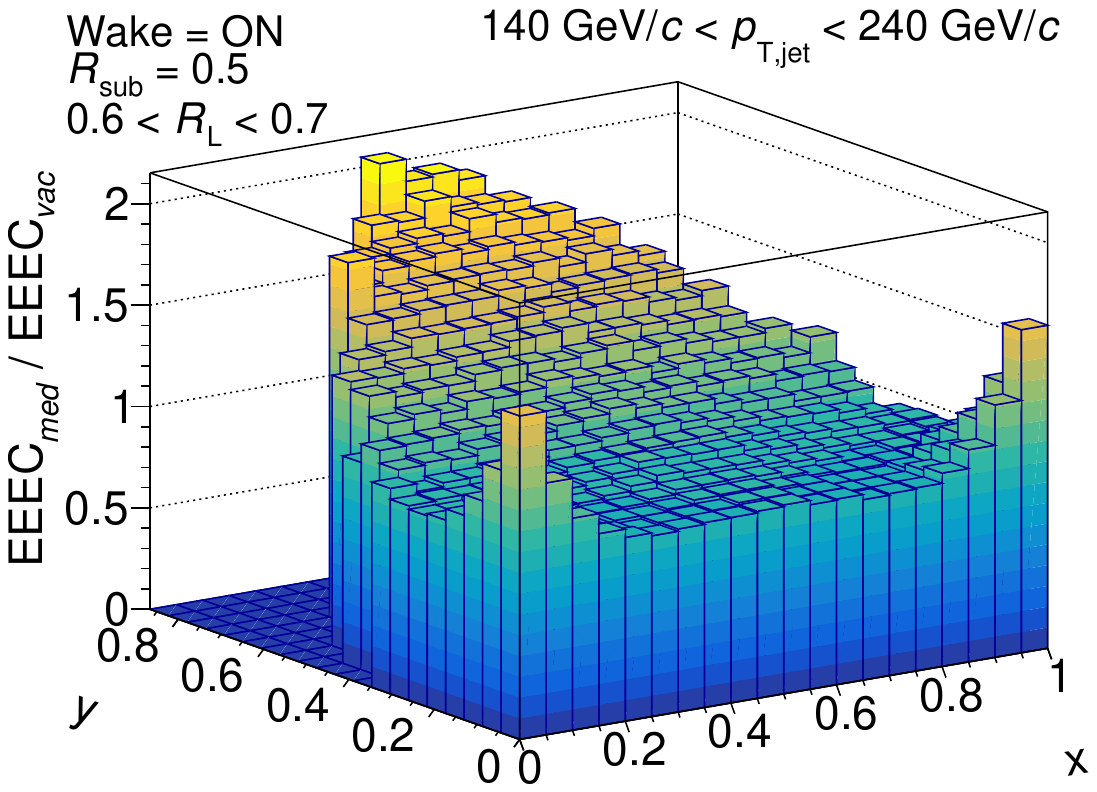}}
\subfloat[$R_\text{sub}=\infty$]{\includegraphics[width = 0.53\textwidth]{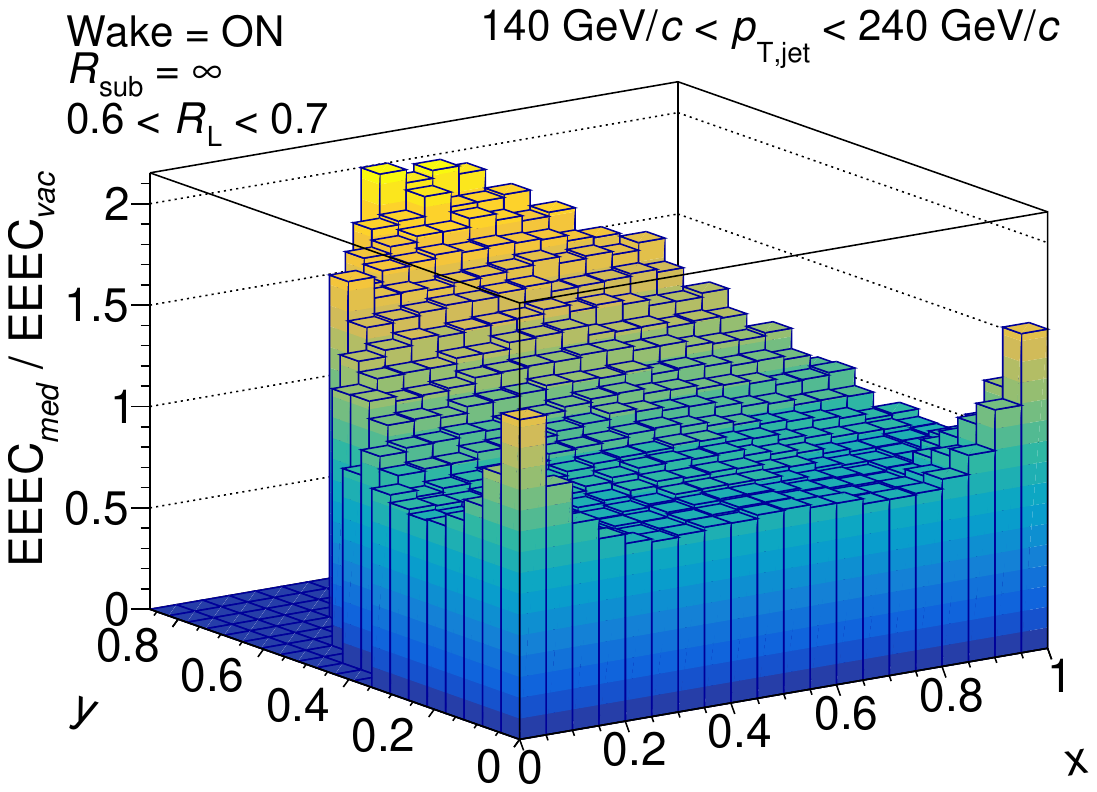}}
\caption{Ratio of EEECs for jets in medium with wake to EEECs for jets in vacuum in $x$-$y$ coordinates for $R$ = 0.8 jets with two different values for the subtraction radius in the subtraction procedure that we use to treat negative wake particles. We see that the EEEC is nearly
identical in the two panels. Since $R_{\rm sub}=\infty$ corresponds to removing all negative wake particles, we see that the subtraction procedure with $R_{\rm sub} = 0.5$ removes almost all of them.}
\label{fig:RSub}
\end{center}
\end{figure}

Note that there may still be some negative wake particles that remain after the above steps are completed. (We shall see below that, with the choice of $R_{\rm sub}=0.5$, in fact very few negative wake particles remain, but some may).
This can happen if a negative wake particle has very few positive energy neighbors with which to neutralize. As designed, this subtraction procedure ensures that all of these straggling negative wake particles will be at least $R_\text{sub}$ away from any 
positive energy particles. Since we cannot locally deplete the energy of these remaining negatives from nearby positive energy particles, we simply remove them from the final ensemble. We shall see below that this makes little difference to the EEEC observable of interest.

\subsection{Sensitivity to the Details of the Subtraction Procedure}\label{app:subCheck}

To investigate the sensitivity of our results for the EEECs to the details of the subtraction procedure, we compared two cases: one where the subtraction radius $R_\text{sub}$ was 0.5 and the other where it was infinity, in which case there are no negative wake particles remaining after the subtraction procedure is carried out. We plot the EEEC ratio for both cases in \Fig{fig:RSub}. We see that our conclusions are almost completely independent of the choice of $R_\text{sub}$.

\subsection{Impact of the Negative Wake on Inclusive and $\gamma$-tagged Jets}

\begin{figure}[t]
\begin{center}
\subfloat[Ignoring negative wake particles]{\includegraphics[width = 0.53\textwidth]{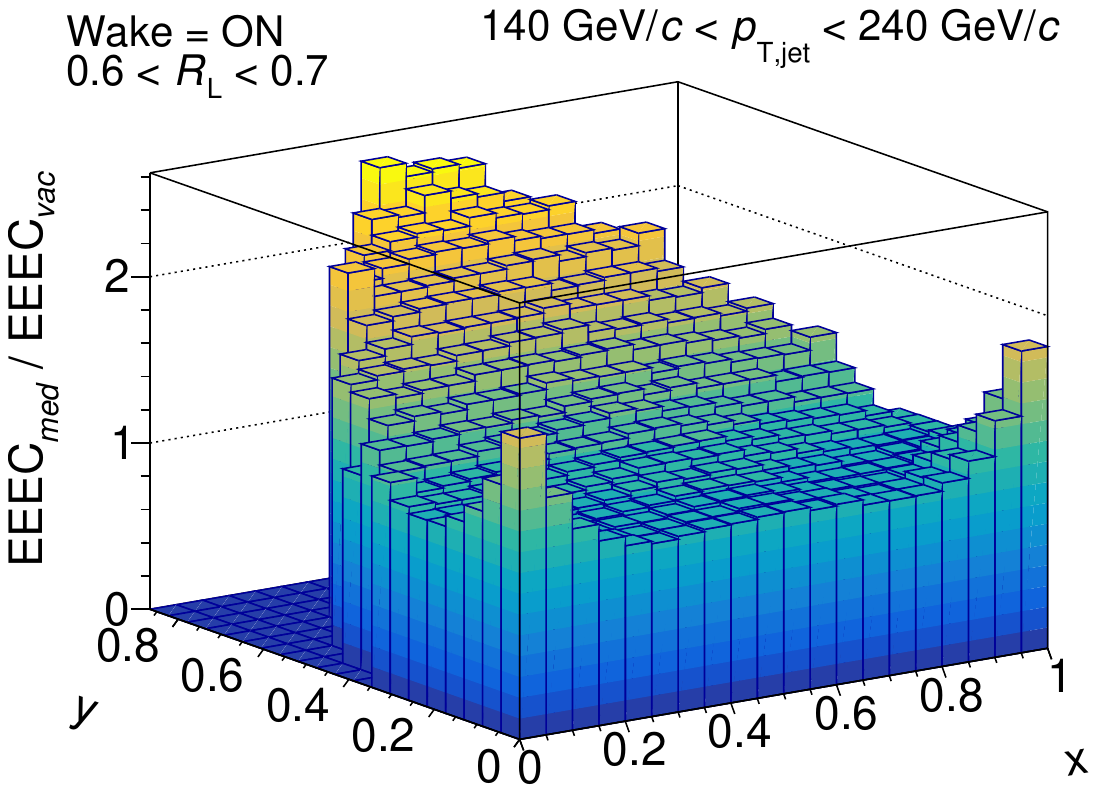}}
\subfloat[Subtracting negative wake particles]{\includegraphics[width = 0.53\textwidth]{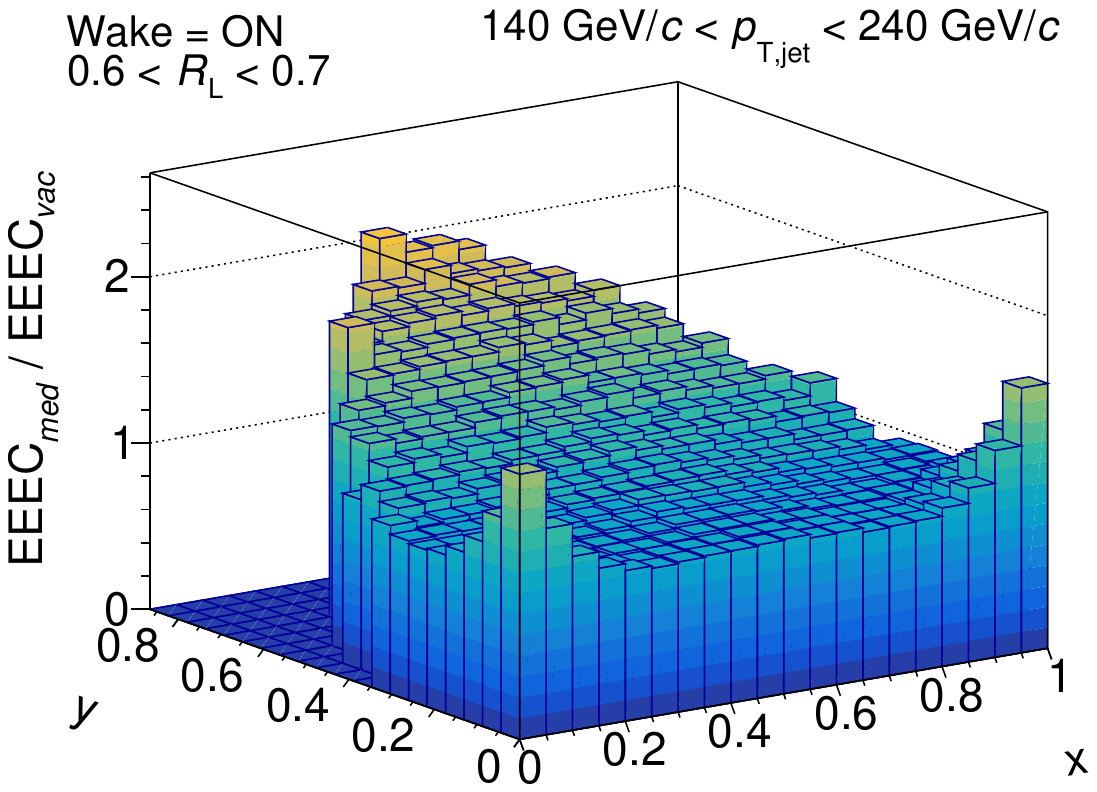}}
\caption{Ratio of EEECs for jets in medium with wake to EEECs for jets in vacuum in $x$-$y$ coordinates for $R$ = 0.8 inclusive jets, ignoring negative wake particles (a) and subtracting negative wake particles using $R_{\rm sub} = 0.5$ (b). Panel (b) is the same as panel (a) in \Fig{fig:wake_ratio}. We see by the comparison between the two panels here that subtracting the negative wake particles (that come from the wakes of away-side jets) makes a significant impact on the magnitude of the wake contribution to the EEEC of the jets in the inclusive sample.}
\label{fig:InclusiveNegSub}
\end{center}
\end{figure}

Although the medium to vacuum EEEC ratio for inclusive jets is largely independent of the subtraction radius, $R_\text{sub}$, it is still sensitive to whether or not we subtract the negative wake particles in the first place. In \Fig{fig:InclusiveNegSub}, we observe that simply ignoring the negative wake particles in each event yields a larger enhancement of 
the EEEC for all triangle shapes, including in both  the 
collinear and equilateral regions, than when the negative wake is subtracted. This is to be expected because, when handled correctly, the negative wake particles from an away-side jet correspond to a depletion of the energy density carried by the particles in the opposite direction, which is to say a depletion of the particles in (the direction of) the jet that we have selected.

\begin{figure}
\begin{center}
\subfloat[Ignoring negative wake particles]{\includegraphics[width = 0.53\textwidth]{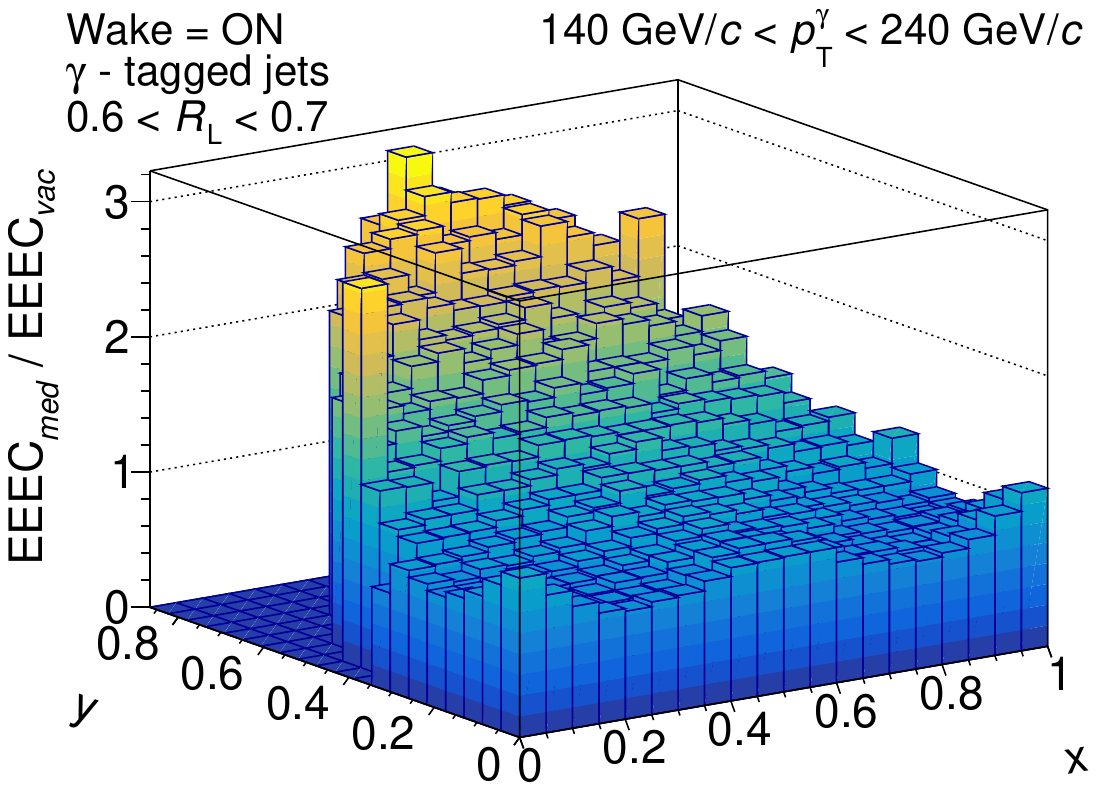}}
\subfloat[Subtracting negative wake particles]{\includegraphics[width = 0.53\textwidth]{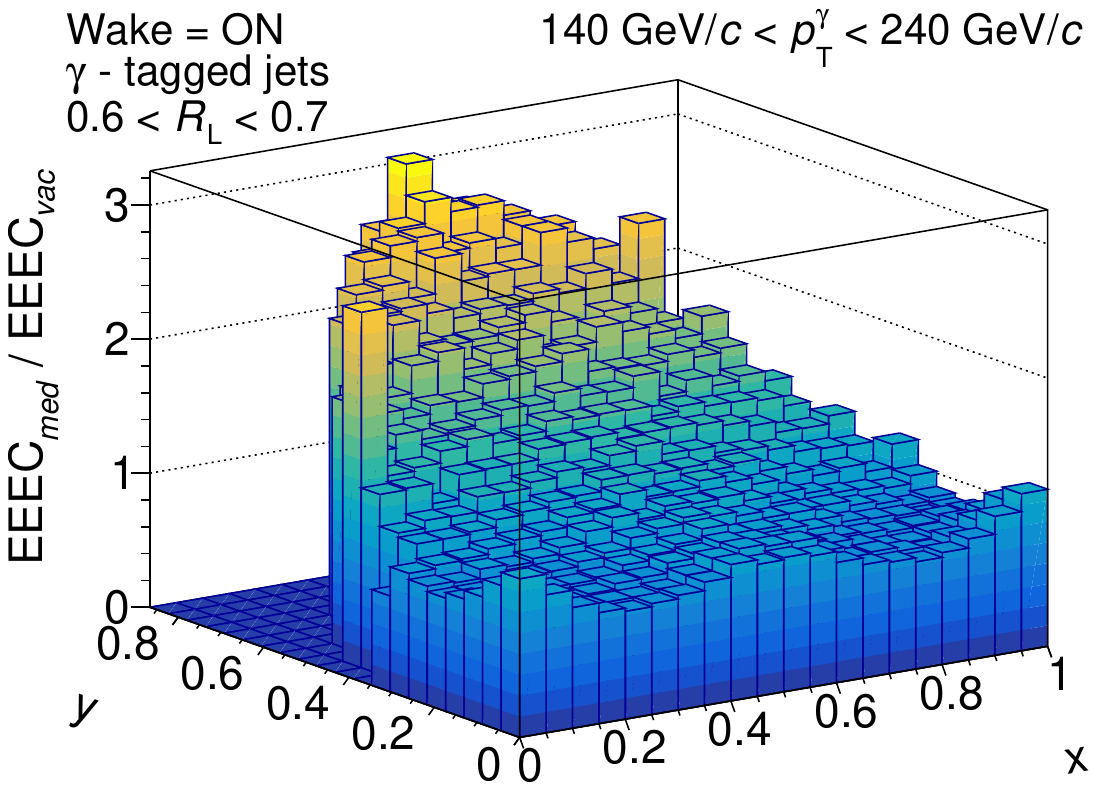}}
\caption{Ratio of EEECs for jets in medium with wake to EEECs for jets in vacuum in $x$-$y$ coordinates for $R$ = 0.8 $\gamma$-tagged jets, ignoring negative wake particles (a) and after subtracting negative wake particles using $R_{\rm sub} = 0.5$ (b). We see that in this case the subtraction makes very little difference, meaning that in the sample of $\gamma$-tagged jets in medium it is very unlikely that the negative wake of a second jet obscures the wake of the selected jet.}
\label{fig:GammaNegSub}
\end{center}
\end{figure}

However, for $\gamma$-tagged jets, we ensure that the leading structure azimuthally opposite to the jets in our sample is an isolated photon. 
There may also be away-side jets in the sample but this is unlikely.
Since the mean free path of a photon in QGP is much longer than the length scale of the QGP droplet, photons do not leave wakes in the plasma as they traverse it. This means that there will be far fewer negative wake particles that superpose with the wakes of $\gamma$-tagged jets than those that superpose with the wakes of inclusive jets. So, we expect $\gamma$-tagged jet EEECs to be insensitive to subtracting the negative wake. This is seen clearly in \Fig{fig:GammaNegSub}. In particular, the in-medium to vacuum EEEC ratio for $\gamma$-tagged jets is almost identical whether we ignore or subtract negative wake particles using $R_{\rm sub} = 0.5$, meaning that there are very few of them.

Thus, we arrive at the following conclusion: negative wakes have a noticeable effect on the full shape-dependent 
three-point energy correlator in a jet whenever the
negative wakes from one or more other jets in the event land on top of the wake of the selected jet.
In principle, one could exploit this fact and use a comparison between the EEECs of inclusive jets and $\gamma$-jets as a means of detecting the presence of negative wakes in the QGP overlapping with the inclusive jets. We leave further investigation of how best to realize this opportunity to future work.

\bibliography{qgp_EEC_ref.bib}

\providecommand{\href}[2]{#2}\begingroup\raggedright\begin{thebibliography}{100}

\bibitem{PHENIX:2004vcz}
{\bf PHENIX} Collaboration, K.~Adcox et~al., {\it {Formation of dense partonic matter in relativistic nucleus-nucleus collisions at RHIC: Experimental evaluation by the PHENIX collaboration}},  {\em Nucl. Phys. A} {\bf 757} (2005) 184--283, [\href{http://arxiv.org/abs/nucl-ex/0410003}{{\tt nucl-ex/0410003}}].

\bibitem{BRAHMS:2004adc}
{\bf BRAHMS} Collaboration, I.~Arsene et~al., {\it {Quark gluon plasma and color glass condensate at RHIC? The Perspective from the BRAHMS experiment}},  {\em Nucl. Phys. A} {\bf 757} (2005) 1--27, [\href{http://arxiv.org/abs/nucl-ex/0410020}{{\tt nucl-ex/0410020}}].

\bibitem{PHOBOS:2004zne}
{\bf PHOBOS} Collaboration, B.~B. Back et~al., {\it {The PHOBOS perspective on discoveries at RHIC}},  {\em Nucl. Phys. A} {\bf 757} (2005) 28--101, [\href{http://arxiv.org/abs/nucl-ex/0410022}{{\tt nucl-ex/0410022}}].

\bibitem{STAR:2005gfr}
{\bf STAR} Collaboration, J.~Adams et~al., {\it {Experimental and theoretical challenges in the search for the quark gluon plasma: The STAR Collaboration's critical assessment of the evidence from RHIC collisions}},  {\em Nucl. Phys. A} {\bf 757} (2005) 102--183, [\href{http://arxiv.org/abs/nucl-ex/0501009}{{\tt nucl-ex/0501009}}].

\bibitem{Gyulassy:2004zy}
M.~Gyulassy and L.~McLerran, {\it {New forms of QCD matter discovered at RHIC}},  {\em Nucl. Phys. A} {\bf 750} (2005) 30--63, [\href{http://arxiv.org/abs/nucl-th/0405013}{{\tt nucl-th/0405013}}].

\bibitem{Muller:2006ee}
B.~Muller and J.~L. Nagle, {\it {Results from the relativistic heavy ion collider}},  {\em Ann. Rev. Nucl. Part. Sci.} {\bf 56} (2006) 93--135, [\href{http://arxiv.org/abs/nucl-th/0602029}{{\tt nucl-th/0602029}}].

\bibitem{Casalderrey-Solana:2007knd}
J.~Casalderrey-Solana and C.~A. Salgado, {\it {Introductory lectures on jet quenching in heavy ion collisions}},  {\em Acta Phys. Polon. B} {\bf 38} (2007) 3731--3794, [\href{http://arxiv.org/abs/0712.3443}{{\tt arXiv:0712.3443}}].

\bibitem{dEnterria:2009xfs}
D.~d'Enterria, {\it {Jet quenching}},  {\em Landolt-Bornstein} {\bf 23} (2010) 471, [\href{http://arxiv.org/abs/0902.2011}{{\tt arXiv:0902.2011}}].

\bibitem{Wiedemann:2009sh}
U.~A. Wiedemann, {\it {Jet Quenching in Heavy Ion Collisions}},  \href{http://arxiv.org/abs/0908.2306}{{\tt arXiv:0908.2306}}.

\bibitem{Jacak:2012dx}
B.~V. Jacak and B.~Muller, {\it {The exploration of hot nuclear matter}},  {\em Science} {\bf 337} (2012) 310--314.

\bibitem{Muller:2012zq}
B.~Muller, J.~Schukraft, and B.~Wyslouch, {\it {First Results from Pb+Pb collisions at the LHC}},  {\em Ann. Rev. Nucl. Part. Sci.} {\bf 62} (2012) 361--386, [\href{http://arxiv.org/abs/1202.3233}{{\tt arXiv:1202.3233}}].

\bibitem{Heinz:2013th}
U.~Heinz and R.~Snellings, {\it {Collective flow and viscosity in relativistic heavy-ion collisions}},  {\em Ann. Rev. Nucl. Part. Sci.} {\bf 63} (2013) 123--151, [\href{http://arxiv.org/abs/1301.2826}{{\tt arXiv:1301.2826}}].

\bibitem{Shuryak:2014zxa}
E.~Shuryak, {\it {Strongly coupled quark-gluon plasma in heavy ion collisions}},  {\em Rev. Mod. Phys.} {\bf 89} (2017) 035001, [\href{http://arxiv.org/abs/1412.8393}{{\tt arXiv:1412.8393}}].

\bibitem{Akiba:2015jwa}
Y.~Akiba et~al., {\it {The Hot QCD White Paper: Exploring the Phases of QCD at RHIC and the LHC}},  \href{http://arxiv.org/abs/1502.02730}{{\tt arXiv:1502.02730}}.

\bibitem{Romatschke:2017ejr}
P.~Romatschke and U.~Romatschke, {\em {Relativistic Fluid Dynamics In and Out of Equilibrium}}.
\newblock Cambridge Monographs on Mathematical Physics. Cambridge University Press, 5, 2019.

\bibitem{Connors:2017ptx}
M.~Connors, C.~Nattrass, R.~Reed, and S.~Salur, {\it {Jet measurements in heavy ion physics}},  {\em Rev. Mod. Phys.} {\bf 90} (2018) 025005, [\href{http://arxiv.org/abs/1705.01974}{{\tt arXiv:1705.01974}}].

\bibitem{Busza:2018rrf}
W.~Busza, K.~Rajagopal, and W.~van~der Schee, {\it {Heavy Ion Collisions: The Big Picture, and the Big Questions}},  {\em Ann. Rev. Nucl. Part. Sci.} {\bf 68} (2018) 339--376, [\href{http://arxiv.org/abs/1802.04801}{{\tt arXiv:1802.04801}}].

\bibitem{Nagle:2018nvi}
J.~L. Nagle and W.~A. Zajc, {\it {Small System Collectivity in Relativistic Hadronic and Nuclear Collisions}},  {\em Ann. Rev. Nucl. Part. Sci.} {\bf 68} (2018) 211--235, [\href{http://arxiv.org/abs/1801.03477}{{\tt arXiv:1801.03477}}].

\bibitem{Cao:2020wlm}
S.~Cao and X.-N. Wang, {\it {Jet quenching and medium response in high-energy heavy-ion collisions: a review}},  {\em Rept. Prog. Phys.} {\bf 84} (2021), no.~2 024301, [\href{http://arxiv.org/abs/2002.04028}{{\tt arXiv:2002.04028}}].

\bibitem{Schenke:2021mxx}
B.~Schenke, {\it {The smallest fluid on Earth}},  {\em Rept. Prog. Phys.} {\bf 84} (2021), no.~8 082301, [\href{http://arxiv.org/abs/2102.11189}{{\tt arXiv:2102.11189}}].

\bibitem{Cunqueiro:2021wls}
L.~Cunqueiro and A.~M. Sickles, {\it {Studying the QGP with Jets at the LHC and RHIC}},  {\em Prog. Part. Nucl. Phys.} {\bf 124} (2022) 103940, [\href{http://arxiv.org/abs/2110.14490}{{\tt arXiv:2110.14490}}].

\bibitem{Apolinario:2022vzg}
L.~Apolin\'ario, Y.-J. Lee, and M.~Winn, {\it {Heavy quarks and jets as probes of the QGP}},  \href{http://arxiv.org/abs/2203.16352}{{\tt arXiv:2203.16352}}.

\bibitem{Harris:2023tti}
J.~W. Harris and B.~M\"uller, {\it {``QGP Signatures'' Revisited}},  \href{http://arxiv.org/abs/2308.05743}{{\tt arXiv:2308.05743}}.

\bibitem{Dasgupta:2013ihk}
M.~Dasgupta, A.~Fregoso, S.~Marzani, and G.~P. Salam, {\it {Towards an understanding of jet substructure}},  {\em JHEP} {\bf 09} (2013) 029, [\href{http://arxiv.org/abs/1307.0007}{{\tt arXiv:1307.0007}}].

\bibitem{Larkoski:2013eya}
A.~J. Larkoski, G.~P. Salam, and J.~Thaler, {\it {Energy Correlation Functions for Jet Substructure}},  {\em JHEP} {\bf 06} (2013) 108, [\href{http://arxiv.org/abs/1305.0007}{{\tt arXiv:1305.0007}}].

\bibitem{Larkoski:2017jix}
A.~J. Larkoski, I.~Moult, and B.~Nachman, {\it {Jet Substructure at the Large Hadron Collider: A Review of Recent Advances in Theory and Machine Learning}},  \href{http://arxiv.org/abs/1709.04464}{{\tt arXiv:1709.04464}}.

\bibitem{Asquith:2018igt}
R.~Kogler et~al., {\it {Jet Substructure at the Large Hadron Collider: Experimental Review}},  {\em Rev. Mod. Phys.} {\bf 91} (2019), no.~4 045003, [\href{http://arxiv.org/abs/1803.06991}{{\tt arXiv:1803.06991}}].

\bibitem{Marzani:2019hun}
S.~Marzani, G.~Soyez, and M.~Spannowsky, {\em {Looking inside jets: an introduction to jet substructure and boosted-object phenomenology}}, vol.~958.
\newblock Springer, 2019.

\bibitem{Heinz:2001xi}
U.~W. Heinz and P.~F. Kolb, {\it {Early thermalization at RHIC}},  {\em Nucl. Phys. A} {\bf 702} (2002) 269--280, [\href{http://arxiv.org/abs/hep-ph/0111075}{{\tt hep-ph/0111075}}].

\bibitem{Heinz:2002un}
U.~W. Heinz and P.~F. Kolb, {\it {Two RHIC puzzles: Early thermalization and the HBT problem}},  in {\em {18th Winter Workshop on Nuclear Dynamics}}, 4, 2002.
\newblock \href{http://arxiv.org/abs/hep-ph/0204061}{{\tt hep-ph/0204061}}.

\bibitem{Kolb:2003dz}
P.~F. Kolb and U.~W. Heinz, {\it {Hydrodynamic description of ultrarelativistic heavy ion collisions}},  \href{http://arxiv.org/abs/nucl-th/0305084}{{\tt nucl-th/0305084}}.

\bibitem{Heinz:2004pj}
U.~W. Heinz, {\it {Thermalization at RHIC}},  {\em AIP Conf. Proc.} {\bf 739} (2004), no.~1 163--180, [\href{http://arxiv.org/abs/nucl-th/0407067}{{\tt nucl-th/0407067}}].

\bibitem{Chesler:2008hg}
P.~M. Chesler and L.~G. Yaffe, {\it {Horizon formation and far-from-equilibrium isotropization in supersymmetric Yang-Mills plasma}},  {\em Phys. Rev. Lett.} {\bf 102} (2009) 211601, [\href{http://arxiv.org/abs/0812.2053}{{\tt arXiv:0812.2053}}].

\bibitem{Chesler:2009cy}
P.~M. Chesler and L.~G. Yaffe, {\it {Boost invariant flow, black hole formation, and far-from-equilibrium dynamics in N = 4 supersymmetric Yang-Mills theory}},  {\em Phys. Rev. D} {\bf 82} (2010) 026006, [\href{http://arxiv.org/abs/0906.4426}{{\tt arXiv:0906.4426}}].

\bibitem{Chesler:2010bi}
P.~M. Chesler and L.~G. Yaffe, {\it {Holography and colliding gravitational shock waves in asymptotically AdS$_{5}$ spacetime}},  {\em Phys. Rev. Lett.} {\bf 106} (2011) 021601, [\href{http://arxiv.org/abs/1011.3562}{{\tt arXiv:1011.3562}}].

\bibitem{Shen:2010uy}
C.~Shen, U.~Heinz, P.~Huovinen, and H.~Song, {\it {Systematic parameter study of hadron spectra and elliptic flow from viscous hydrodynamic simulations of Au+Au collisions at $\sqrt{s_{NN}}=200$ GeV}},  {\em Phys. Rev. C} {\bf 82} (2010) 054904, [\href{http://arxiv.org/abs/1010.1856}{{\tt arXiv:1010.1856}}].

\bibitem{Heller:2011ju}
M.~P. Heller, R.~A. Janik, and P.~Witaszczyk, {\it {The characteristics of thermalization of boost-invariant plasma from holography}},  {\em Phys. Rev. Lett.} {\bf 108} (2012) 201602, [\href{http://arxiv.org/abs/1103.3452}{{\tt arXiv:1103.3452}}].

\bibitem{Shen:2012vn}
C.~Shen and U.~Heinz, {\it {Collision Energy Dependence of Viscous Hydrodynamic Flow in Relativistic Heavy-Ion Collisions}},  {\em Phys. Rev. C} {\bf 85} (2012) 054902, [\href{http://arxiv.org/abs/1202.6620}{{\tt arXiv:1202.6620}}]. [Erratum: Phys.Rev.C 86, 049903 (2012)].

\bibitem{Heller:2012je}
M.~P. Heller, R.~A. Janik, and P.~Witaszczyk, {\it {A numerical relativity approach to the initial value problem in asymptotically Anti-de Sitter spacetime for plasma thermalization - an ADM formulation}},  {\em Phys. Rev. D} {\bf 85} (2012) 126002, [\href{http://arxiv.org/abs/1203.0755}{{\tt arXiv:1203.0755}}].

\bibitem{Heller:2012km}
M.~P. Heller, D.~Mateos, W.~van~der Schee, and D.~Trancanelli, {\it {Strong Coupling Isotropization of Non-Abelian Plasmas Simplified}},  {\em Phys. Rev. Lett.} {\bf 108} (2012) 191601, [\href{http://arxiv.org/abs/1202.0981}{{\tt arXiv:1202.0981}}].

\bibitem{vanderSchee:2012qj}
W.~van~der Schee, {\it {Holographic thermalization with radial flow}},  {\em Phys. Rev. D} {\bf 87} (2013), no.~6 061901, [\href{http://arxiv.org/abs/1211.2218}{{\tt arXiv:1211.2218}}].

\bibitem{Heller:2013oxa}
M.~P. Heller, D.~Mateos, W.~van~der Schee, and M.~Triana, {\it {Holographic isotropization linearized}},  {\em JHEP} {\bf 09} (2013) 026, [\href{http://arxiv.org/abs/1304.5172}{{\tt arXiv:1304.5172}}].

\bibitem{Kurkela:2015qoa}
A.~Kurkela and Y.~Zhu, {\it {Isotropization and hydrodynamization in weakly coupled heavy-ion collisions}},  {\em Phys. Rev. Lett.} {\bf 115} (2015), no.~18 182301, [\href{http://arxiv.org/abs/1506.06647}{{\tt arXiv:1506.06647}}].

\bibitem{Chesler:2015lsa}
P.~M. Chesler and W.~van~der Schee, {\it {Early thermalization, hydrodynamics and energy loss in AdS/CFT}},  {\em Int. J. Mod. Phys. E} {\bf 24} (2015), no.~10 1530011, [\href{http://arxiv.org/abs/1501.04952}{{\tt arXiv:1501.04952}}].

\bibitem{Chesler:2015fpa}
P.~M. Chesler, N.~Kilbertus, and W.~van~der Schee, {\it {Universal hydrodynamic flow in holographic planar shock collisions}},  {\em JHEP} {\bf 11} (2015) 135, [\href{http://arxiv.org/abs/1507.02548}{{\tt arXiv:1507.02548}}].

\bibitem{Heller:2016gbp}
M.~P. Heller, {\it {Holography, Hydrodynamization and Heavy-Ion Collisions}},  {\em Acta Phys. Polon. B} {\bf 47} (2016) 2581, [\href{http://arxiv.org/abs/1610.02023}{{\tt arXiv:1610.02023}}].

\bibitem{Kurkela:2018vqr}
A.~Kurkela, A.~Mazeliauskas, J.-F. Paquet, S.~Schlichting, and D.~Teaney, {\it {Effective kinetic description of event-by-event pre-equilibrium dynamics in high-energy heavy-ion collisions}},  {\em Phys. Rev. C} {\bf 99} (2019), no.~3 034910, [\href{http://arxiv.org/abs/1805.00961}{{\tt arXiv:1805.00961}}].

\bibitem{Kurkela:2018wud}
A.~Kurkela, A.~Mazeliauskas, J.-F. Paquet, S.~Schlichting, and D.~Teaney, {\it {Matching the Nonequilibrium Initial Stage of Heavy Ion Collisions to Hydrodynamics with QCD Kinetic Theory}},  {\em Phys. Rev. Lett.} {\bf 122} (2019), no.~12 122302, [\href{http://arxiv.org/abs/1805.01604}{{\tt arXiv:1805.01604}}].

\bibitem{Brewer:2019oha}
J.~Brewer, L.~Yan, and Y.~Yin, {\it {Adiabatic hydrodynamization in rapidly-expanding quark\textendash{}gluon plasma}},  {\em Phys. Lett. B} {\bf 816} (2021) 136189, [\href{http://arxiv.org/abs/1910.00021}{{\tt arXiv:1910.00021}}].

\bibitem{Brewer:2022vkq}
J.~Brewer, B.~Scheihing-Hitschfeld, and Y.~Yin, {\it {Scaling and adiabaticity in a rapidly expanding gluon plasma}},  {\em JHEP} {\bf 05} (2022) 145, [\href{http://arxiv.org/abs/2203.02427}{{\tt arXiv:2203.02427}}].

\bibitem{Rajagopal:2024lou}
K.~Rajagopal, B.~Scheihing-Hitschfeld, and R.~Steinhorst, {\it {Adiabatic Hydrodynamization and the Emergence of Attractors: a Unified Description of Hydrodynamization in Kinetic Theory}},  \href{http://arxiv.org/abs/2405.17545}{{\tt arXiv:2405.17545}}.

\bibitem{Casalderrey-Solana:2011dxg}
J.~Casalderrey-Solana, H.~Liu, D.~Mateos, K.~Rajagopal, and U.~A. Wiedemann, {\em {Gauge/String Duality, Hot QCD and Heavy Ion Collisions}}.
\newblock Cambridge University Press, 2014.

\bibitem{Schlichting:2019abc}
S.~Schlichting and D.~Teaney, {\it {The First fm/c of Heavy-Ion Collisions}},  {\em Ann. Rev. Nucl. Part. Sci.} {\bf 69} (2019) 447--476, [\href{http://arxiv.org/abs/1908.02113}{{\tt arXiv:1908.02113}}].

\bibitem{Berges:2020fwq}
J.~Berges, M.~P. Heller, A.~Mazeliauskas, and R.~Venugopalan, {\it {QCD thermalization: Ab initio approaches and interdisciplinary connections}},  {\em Rev. Mod. Phys.} {\bf 93} (2021), no.~3 035003, [\href{http://arxiv.org/abs/2005.12299}{{\tt arXiv:2005.12299}}].

\bibitem{Casalderrey-Solana:2004fdk}
J.~Casalderrey-Solana, E.~V. Shuryak, and D.~Teaney, {\it {Conical flow induced by quenched QCD jets}},  {\em J. Phys. Conf. Ser.} {\bf 27} (2005) 22--31, [\href{http://arxiv.org/abs/hep-ph/0411315}{{\tt hep-ph/0411315}}].

\bibitem{Chesler:2007an}
P.~M. Chesler and L.~G. Yaffe, {\it {The Wake of a quark moving through a strongly-coupled plasma}},  {\em Phys. Rev. Lett.} {\bf 99} (2007) 152001, [\href{http://arxiv.org/abs/0706.0368}{{\tt arXiv:0706.0368}}].

\bibitem{Gubser:2007ga}
S.~S. Gubser, S.~S. Pufu, and A.~Yarom, {\it {Sonic booms and diffusion wakes generated by a heavy quark in thermal AdS/CFT}},  {\em Phys. Rev. Lett.} {\bf 100} (2008) 012301, [\href{http://arxiv.org/abs/0706.4307}{{\tt arXiv:0706.4307}}].

\bibitem{Gubser:2007ni}
S.~S. Gubser and A.~Yarom, {\it {Universality of the diffusion wake in the gauge-string duality}},  {\em Phys. Rev. D} {\bf 77} (2008) 066007, [\href{http://arxiv.org/abs/0709.1089}{{\tt arXiv:0709.1089}}].

\bibitem{Gubser:2007xz}
S.~S. Gubser, S.~S. Pufu, and A.~Yarom, {\it {Energy disturbances due to a moving quark from gauge-string duality}},  {\em JHEP} {\bf 09} (2007) 108, [\href{http://arxiv.org/abs/0706.0213}{{\tt arXiv:0706.0213}}].

\bibitem{Chesler:2007sv}
P.~M. Chesler and L.~G. Yaffe, {\it {The Stress-energy tensor of a quark moving through a strongly-coupled N=4 supersymmetric Yang-Mills plasma: Comparing hydrodynamics and AdS/CFT}},  {\em Phys. Rev. D} {\bf 78} (2008) 045013, [\href{http://arxiv.org/abs/0712.0050}{{\tt arXiv:0712.0050}}].

\bibitem{Chesler:2014jva}
P.~M. Chesler and K.~Rajagopal, {\it {Jet quenching in strongly coupled plasma}},  {\em Phys. Rev. D} {\bf 90} (2014), no.~2 025033, [\href{http://arxiv.org/abs/1402.6756}{{\tt arXiv:1402.6756}}].

\bibitem{Chesler:2015nqz}
P.~M. Chesler and K.~Rajagopal, {\it {On the Evolution of Jet Energy and Opening Angle in Strongly Coupled Plasma}},  {\em JHEP} {\bf 05} (2016) 098, [\href{http://arxiv.org/abs/1511.07567}{{\tt arXiv:1511.07567}}].

\bibitem{Rajagopal:2016uip}
K.~Rajagopal, A.~V. Sadofyev, and W.~van~der Schee, {\it {Evolution of the jet opening angle distribution in holographic plasma}},  {\em Phys. Rev. Lett.} {\bf 116} (2016), no.~21 211603, [\href{http://arxiv.org/abs/1602.04187}{{\tt arXiv:1602.04187}}].

\bibitem{Brewer:2017fqy}
J.~Brewer, K.~Rajagopal, A.~Sadofyev, and W.~Van Der~Schee, {\it {Evolution of the Mean Jet Shape and Dijet Asymmetry Distribution of an Ensemble of Holographic Jets in Strongly Coupled Plasma}},  {\em JHEP} {\bf 02} (2018) 015, [\href{http://arxiv.org/abs/1710.03237}{{\tt arXiv:1710.03237}}].

\bibitem{Brewer:2018mpk}
J.~Brewer, A.~Sadofyev, and W.~van~der Schee, {\it {Jet shape modifications in holographic dijet systems}},  {\em Phys. Lett. B} {\bf 820} (2021) 136492, [\href{http://arxiv.org/abs/1809.10695}{{\tt arXiv:1809.10695}}].

\bibitem{Ruppert:2005uz}
J.~Ruppert and B.~Muller, {\it {Waking the colored plasma}},  {\em Phys. Lett. B} {\bf 618} (2005) 123--130, [\href{http://arxiv.org/abs/hep-ph/0503158}{{\tt hep-ph/0503158}}].

\bibitem{Renk:2005si}
T.~Renk and J.~Ruppert, {\it {Mach cones in an evolving medium}},  {\em Phys. Rev. C} {\bf 73} (2006) 011901, [\href{http://arxiv.org/abs/hep-ph/0509036}{{\tt hep-ph/0509036}}].

\bibitem{Casalderrey-Solana:2006lmc}
J.~Casalderrey-Solana, E.~V. Shuryak, and D.~Teaney, {\it {Hydrodynamic flow from fast particles}},  \href{http://arxiv.org/abs/hep-ph/0602183}{{\tt hep-ph/0602183}}.

\bibitem{Betz:2008ka}
B.~Betz, J.~Noronha, G.~Torrieri, M.~Gyulassy, I.~Mishustin, and D.~H. Rischke, {\it {Universality of the Diffusion Wake from Stopped and Punch-Through Jets in Heavy-Ion Collisions}},  {\em Phys. Rev. C} {\bf 79} (2009) 034902, [\href{http://arxiv.org/abs/0812.4401}{{\tt arXiv:0812.4401}}].

\bibitem{Neufeld:2008fi}
R.~B. Neufeld, B.~Muller, and J.~Ruppert, {\it {Sonic Mach Cones Induced by Fast Partons in a Perturbative Quark-Gluon Plasma}},  {\em Phys. Rev. C} {\bf 78} (2008) 041901, [\href{http://arxiv.org/abs/0802.2254}{{\tt arXiv:0802.2254}}].

\bibitem{Li:2010ts}
H.~Li, F.~Liu, G.-l. Ma, X.-N. Wang, and Y.~Zhu, {\it {Mach cone induced by $\gamma$-triggered jets in high-energy heavy-ion collisions}},  {\em Phys. Rev. Lett.} {\bf 106} (2011) 012301, [\href{http://arxiv.org/abs/1006.2893}{{\tt arXiv:1006.2893}}].

\bibitem{He:2015pra}
Y.~He, T.~Luo, X.-N. Wang, and Y.~Zhu, {\it {Linear Boltzmann Transport for Jet Propagation in the Quark-Gluon Plasma: Elastic Processes and Medium Recoil}},  {\em Phys. Rev. C} {\bf 91} (2015) 054908, [\href{http://arxiv.org/abs/1503.03313}{{\tt arXiv:1503.03313}}]. [Erratum: Phys.Rev.C 97, 019902 (2018)].

\bibitem{Casalderrey-Solana:2016jvj}
J.~Casalderrey-Solana, D.~Gulhan, G.~Milhano, D.~Pablos, and K.~Rajagopal, {\it {Angular Structure of Jet Quenching Within a Hybrid Strong/Weak Coupling Model}},  {\em JHEP} {\bf 03} (2017) 135, [\href{http://arxiv.org/abs/1609.05842}{{\tt arXiv:1609.05842}}].

\bibitem{Cao:2016gvr}
S.~Cao, T.~Luo, G.-Y. Qin, and X.-N. Wang, {\it {Linearized Boltzmann transport model for jet propagation in the quark-gluon plasma: Heavy quark evolution}},  {\em Phys. Rev. C} {\bf 94} (2016), no.~1 014909, [\href{http://arxiv.org/abs/1605.06447}{{\tt arXiv:1605.06447}}].

\bibitem{Chen:2017zte}
W.~Chen, S.~Cao, T.~Luo, L.-G. Pang, and X.-N. Wang, {\it {Effects of jet-induced medium excitation in $\gamma$-hadron correlation in A+A collisions}},  {\em Phys. Lett. B} {\bf 777} (2018) 86--90, [\href{http://arxiv.org/abs/1704.03648}{{\tt arXiv:1704.03648}}].

\bibitem{Tachibana:2017syd}
Y.~Tachibana, N.-B. Chang, and G.-Y. Qin, {\it {Full jet in quark-gluon plasma with hydrodynamic medium response}},  {\em Phys. Rev. C} {\bf 95} (2017), no.~4 044909, [\href{http://arxiv.org/abs/1701.07951}{{\tt arXiv:1701.07951}}].

\bibitem{He:2018xjv}
Y.~He, S.~Cao, W.~Chen, T.~Luo, L.-G. Pang, and X.-N. Wang, {\it {Interplaying mechanisms behind single inclusive jet suppression in heavy-ion collisions}},  {\em Phys. Rev. C} {\bf 99} (2019), no.~5 054911, [\href{http://arxiv.org/abs/1809.02525}{{\tt arXiv:1809.02525}}].

\bibitem{Park:2018acg}
C.~Park, S.~Jeon, and C.~Gale, {\it {Jet modification with medium recoil in quark-gluon plasma}},  {\em Nucl. Phys. A} {\bf 982} (2019) 643--646, [\href{http://arxiv.org/abs/1807.06550}{{\tt arXiv:1807.06550}}].

\bibitem{Chang:2019sae}
N.-B. Chang, Y.~Tachibana, and G.-Y. Qin, {\it {Nuclear modification of jet shape for inclusive jets and $\gamma$-jets at the LHC energies}},  {\em Phys. Lett. B} {\bf 801} (2020) 135181, [\href{http://arxiv.org/abs/1906.09562}{{\tt arXiv:1906.09562}}].

\bibitem{Casalderrey-Solana:2020rsj}
J.~Casalderrey-Solana, J.~G. Milhano, D.~Pablos, K.~Rajagopal, and X.~Yao, {\it {Jet Wake from Linearized Hydrodynamics}},  {\em JHEP} {\bf 05} (2021) 230, [\href{http://arxiv.org/abs/2010.01140}{{\tt arXiv:2010.01140}}].

\bibitem{Chen:2020tbl}
W.~Chen, S.~Cao, T.~Luo, L.-G. Pang, and X.-N. Wang, {\it {Medium modification of $\gamma$-jet fragmentation functions in Pb+Pb collisions at LHC}},  {\em Phys. Lett. B} {\bf 810} (2020) 135783, [\href{http://arxiv.org/abs/2005.09678}{{\tt arXiv:2005.09678}}].

\bibitem{Yang:2021qtl}
Z.~Yang, W.~Chen, Y.~He, W.~Ke, L.~Pang, and X.-N. Wang, {\it {Search for the Elusive Jet-Induced Diffusion Wake in $Z/\gamma$-Jets with 2D Jet Tomography in High-Energy Heavy-Ion Collisions}},  {\em Phys. Rev. Lett.} {\bf 127} (2021), no.~8 082301, [\href{http://arxiv.org/abs/2101.05422}{{\tt arXiv:2101.05422}}].

\bibitem{Yang:2022nei}
Z.~Yang, T.~Luo, W.~Chen, L.-G. Pang, and X.-N. Wang, {\it {3D Structure of Jet-Induced Diffusion Wake in an Expanding Quark-Gluon Plasma}},  {\em Phys. Rev. Lett.} {\bf 130} (2023), no.~5 052301, [\href{http://arxiv.org/abs/2203.03683}{{\tt arXiv:2203.03683}}].

\bibitem{Luo:2023nsi}
T.~Luo, Y.~He, S.~Cao, and X.-N. Wang, {\it {Linear Boltzmann transport for jet propagation in the quark-gluon plasma: Inelastic processes and jet modification}},  {\em Phys. Rev. C} {\bf 109} (2024), no.~3 034919, [\href{http://arxiv.org/abs/2306.13742}{{\tt arXiv:2306.13742}}].

\bibitem{Maldacena:2002vr}
J.~M. Maldacena, {\it {Non-Gaussian features of primordial fluctuations in single field inflationary models}},  {\em JHEP} {\bf 05} (2003) 013, [\href{http://arxiv.org/abs/astro-ph/0210603}{{\tt astro-ph/0210603}}].

\bibitem{Babich:2004gb}
D.~Babich, P.~Creminelli, and M.~Zaldarriaga, {\it {The Shape of non-Gaussianities}},  {\em JCAP} {\bf 08} (2004) 009, [\href{http://arxiv.org/abs/astro-ph/0405356}{{\tt astro-ph/0405356}}].

\bibitem{Arkani-Hamed:2015bza}
N.~Arkani-Hamed and J.~Maldacena, {\it {Cosmological Collider Physics}},  \href{http://arxiv.org/abs/1503.08043}{{\tt arXiv:1503.08043}}.

\bibitem{Arkani-Hamed:2018kmz}
N.~Arkani-Hamed, D.~Baumann, H.~Lee, and G.~L. Pimentel, {\it {The Cosmological Bootstrap: Inflationary Correlators from Symmetries and Singularities}},  {\em JHEP} {\bf 04} (2020) 105, [\href{http://arxiv.org/abs/1811.00024}{{\tt arXiv:1811.00024}}].

\bibitem{Basham:1978bw}
C.~Basham, L.~S. Brown, S.~D. Ellis, and S.~T. Love, {\it {Energy Correlations in electron - Positron Annihilation: Testing QCD}},  {\em Phys. Rev. Lett.} {\bf 41} (1978) 1585.

\bibitem{Basham:1977iq}
C.~L. Basham, L.~S. Brown, S.~D. Ellis, and S.~T. Love, {\it {Electron - Positron Annihilation Energy Pattern in Quantum Chromodynamics: Asymptotically Free Perturbation Theory}},  {\em Phys. Rev. D} {\bf 17} (1978) 2298.

\bibitem{Ellis:1978ty}
R.~K. Ellis, H.~Georgi, M.~Machacek, H.~D. Politzer, and G.~G. Ross, {\it {Perturbation Theory and the Parton Model in QCD}},  {\em Nucl. Phys. B} {\bf 152} (1979) 285--329.

\bibitem{Basham:1979gh}
C.~L. Basham, L.~S. Brown, S.~D. Ellis, and S.~T. Love, {\it {Energy Correlations in Perturbative Quantum Chromodynamics: A Conjecture for All Orders}},  {\em Phys. Lett. B} {\bf 85} (1979) 297--299.

\bibitem{Basham:1978zq}
C.~Basham, L.~Brown, S.~Ellis, and S.~Love, {\it {Energy Correlations in electron-Positron Annihilation in Quantum Chromodynamics: Asymptotically Free Perturbation Theory}},  {\em Phys. Rev. D} {\bf 19} (1979) 2018.

\bibitem{Hofman:2008ar}
D.~M. Hofman and J.~Maldacena, {\it {Conformal collider physics: Energy and charge correlations}},  {\em JHEP} {\bf 05} (2008) 012, [\href{http://arxiv.org/abs/0803.1467}{{\tt arXiv:0803.1467}}].

\bibitem{Dixon:2019uzg}
L.~J. Dixon, I.~Moult, and H.~X. Zhu, {\it {Collinear limit of the energy-energy correlator}},  {\em Phys. Rev. D} {\bf 100} (2019), no.~1 014009, [\href{http://arxiv.org/abs/1905.01310}{{\tt arXiv:1905.01310}}].

\bibitem{Chen:2020vvp}
H.~Chen, I.~Moult, X.~Zhang, and H.~X. Zhu, {\it {Rethinking jets with energy correlators: Tracks, resummation, and analytic continuation}},  {\em Phys. Rev. D} {\bf 102} (2020), no.~5 054012, [\href{http://arxiv.org/abs/2004.11381}{{\tt arXiv:2004.11381}}].

\bibitem{Komiske:2022enw}
P.~T. Komiske, I.~Moult, J.~Thaler, and H.~X. Zhu, {\it {Analyzing N-point Energy Correlators Inside Jets with CMS Open Data}},  \href{http://arxiv.org/abs/2201.07800}{{\tt arXiv:2201.07800}}.

\bibitem{Chen:2019bpb}
H.~Chen, M.-X. Luo, I.~Moult, T.-Z. Yang, X.~Zhang, and H.~X. Zhu, {\it {Three point energy correlators in the collinear limit: symmetries, dualities and analytic results}},  {\em JHEP} {\bf 08} (2020), no.~08 028, [\href{http://arxiv.org/abs/1912.11050}{{\tt arXiv:1912.11050}}].

\bibitem{Chen:2020adz}
H.~Chen, I.~Moult, and H.~X. Zhu, {\it {Quantum Interference in Jet Substructure from Spinning Gluons}},  \href{http://arxiv.org/abs/2011.02492}{{\tt arXiv:2011.02492}}.

\bibitem{Chen:2021gdk}
H.~Chen, I.~Moult, and H.~X. Zhu, {\it {Spinning Gluons from the QCD Light-Ray OPE}},  \href{http://arxiv.org/abs/2104.00009}{{\tt arXiv:2104.00009}}.

\bibitem{Lee:2022ige}
K.~Lee, B.~Me\c{c}aj, and I.~Moult, {\it {Conformal Colliders Meet the LHC}},  \href{http://arxiv.org/abs/2205.03414}{{\tt arXiv:2205.03414}}.

\bibitem{Chen:2022swd}
H.~Chen, I.~Moult, J.~Thaler, and H.~X. Zhu, {\it {Non-Gaussianities in Collider Energy Flux}},  \href{http://arxiv.org/abs/2205.02857}{{\tt arXiv:2205.02857}}.

\bibitem{Craft:2022kdo}
E.~Craft, K.~Lee, B.~Me\c{c}aj, and I.~Moult, {\it {Beautiful and Charming Energy Correlators}},  \href{http://arxiv.org/abs/2210.09311}{{\tt arXiv:2210.09311}}.

\bibitem{Lee:2023npz}
K.~Lee and I.~Moult, {\it {Energy Correlators Taking Charge}},  \href{http://arxiv.org/abs/2308.00746}{{\tt arXiv:2308.00746}}.

\bibitem{Lee:2023tkr}
K.~Lee and I.~Moult, {\it {Joint Track Functions: Expanding the Space of Calculable Correlations at Colliders}},  \href{http://arxiv.org/abs/2308.01332}{{\tt arXiv:2308.01332}}.

\bibitem{Lin:2024lsj}
Z.~Lin, M.~Ruan, M.~Xiao, and Z.~Xu, {\it {Extracting $\alpha_\mathrm{S}$ at future $e^+e^{-}$ Higgs factory with energy correlators}},  \href{http://arxiv.org/abs/2406.10946}{{\tt arXiv:2406.10946}}.

\bibitem{Holguin:2022epo}
J.~Holguin, I.~Moult, A.~Pathak, and M.~Procura, {\it {A New Paradigm for Precision Top Physics: Weighing the Top with Energy Correlators}},  \href{http://arxiv.org/abs/2201.08393}{{\tt arXiv:2201.08393}}.

\bibitem{Holguin:2023bjf}
J.~Holguin, I.~Moult, A.~Pathak, M.~Procura, R.~Sch\"ofbeck, and D.~Schwarz, {\it {Using the $W$ as a Standard Candle to Reach the Top: Calibrating Energy Correlator Based Top Mass Measurements}},  \href{http://arxiv.org/abs/2311.02157}{{\tt arXiv:2311.02157}}.

\bibitem{Xiao:2024rol}
M.~Xiao, Y.~Ye, and X.~Zhu, {\it {Prospect of measuring the top quark mass through energy correlators}},  \href{http://arxiv.org/abs/2405.20001}{{\tt arXiv:2405.20001}}.

\bibitem{Liu:2022wop}
X.~Liu and H.~X. Zhu, {\it {Nucleon Energy Correlators}},  {\em Phys. Rev. Lett.} {\bf 130} (2023), no.~9 091901, [\href{http://arxiv.org/abs/2209.02080}{{\tt arXiv:2209.02080}}].

\bibitem{Liu:2023aqb}
H.-Y. Liu, X.~Liu, J.-C. Pan, F.~Yuan, and H.~X. Zhu, {\it {Nucleon Energy Correlators for the Color Glass Condensate}},  {\em Phys. Rev. Lett.} {\bf 130} (2023), no.~18 181901, [\href{http://arxiv.org/abs/2301.01788}{{\tt arXiv:2301.01788}}].

\bibitem{Cao:2023oef}
H.~Cao, X.~Liu, and H.~X. Zhu, {\it {Toward precision measurements of nucleon energy correlators in lepton-nucleon collisions}},  {\em Phys. Rev. D} {\bf 107} (2023), no.~11 114008, [\href{http://arxiv.org/abs/2303.01530}{{\tt arXiv:2303.01530}}].

\bibitem{Li:2023gkh}
X.~L. Li, X.~Liu, F.~Yuan, and H.~X. Zhu, {\it {Illuminating nucleon-gluon interference via calorimetric asymmetry}},  {\em Phys. Rev. D} {\bf 108} (2023), no.~9 L091502, [\href{http://arxiv.org/abs/2308.10942}{{\tt arXiv:2308.10942}}].

\bibitem{Liu:2024kqt}
X.~Liu and H.~X. Zhu, {\it {TMDs from Semi-inclusive Energy Correlators}},  \href{http://arxiv.org/abs/2403.08874}{{\tt arXiv:2403.08874}}.

\bibitem{Guo:2024jch}
Y.~Guo, X.~Liu, F.~Yuan, and H.~X. Zhu, {\it {Long Range Azimuthal Correlation, Entanglement and Bell Inequality Violation by Spinning Gluons at the LHC}},  \href{http://arxiv.org/abs/2406.05880}{{\tt arXiv:2406.05880}}.

\bibitem{Devereaux:2023vjz}
K.~Devereaux, W.~Fan, W.~Ke, K.~Lee, and I.~Moult, {\it {Imaging Cold Nuclear Matter with Energy Correlators}},  \href{http://arxiv.org/abs/2303.08143}{{\tt arXiv:2303.08143}}.

\bibitem{Chen:2024nfl}
A.-P. Chen, X.~Liu, and Y.-Q. Ma, {\it {Shedding Light on Hadronization by Quarkonium Energy Correlator}},  \href{http://arxiv.org/abs/2405.10056}{{\tt arXiv:2405.10056}}.

\bibitem{Chen:2024bpj}
K.-B. Chen, J.-P. Ma, and X.-B. Tong, {\it {The connection between Nucleon Energy Correlators and Fracture Functions}},  \href{http://arxiv.org/abs/2406.08559}{{\tt arXiv:2406.08559}}.

\bibitem{Andres:2022ovj}
C.~Andres, F.~Dominguez, R.~Kunnawalkam~Elayavalli, J.~Holguin, C.~Marquet, and I.~Moult, {\it {Resolving the Scales of the Quark-Gluon Plasma with Energy Correlators}},  \href{http://arxiv.org/abs/2209.11236}{{\tt arXiv:2209.11236}}.

\bibitem{Andres:2023xwr}
C.~Andres, F.~Dominguez, J.~Holguin, C.~Marquet, and I.~Moult, {\it {A Coherent View of the Quark-Gluon Plasma from Energy Correlators}},  \href{http://arxiv.org/abs/2303.03413}{{\tt arXiv:2303.03413}}.

\bibitem{Andres:2023ymw}
C.~Andres, F.~Dominguez, J.~Holguin, C.~Marquet, and I.~Moult, {\it {Seeing Beauty in the Quark-Gluon Plasma with Energy Correlators}},  \href{http://arxiv.org/abs/2307.15110}{{\tt arXiv:2307.15110}}.

\bibitem{Barata:2023vnl}
J.~a. Barata and Y.~Mehtar-Tani, {\it {Energy loss effects in EECs at LO}},  in {\em {11th International Conference on Hard and Electromagnetic Probes of High-Energy Nuclear Collisions}: {Hard Probes 2023}}, 7, 2023.
\newblock \href{http://arxiv.org/abs/2307.08943}{{\tt arXiv:2307.08943}}.

\bibitem{Barata:2023zqg}
J.~a. Barata, J.~G. Milhano, and A.~V. Sadofyev, {\it {Picturing QCD jets in anisotropic matter: from jet shapes to Energy Energy Correlators}},  \href{http://arxiv.org/abs/2308.01294}{{\tt arXiv:2308.01294}}.

\bibitem{Yang:2023dwc}
Z.~Yang, Y.~He, I.~Moult, and X.-N. Wang, {\it {Probing the Short-Distance Structure of the Quark-Gluon Plasma with Energy Correlators}},  \href{http://arxiv.org/abs/2310.01500}{{\tt arXiv:2310.01500}}.

\bibitem{Barata:2023bhh}
J.~a. Barata, P.~Caucal, A.~Soto-Ontoso, and R.~Szafron, {\it {Advancing the understanding of energy-energy correlators in heavy-ion collisions}},  \href{http://arxiv.org/abs/2312.12527}{{\tt arXiv:2312.12527}}.

\bibitem{Barata:2024nqo}
J.~a. Barata and R.~Szafron, {\it {Leading order track functions in a hot and dense QGP}},  \href{http://arxiv.org/abs/2401.04164}{{\tt arXiv:2401.04164}}.

\bibitem{Andres:2024ksi}
C.~Andres, F.~Dominguez, J.~Holguin, C.~Marquet, and I.~Moult, {\it {Towards an Interpretation of the First Measurements of Energy Correlators in the Quark-Gluon Plasma}},  \href{http://arxiv.org/abs/2407.07936}{{\tt arXiv:2407.07936}}.

\bibitem{Fan:2023}
W.~Fan, {\it {First Energy-Energy Correlator Measurements for Inclusive and Heavy-Flavour Tagged Jets with ALICE, Quark Matter 2023}}, .

\bibitem{Tamis:2023guc}
A.~Tamis, {\it {Measurement of Two-Point Energy Correlators Within Jets in $pp$ Collisions at $\sqrt{s}$ = 200 GeV at STAR}},  in {\em {11th International Conference on Hard and Electromagnetic Probes of High-Energy Nuclear Collisions}: {Hard Probes 2023}}, 9, 2023.
\newblock \href{http://arxiv.org/abs/2309.05761}{{\tt arXiv:2309.05761}}.

\bibitem{CMS:2024mlf}
{\bf CMS} Collaboration, A.~Hayrapetyan et~al., {\it {Measurement of energy correlators inside jets and determination of the strong coupling $\alpha_\mathrm{S}(m_\mathrm{Z})$}},  \href{http://arxiv.org/abs/2402.13864}{{\tt arXiv:2402.13864}}.

\bibitem{Casalderrey-Solana:2014bpa}
J.~Casalderrey-Solana, D.~C. Gulhan, J.~G. Milhano, D.~Pablos, and K.~Rajagopal, {\it {A Hybrid Strong/Weak Coupling Approach to Jet Quenching}},  {\em JHEP} {\bf 10} (2014) 019, [\href{http://arxiv.org/abs/1405.3864}{{\tt arXiv:1405.3864}}]. [Erratum: JHEP 09, 175 (2015)].

\bibitem{Casalderrey-Solana:2015vaa}
J.~Casalderrey-Solana, D.~C. Gulhan, J.~G. Milhano, D.~Pablos, and K.~Rajagopal, {\it {Predictions for Boson-Jet Observables and Fragmentation Function Ratios from a Hybrid Strong/Weak Coupling Model for Jet Quenching}},  {\em JHEP} {\bf 03} (2016) 053, [\href{http://arxiv.org/abs/1508.00815}{{\tt arXiv:1508.00815}}].

\bibitem{Hulcher:2017cpt}
Z.~Hulcher, D.~Pablos, and K.~Rajagopal, {\it {Resolution Effects in the Hybrid Strong/Weak Coupling Model}},  {\em JHEP} {\bf 03} (2018) 010, [\href{http://arxiv.org/abs/1707.05245}{{\tt arXiv:1707.05245}}].

\bibitem{Casalderrey-Solana:2018wrw}
J.~Casalderrey-Solana, Z.~Hulcher, G.~Milhano, D.~Pablos, and K.~Rajagopal, {\it {Simultaneous description of hadron and jet suppression in heavy-ion collisions}},  {\em Phys. Rev. C} {\bf 99} (2019), no.~5 051901, [\href{http://arxiv.org/abs/1808.07386}{{\tt arXiv:1808.07386}}].

\bibitem{Casalderrey-Solana:2019ubu}
J.~Casalderrey-Solana, G.~Milhano, D.~Pablos, and K.~Rajagopal, {\it {Modification of Jet Substructure in Heavy Ion Collisions as a Probe of the Resolution Length of Quark-Gluon Plasma}},  {\em JHEP} {\bf 01} (2020) 044, [\href{http://arxiv.org/abs/1907.11248}{{\tt arXiv:1907.11248}}].

\bibitem{Hulcher:2022kmn}
Z.~Hulcher, D.~Pablos, and K.~Rajagopal, {\it {Sensitivity of Jet Observables to the Presence of Quasi-particles in QGP}},  {\em Acta Phys. Polon. Supp.} {\bf 16} (2023), no.~1 1--A57, [\href{http://arxiv.org/abs/2208.13593}{{\tt arXiv:2208.13593}}].

\bibitem{Sveshnikov:1995vi}
N.~Sveshnikov and F.~Tkachov, {\it {Jets and quantum field theory}},  {\em Phys. Lett. B} {\bf 382} (1996) 403--408, [\href{http://arxiv.org/abs/hep-ph/9512370}{{\tt hep-ph/9512370}}].

\bibitem{Tkachov:1995kk}
F.~V. Tkachov, {\it {Measuring multi - jet structure of hadronic energy flow or What is a jet?}},  {\em Int. J. Mod. Phys. A} {\bf 12} (1997) 5411--5529, [\href{http://arxiv.org/abs/hep-ph/9601308}{{\tt hep-ph/9601308}}].

\bibitem{Korchemsky:1999kt}
G.~P. Korchemsky and G.~F. Sterman, {\it {Power corrections to event shapes and factorization}},  {\em Nucl. Phys. B} {\bf 555} (1999) 335--351, [\href{http://arxiv.org/abs/hep-ph/9902341}{{\tt hep-ph/9902341}}].

\bibitem{Bauer:2008dt}
C.~W. Bauer, S.~P. Fleming, C.~Lee, and G.~F. Sterman, {\it {Factorization of e+e- Event Shape Distributions with Hadronic Final States in Soft Collinear Effective Theory}},  {\em Phys. Rev. D} {\bf 78} (2008) 034027, [\href{http://arxiv.org/abs/0801.4569}{{\tt arXiv:0801.4569}}].

\bibitem{Belitsky:2013xxa}
A.~Belitsky, S.~Hohenegger, G.~Korchemsky, E.~Sokatchev, and A.~Zhiboedov, {\it {From correlation functions to event shapes}},  {\em Nucl. Phys. B} {\bf 884} (2014) 305--343, [\href{http://arxiv.org/abs/1309.0769}{{\tt arXiv:1309.0769}}].

\bibitem{Belitsky:2013bja}
A.~Belitsky, S.~Hohenegger, G.~Korchemsky, E.~Sokatchev, and A.~Zhiboedov, {\it {Event shapes in $\mathcal{N} = 4$ super-Yang-Mills theory}},  {\em Nucl. Phys. B} {\bf 884} (2014) 206--256, [\href{http://arxiv.org/abs/1309.1424}{{\tt arXiv:1309.1424}}].

\bibitem{Kravchuk:2018htv}
P.~Kravchuk and D.~Simmons-Duffin, {\it {Light-ray operators in conformal field theory}},  {\em JHEP} {\bf 11} (2018) 102, [\href{http://arxiv.org/abs/1805.00098}{{\tt arXiv:1805.00098}}].

\bibitem{Dixon:2018qgp}
L.~J. Dixon, M.-X. Luo, V.~Shtabovenko, T.-Z. Yang, and H.~X. Zhu, {\it {Analytical Computation of Energy-Energy Correlation at Next-to-Leading Order in QCD}},  {\em Phys. Rev. Lett.} {\bf 120} (2018), no.~10 102001, [\href{http://arxiv.org/abs/1801.03219}{{\tt arXiv:1801.03219}}].

\bibitem{Luo:2019nig}
M.-X. Luo, V.~Shtabovenko, T.-Z. Yang, and H.~X. Zhu, {\it {Analytic Next-To-Leading Order Calculation of Energy-Energy Correlation in Gluon-Initiated Higgs Decays}},  {\em JHEP} {\bf 06} (2019) 037, [\href{http://arxiv.org/abs/1903.07277}{{\tt arXiv:1903.07277}}].

\bibitem{Belitsky:2013ofa}
A.~Belitsky, S.~Hohenegger, G.~Korchemsky, E.~Sokatchev, and A.~Zhiboedov, {\it {Energy-Energy Correlations in N=4 Supersymmetric Yang-Mills Theory}},  {\em Phys. Rev. Lett.} {\bf 112} (2014), no.~7 071601, [\href{http://arxiv.org/abs/1311.6800}{{\tt arXiv:1311.6800}}].

\bibitem{Henn:2019gkr}
J.~Henn, E.~Sokatchev, K.~Yan, and A.~Zhiboedov, {\it {Energy-energy correlation in $N$=4 super Yang-Mills theory at next-to-next-to-leading order}},  {\em Phys. Rev. D} {\bf 100} (2019), no.~3 036010, [\href{http://arxiv.org/abs/1903.05314}{{\tt arXiv:1903.05314}}].

\bibitem{Yan:2022cye}
K.~Yan and X.~Zhang, {\it {Three-point energy correlator in $\mathcal{N}=4$ super Yang-Mills Theory}},  \href{http://arxiv.org/abs/2203.04349}{{\tt arXiv:2203.04349}}.

\bibitem{Yang:2022tgm}
T.-Z. Yang and X.~Zhang, {\it {Analytic Computation of three-point energy correlator in QCD}},  {\em JHEP} {\bf 09} (2022) 006, [\href{http://arxiv.org/abs/2208.01051}{{\tt arXiv:2208.01051}}].

\bibitem{Yang:2024gcn}
T.-Z. Yang and X.~Zhang, {\it {Three-point energy correlators in hadronic Higgs boson decays}},  {\em Phys. Rev. D} {\bf 109} (2024), no.~11 114036, [\href{http://arxiv.org/abs/2402.05174}{{\tt arXiv:2402.05174}}].

\bibitem{Chicherin:2024ifn}
D.~Chicherin, I.~Moult, E.~Sokatchev, K.~Yan, and Y.~Zhu, {\it {The Collinear Limit of the Four-Point Energy Correlator in $\mathcal{N} = 4$ Super Yang-Mills Theory}},  \href{http://arxiv.org/abs/2401.06463}{{\tt arXiv:2401.06463}}.

\bibitem{Maldacena:1997re}
J.~M. Maldacena, {\it {The Large N limit of superconformal field theories and supergravity}},  {\em Adv. Theor. Math. Phys.} {\bf 2} (1998) 231--252, [\href{http://arxiv.org/abs/hep-th/9711200}{{\tt hep-th/9711200}}].

\bibitem{Gubser:1998bc}
S.~S. Gubser, I.~R. Klebanov, and A.~M. Polyakov, {\it {Gauge theory correlators from noncritical string theory}},  {\em Phys. Lett. B} {\bf 428} (1998) 105--114, [\href{http://arxiv.org/abs/hep-th/9802109}{{\tt hep-th/9802109}}].

\bibitem{Witten:1998qj}
E.~Witten, {\it {Anti-de Sitter space and holography}},  {\em Adv. Theor. Math. Phys.} {\bf 2} (1998) 253--291, [\href{http://arxiv.org/abs/hep-th/9802150}{{\tt hep-th/9802150}}].

\bibitem{Csaki:2024joe}
C.~Cs\'aki and A.~Ismail, {\it {Holographic Energy Correlators for Confining Theories}},  \href{http://arxiv.org/abs/2403.12123}{{\tt arXiv:2403.12123}}.

\bibitem{Chicherin:2023gxt}
D.~Chicherin, G.~P. Korchemsky, E.~Sokatchev, and A.~Zhiboedov, {\it {Energy correlations in heavy states}},  {\em JHEP} {\bf 11} (2023) 134, [\href{http://arxiv.org/abs/2306.14330}{{\tt arXiv:2306.14330}}].

\bibitem{Firat:2023lbp}
E.~Firat, A.~Monin, R.~Rattazzi, and M.~T. Walters, {\it {Flux correlators and semiclassics}},  {\em JHEP} {\bf 03} (2024) 067, [\href{http://arxiv.org/abs/2309.14428}{{\tt arXiv:2309.14428}}].

\bibitem{Kologlu:2019mfz}
M.~Kologlu, P.~Kravchuk, D.~Simmons-Duffin, and A.~Zhiboedov, {\it {The light-ray OPE and conformal colliders}},  {\em JHEP} {\bf 01} (2021) 128, [\href{http://arxiv.org/abs/1905.01311}{{\tt arXiv:1905.01311}}].

\bibitem{Chang:2020qpj}
C.-H. Chang, M.~Kologlu, P.~Kravchuk, D.~Simmons-Duffin, and A.~Zhiboedov, {\it {Transverse spin in the light-ray OPE}},  \href{http://arxiv.org/abs/2010.04726}{{\tt arXiv:2010.04726}}.

\bibitem{Cacciari:2008gp}
M.~Cacciari, G.~P. Salam, and G.~Soyez, {\it {The anti-$k_t$ jet clustering algorithm}},  {\em JHEP} {\bf 04} (2008) 063, [\href{http://arxiv.org/abs/0802.1189}{{\tt arXiv:0802.1189}}].

\bibitem{Cacciari:2011ma}
M.~Cacciari, G.~P. Salam, and G.~Soyez, {\it {FastJet User Manual}},  {\em Eur. Phys. J. C} {\bf 72} (2012) 1896, [\href{http://arxiv.org/abs/1111.6097}{{\tt arXiv:1111.6097}}].

\bibitem{Sjostrand:2014zea}
T.~Sj\"ostrand, S.~Ask, J.~R. Christiansen, R.~Corke, N.~Desai, P.~Ilten, S.~Mrenna, S.~Prestel, C.~O. Rasmussen, and P.~Z. Skands, {\it {An introduction to PYTHIA 8.2}},  {\em Comput. Phys. Commun.} {\bf 191} (2015) 159--177, [\href{http://arxiv.org/abs/1410.3012}{{\tt arXiv:1410.3012}}].

\bibitem{Eskola:2009uj}
K.~J. Eskola, H.~Paukkunen, and C.~A. Salgado, {\it {EPS09: A New Generation of NLO and LO Nuclear Parton Distribution Functions}},  {\em JHEP} {\bf 04} (2009) 065, [\href{http://arxiv.org/abs/0902.4154}{{\tt arXiv:0902.4154}}].

\bibitem{Shen:2014vra}
C.~Shen, Z.~Qiu, H.~Song, J.~Bernhard, S.~Bass, and U.~Heinz, {\it {The iEBE-VISHNU code package for relativistic heavy-ion collisions}},  {\em Comput. Phys. Commun.} {\bf 199} (2016) 61--85, [\href{http://arxiv.org/abs/1409.8164}{{\tt arXiv:1409.8164}}].

\bibitem{Casalderrey-Solana:2011fza}
J.~Casalderrey-Solana, J.~G. Milhano, and P.~Quiroga-Arias, {\it {Out of Medium Fragmentation from Long-Lived Jet Showers}},  {\em Phys. Lett. B} {\bf 710} (2012) 175--181, [\href{http://arxiv.org/abs/1111.0310}{{\tt arXiv:1111.0310}}].

\bibitem{Caucal:2018dla}
P.~Caucal, E.~Iancu, A.~H. Mueller, and G.~Soyez, {\it {Vacuum-like jet fragmentation in a dense QCD medium}},  {\em Phys. Rev. Lett.} {\bf 120} (2018) 232001, [\href{http://arxiv.org/abs/1801.09703}{{\tt arXiv:1801.09703}}].

\bibitem{Gubser:2008as}
S.~S. Gubser, D.~R. Gulotta, S.~S. Pufu, and F.~D. Rocha, {\it {Gluon energy loss in the gauge-string duality}},  {\em JHEP} {\bf 10} (2008) 052, [\href{http://arxiv.org/abs/0803.1470}{{\tt arXiv:0803.1470}}].

\bibitem{PhysRevD.10.186}
F.~Cooper and G.~Frye, {\it Single-particle distribution in the hydrodynamic and statistical thermodynamic models of multiparticle production},  {\em Phys. Rev. D} {\bf 10} (Jul, 1974) 186--189.

\bibitem{CMS:2015hkr}
{\bf CMS} Collaboration, V.~Khachatryan et~al., {\it {Measurement of transverse momentum relative to dijet systems in PbPb and pp collisions at $ \sqrt{s_{\mathrm{NN}}}=2.76 $ TeV}},  {\em JHEP} {\bf 01} (2016) 006, [\href{http://arxiv.org/abs/1509.09029}{{\tt arXiv:1509.09029}}].

\bibitem{Yan:2017rku}
L.~Yan, S.~Jeon, and C.~Gale, {\it {Jet-medium interaction and conformal relativistic fluid dynamics}},  {\em Phys. Rev. C} {\bf 97} (2018), no.~3 034914, [\href{http://arxiv.org/abs/1707.09519}{{\tt arXiv:1707.09519}}].

\bibitem{Tachibana:2020mtb}
Y.~Tachibana, C.~Shen, and A.~Majumder, {\it {Bulk medium evolution has considerable effects on jet observables}},  {\em Phys. Rev. C} {\bf 106} (2022), no.~2 L021902, [\href{http://arxiv.org/abs/2001.08321}{{\tt arXiv:2001.08321}}].

\bibitem{wakeprep}
J.~Casalderrey-Solana, J.~G. Milhano, D.~Pablos, K.~Rajagopal, and X.~Yao In preparation.

\bibitem{Chang:2013rca}
H.-M. Chang, M.~Procura, J.~Thaler, and W.~J. Waalewijn, {\it {Calculating Track-Based Observables for the LHC}},  {\em Phys. Rev. Lett.} {\bf 111} (2013) 102002, [\href{http://arxiv.org/abs/1303.6637}{{\tt arXiv:1303.6637}}].

\bibitem{Chang:2013iba}
H.-M. Chang, M.~Procura, J.~Thaler, and W.~J. Waalewijn, {\it {Calculating Track Thrust with Track Functions}},  {\em Phys. Rev.} {\bf D88} (2013) 034030, [\href{http://arxiv.org/abs/1306.6630}{{\tt arXiv:1306.6630}}].

\bibitem{Jaarsma:2023ell}
M.~Jaarsma, Y.~Li, I.~Moult, W.~J. Waalewijn, and H.~X. Zhu, {\it {Energy Correlators on Tracks: Resummation and Non-Perturbative Effects}},  \href{http://arxiv.org/abs/2307.15739}{{\tt arXiv:2307.15739}}.

\bibitem{Chen:2022pdu}
H.~Chen, M.~Jaarsma, Y.~Li, I.~Moult, W.~J. Waalewijn, and H.~X. Zhu, {\it {Multi-collinear splitting kernels for track function evolution}},  {\em JHEP} {\bf 07} (2023) 185, [\href{http://arxiv.org/abs/2210.10058}{{\tt arXiv:2210.10058}}].

\bibitem{Chen:2022muj}
H.~Chen, M.~Jaarsma, Y.~Li, I.~Moult, W.~J. Waalewijn, and H.~X. Zhu, {\it {Collinear Parton Dynamics Beyond DGLAP}},  \href{http://arxiv.org/abs/2210.10061}{{\tt arXiv:2210.10061}}.

\bibitem{Jaarsma:2022kdd}
M.~Jaarsma, Y.~Li, I.~Moult, W.~Waalewijn, and H.~X. Zhu, {\it {Renormalization Group Flows for Track Function Moments}},  \href{http://arxiv.org/abs/2201.05166}{{\tt arXiv:2201.05166}}.

\bibitem{Li:2021zcf}
Y.~Li, I.~Moult, S.~S. van Velzen, W.~J. Waalewijn, and H.~X. Zhu, {\it {Extending Precision Perturbative QCD with Track Functions}},  \href{http://arxiv.org/abs/2108.01674}{{\tt arXiv:2108.01674}}.

\bibitem{ALICE:2020atx}
{\bf ALICE} Collaboration, S.~Acharya et~al., {\it {Measurement of isolated photon-hadron correlations in $\sqrt{s_{\rm{NN}}}$ = 5.02 TeV $pp$ and $p$-Pb collisions}},  {\em Phys. Rev. C} {\bf 102} (2020), no.~4 044908, [\href{http://arxiv.org/abs/2005.14637}{{\tt arXiv:2005.14637}}].

\bibitem{Zhao:2021vmu}
W.~Zhao, W.~Ke, W.~Chen, T.~Luo, and X.-N. Wang, {\it {From Hydrodynamics to Jet Quenching, Coalescence, and Hadron Cascade: A Coupled Approach to Solving the RAA\ensuremath{\otimes}v2 Puzzle}},  {\em Phys. Rev. Lett.} {\bf 128} (2022), no.~2 022302, [\href{http://arxiv.org/abs/2103.14657}{{\tt arXiv:2103.14657}}].

\bibitem{DEramo:2012uzl}
F.~D'Eramo, M.~Lekaveckas, H.~Liu, and K.~Rajagopal, {\it {Momentum Broadening in Weakly Coupled Quark-Gluon Plasma (with a view to finding the quasiparticles within liquid quark-gluon plasma)}},  {\em JHEP} {\bf 05} (2013) 031, [\href{http://arxiv.org/abs/1211.1922}{{\tt arXiv:1211.1922}}].

\bibitem{DEramo:2018eoy}
F.~D'Eramo, K.~Rajagopal, and Y.~Yin, {\it {Moli\`ere scattering in quark-gluon plasma: finding point-like scatterers in a liquid}},  {\em JHEP} {\bf 01} (2019) 172, [\href{http://arxiv.org/abs/1808.03250}{{\tt arXiv:1808.03250}}].

\bibitem{Pablos:2024muu}
D.~Pablos and S.~Sanjurjo, {\it {Color Coherence Effects in Dipole-Quark Scattering in the Soft Limit}},  \href{http://arxiv.org/abs/2406.08550}{{\tt arXiv:2406.08550}}.

\bibitem{Renk:2012ve}
T.~Renk, {\it {Biased showers: A common conceptual framework for the interpretation of high-$P_T$ observables in heavy-ion collisions}},  {\em Phys. Rev. C} {\bf 88} (2013), no.~5 054902, [\href{http://arxiv.org/abs/1212.0646}{{\tt arXiv:1212.0646}}].

\bibitem{Spousta:2015fca}
M.~Spousta and B.~Cole, {\it {Interpreting single jet measurements in Pb $+$ Pb collisions at the LHC}},  {\em Eur. Phys. J. C} {\bf 76} (2016), no.~2 50, [\href{http://arxiv.org/abs/1504.05169}{{\tt arXiv:1504.05169}}].

\bibitem{Milhano:2015mng}
J.~G. Milhano and K.~C. Zapp, {\it {Origins of the di-jet asymmetry in heavy ion collisions}},  {\em Eur. Phys. J. C} {\bf 76} (2016), no.~5 288, [\href{http://arxiv.org/abs/1512.08107}{{\tt arXiv:1512.08107}}].

\bibitem{Caucal:2020xad}
P.~Caucal, E.~Iancu, A.~H. Mueller, and G.~Soyez, {\it {Nuclear modification factors for jet fragmentation}},  {\em JHEP} {\bf 10} (2020) 204, [\href{http://arxiv.org/abs/2005.05852}{{\tt arXiv:2005.05852}}].

\bibitem{Du:2020pmp}
Y.-L. Du, D.~Pablos, and K.~Tywoniuk, {\it {Deep learning jet modifications in heavy-ion collisions}},  {\em JHEP} {\bf 21} (2020) 206, [\href{http://arxiv.org/abs/2012.07797}{{\tt arXiv:2012.07797}}].

\bibitem{Brewer:2021hmh}
J.~Brewer, Q.~Brodsky, and K.~Rajagopal, {\it {Disentangling jet modification in jet simulations and in Z+jet data}},  {\em JHEP} {\bf 02} (2022) 175, [\href{http://arxiv.org/abs/2110.13159}{{\tt arXiv:2110.13159}}].

\bibitem{Caucal:2021cfb}
P.~Caucal, A.~Soto-Ontoso, and A.~Takacs, {\it {Dynamically groomed jet radius in heavy-ion collisions}},  {\em Phys. Rev. D} {\bf 105} (2022), no.~11 114046, [\href{http://arxiv.org/abs/2111.14768}{{\tt arXiv:2111.14768}}].

\bibitem{Pablos:2022mrx}
D.~Pablos and A.~Soto-Ontoso, {\it {Pushing forward jet substructure measurements in heavy-ion collisions}},  {\em Phys. Rev. D} {\bf 107} (2023), no.~9 094003, [\href{http://arxiv.org/abs/2210.07901}{{\tt arXiv:2210.07901}}].

\bibitem{Kang:2023ycg}
J.-W. Kang, S.~Wang, L.~Wang, and B.-W. Zhang, {\it {Phenomenological study of the angle between jet axes in heavy-ion collisions}},  \href{http://arxiv.org/abs/2312.15518}{{\tt arXiv:2312.15518}}.

\bibitem{CMS:2024zjn}
{\bf CMS} Collaboration, A.~Hayrapetyan et~al., {\it {Girth and groomed radius of jets recoiling against isolated photons in lead-lead and proton-proton collisions at $\sqrt{s_\mathrm{NN}}$ = 5.02 TeV}},  \href{http://arxiv.org/abs/2405.02737}{{\tt arXiv:2405.02737}}.

\bibitem{Chen:2022jhb}
H.~Chen, I.~Moult, J.~Sandor, and H.~X. Zhu, {\it {Celestial Blocks and Transverse Spin in the Three-Point Energy Correlator}},  \href{http://arxiv.org/abs/2202.04085}{{\tt arXiv:2202.04085}}.

\bibitem{Chang:2022ryc}
C.-H. Chang and D.~Simmons-Duffin, {\it {Three-point energy correlators and the celestial block expansion}},  \href{http://arxiv.org/abs/2202.04090}{{\tt arXiv:2202.04090}}.

\bibitem{CMS-PAS-HIN-23-004}
{\bf CMS} Collaboration, {\it {Energy-energy correlators from PbPb and pp collisions at 5.02 TeV, CMS-PAS-HIN-23-004}},  tech. rep., CERN, Geneva, 2023.

\bibitem{DAgostini:1994fjx}
G.~D'Agostini, {\it {A Multidimensional unfolding method based on Bayes' theorem}},  {\em Nucl. Instrum. Meth. A} {\bf 362} (1995) 487--498.

\bibitem{Hocker:1995kb}
A.~Hocker and V.~Kartvelishvili, {\it {SVD approach to data unfolding}},  {\em Nucl. Instrum. Meth. A} {\bf 372} (1996) 469--481, [\href{http://arxiv.org/abs/hep-ph/9509307}{{\tt hep-ph/9509307}}].

\bibitem{Andreassen:2019cjw}
A.~Andreassen, P.~T. Komiske, E.~M. Metodiev, B.~Nachman, and J.~Thaler, {\it {OmniFold: A Method to Simultaneously Unfold All Observables}},  {\em Phys. Rev. Lett.} {\bf 124} (2020), no.~18 182001, [\href{http://arxiv.org/abs/1911.09107}{{\tt arXiv:1911.09107}}].

\bibitem{CMS:2021vui}
{\bf CMS} Collaboration, A.~M. Sirunyan et~al., {\it {First measurement of large area jet transverse momentum spectra in heavy-ion collisions}},  {\em JHEP} {\bf 05} (2021) 284, [\href{http://arxiv.org/abs/2102.13080}{{\tt arXiv:2102.13080}}].

\bibitem{Dasgupta:2007wa}
M.~Dasgupta, L.~Magnea, and G.~P. Salam, {\it {Non-perturbative QCD effects in jets at hadron colliders}},  {\em JHEP} {\bf 02} (2008) 055, [\href{http://arxiv.org/abs/0712.3014}{{\tt arXiv:0712.3014}}].

\bibitem{CMS:2020caw}
{\bf CMS} Collaboration, A.~M. Sirunyan et~al., {\it {Dependence of inclusive jet production on the anti-k$_{T}$ distance parameter in pp collisions at $ \sqrt{\mathrm{s}} $ = 13 TeV}},  {\em JHEP} {\bf 12} (2020) 082, [\href{http://arxiv.org/abs/2005.05159}{{\tt arXiv:2005.05159}}].

\bibitem{CMS:2011iwn}
{\bf CMS} Collaboration, S.~Chatrchyan et~al., {\it {Observation and studies of jet quenching in PbPb collisions at nucleon-nucleon center-of-mass energy = 2.76 TeV}},  {\em Phys. Rev. C} {\bf 84} (2011) 024906, [\href{http://arxiv.org/abs/1102.1957}{{\tt arXiv:1102.1957}}].

\bibitem{ALICE:2022wpn}
{\bf ALICE} Collaboration, {\it {The ALICE experiment -- A journey through QCD}},  \href{http://arxiv.org/abs/2211.04384}{{\tt arXiv:2211.04384}}.

\bibitem{STAR:2020xiv}
{\bf STAR} Collaboration, J.~Adam et~al., {\it {Measurement of inclusive charged-particle jet production in Au + Au collisions at $\sqrt{s_{NN}}=$200 GeV}},  {\em Phys. Rev. C} {\bf 102} (2020), no.~5 054913, [\href{http://arxiv.org/abs/2006.00582}{{\tt arXiv:2006.00582}}].

\bibitem{PHENIX:2020alr}
{\bf PHENIX} Collaboration, U.~Acharya et~al., {\it {Measurement of jet-medium interactions via direct photon-hadron correlations in Au$+$Au and $d$ $+$Au collisions at $\sqrt{s_{_{NN}}}=200$ GeV}},  {\em Phys. Rev. C} {\bf 102} (2020), no.~5 054910, [\href{http://arxiv.org/abs/2005.14270}{{\tt arXiv:2005.14270}}].

\bibitem{ATLAS:2023iad}
{\bf ATLAS} Collaboration, G.~Aad et~al., {\it {Comparison of inclusive and photon-tagged jet suppression in 5.02 TeV Pb+Pb collisions with ATLAS}},  {\em Phys. Lett. B} {\bf 846} (2023) 138154, [\href{http://arxiv.org/abs/2303.10090}{{\tt arXiv:2303.10090}}].

\bibitem{He:2024roj}
{\bf STAR} Collaboration, Y.~He, {\it {Measurements of semi-inclusive $\gamma$+jet and hadron+jet distributions in heavy-ion collisions at $\sqrt{s_\mathrm{NN}}$ = 200 GeV with STAR}},  {\em PoS} {\bf HardProbes2023} (2024) 174.

\bibitem{STAR:2023ksv}
{\bf STAR} Collaboration, {\it {Semi-inclusive direct photon+jet and $\pi^{0}$+jet correlations measured in $p+p$ and central Au+Au collisions at $\sqrt{s_\mathrm{NN}} = 200$ GeV}},  \href{http://arxiv.org/abs/2309.00145}{{\tt arXiv:2309.00145}}.

\bibitem{OPAL:1993pnw}
{\bf OPAL} Collaboration, P.~D. Acton et~al., {\it {A Determination of alpha-s (M (Z0)) at LEP using resummed QCD calculations}},  {\em Z. Phys. C} {\bf 59} (1993) 1--20.

\bibitem{ALICE:2023waz}
{\bf ALICE} Collaboration, S.~Acharya et~al., {\it {Measurement of the radius dependence of charged-particle jet suppression in Pb\textendash{}Pb collisions at sNN=5.02TeV}},  {\em Phys. Lett. B} {\bf 849} (2024) 138412, [\href{http://arxiv.org/abs/2303.00592}{{\tt arXiv:2303.00592}}].

\bibitem{Fickinger:2013xwa}
M.~Fickinger, G.~Ovanesyan, and I.~Vitev, {\it {Angular distributions of higher order splitting functions in the vacuum and in dense QCD matter}},  {\em JHEP} {\bf 07} (2013) 059, [\href{http://arxiv.org/abs/1304.3497}{{\tt arXiv:1304.3497}}].

\bibitem{Larkoski:2014uqa}
A.~J. Larkoski, D.~Neill, and J.~Thaler, {\it {Jet Shapes with the Broadening Axis}},  {\em JHEP} {\bf 04} (2014) 017, [\href{http://arxiv.org/abs/1401.2158}{{\tt arXiv:1401.2158}}].

\bibitem{Neill:2018wtk}
D.~Neill, A.~Papaefstathiou, W.~J. Waalewijn, and L.~Zoppi, {\it {Phenomenology with a recoil-free jet axis: TMD fragmentation and the jet shape}},  {\em JHEP} {\bf 01} (2019) 067, [\href{http://arxiv.org/abs/1810.12915}{{\tt arXiv:1810.12915}}].

\bibitem{Xiao:2024ffk}
Y.-X. Xiao, Y.~He, L.-G. Pang, H.~Zhang, and X.-N. Wang, {\it {Asymmetric jet shapes with two-dimensional jet tomography}},  {\em Phys. Rev. C} {\bf 109} (2024), no.~5 054906, [\href{http://arxiv.org/abs/2402.00264}{{\tt arXiv:2402.00264}}].

\bibitem{Pablos:2019ngg}
D.~Pablos, {\it {Jet Suppression From a Small to Intermediate to Large Radius}},  {\em Phys. Rev. Lett.} {\bf 124} (2020), no.~5 052301, [\href{http://arxiv.org/abs/1907.12301}{{\tt arXiv:1907.12301}}].

\bibitem{Milhano:2022kzx}
J.~G. Milhano and K.~Zapp, {\it {Improved background subtraction and a fresh look at jet sub-structure in JEWEL}},  {\em Eur. Phys. J. C} {\bf 82} (2022), no.~11 1010, [\href{http://arxiv.org/abs/2207.14814}{{\tt arXiv:2207.14814}}].

\bibitem{Zapp:2012ak}
K.~C. Zapp, F.~Krauss, and U.~A. Wiedemann, {\it {A perturbative framework for jet quenching}},  {\em JHEP} {\bf 03} (2013) 080, [\href{http://arxiv.org/abs/1212.1599}{{\tt arXiv:1212.1599}}].

\bibitem{Zapp:2013vla}
K.~C. Zapp, {\it {JEWEL 2.0.0: directions for use}},  {\em Eur. Phys. J. C} {\bf 74} (2014), no.~2 2762, [\href{http://arxiv.org/abs/1311.0048}{{\tt arXiv:1311.0048}}].

\end{thebibliography}\endgroup
\bibliographystyle{JHEP}

\end{document}